\documentclass[11pt, oneside]{article}   	

\usepackage{jheppub}

\usepackage{amsmath}
\usepackage{amssymb}
\usepackage{amsfonts}
\usepackage{amsthm}
\usepackage{amsbsy}
\usepackage{array}
\usepackage[abs]{overpic}
\usepackage{pict2e}

\usepackage[OT2,OT1]{fontenc}
\newcommand\cyr
{
\renewcommand\rmdefault{wncyr}
\renewcommand\sfdefault{wncyss}
\renewcommand\encodingdefault{OT2}
\normalfont
\selectfont
}
\DeclareTextFontCommand{\textcyr}{\cyr}
\def\cprime{\char"7E }
\def\cdprime{\char"7F }
\def\eoborotnoye{\char'013}

\usepackage{mathtools}
\usepackage[usenames,dvipsnames]{xcolor}

\usepackage{subcaption}

\usepackage{dsdshorthand}

\usepackage{graphicx}

\usepackage[vcentermath]{youngtab}
\newcommand{\myng}[1]{\,{\tiny\yng #1}\,}

\newcommand\wL{\mathbf{L}}

\renewcommand\vol{\mathop{\mathrm{vol}}}

\usepackage{tikzit}

\tikzstyle{small circle}=[shape=circle, fill=white, draw=black]
\tikzstyle{medium circle}=[fill=white, draw=black, shape=circle, minimum width=1cm, minimum height=1cm]
\tikzstyle{twoPoint}=[fill=white, draw=black, shape=circle, twopt]
\tikzstyle{threePoint}=[fill=white, draw=black, shape=circle, threept]
\tikzstyle{m vert rect}=[fill=white, draw=black, shape=rectangle, minimum height=2cm, minimum width=.5cm]
\tikzstyle{s box}=[fill=white, draw=black, shape=rectangle]
\tikzstyle{celestialThreePoint}=[fill=white, draw=red, shape=circle, threept]
\tikzstyle{large circle}=[fill=white, draw=black, shape=circle, minimum height=1.5cm, minimum width=1.5cm]
\tikzstyle{green circle of inversion}=[draw={black!20!green}, shape=circle, minimum size=2cm, dashed, tikzit draw={black!20!green}]
\tikzstyle{large green circle}=[draw={black!20!green}, shape=circle, dashed, minimum size=3cm, semithick]
\tikzstyle{small triangle}=[fill=white, draw=black, regular polygon, regular polygon sides=3, inner sep=1.5pt, rotate=90]
\tikzstyle{small blue triangle}=[fill=white, draw={black!20!blue}, regular polygon, regular polygon sides=3, inner sep=1.5pt, rotate=-90, tikzit draw=blue]
\tikzstyle{small green circle}=[draw={black!20!green}, shape=circle, tikzit draw={black!20!green}, dashed, minimum size=.5cm]
\tikzstyle{celestialTwoPoint}=[fill=red, draw=red, shape=circle, inner sep=1pt, minimum size=1pt]

\tikzstyle{arrow}=[->]
\tikzstyle{op}=[-, spinning]
\tikzstyle{dashedOp}=[-, scalar]
\tikzstyle{celestial}=[-, color=red, scalar, tikzit draw=red]
\tikzstyle{minkowski}=[-, spinning, color={black!20!blue}, tikzit draw=blue]

\input{harmonic_analysis.tikzdefs}

\tikzstyle{pole}=[fill=black, draw=black, shape=circle, inner sep=0pt, minimum size=2pt]

\tikzstyle{Arrow}=[->, draw=blue]
\tikzstyle{anti arrow}=[<-, draw=blue]
\tikzstyle{axis}=[->]
\tikzstyle{dashed_arrow}=[->, draw=red]
\tikzstyle{dashed_only}=[-, draw=red]
\tikzstyle{dashed_anti}=[<-, draw=red]

\tikzstyle{point}=[fill=black, draw=black, shape=circle, inner sep=0pt, minimum size=3pt]
\tikzstyle{lr_point}=[fill=red, draw=red, shape=circle, inner sep=0pt, minimum size=2pt]
\tikzstyle{sh_point}=[fill={rgb,255: red,128; green,128; blue,128}, draw={rgb,255: red,128; green,128; blue,128}, shape=circle, inner sep=0pt, minimum size=3pt]
\tikzstyle{sh_lr_point}=[fill={red!40}, draw={red!40}, shape=circle, inner sep=0pt, minimum size=2pt, tikzit fill={rgb,255: red,255; green,128; blue,0}, tikzit draw={rgb,255: red,255; green,128; blue,0}]

\tikzstyle{dashed_st}=[-, dashed]
\tikzstyle{arrow}=[->]
\tikzstyle{dashed_grey}=[-, draw=black, dotted]
\tikzstyle{blue_line}=[draw=blue, ->]
\tikzstyle{energy}=[draw={rgb,255: red,255; green,128; blue,0}, decoration={{snake,amplitude=1pt,segment length=6pt,post length=1pt}}, decorate, ->]
\tikzstyle{blue_line_0}=[-, draw=blue]
\tikzstyle{redline}=[-, draw=red]

\usetikzlibrary{math}
\usetikzlibrary{positioning,decorations.pathreplacing}

\newcommand{\tsym}{\cT}

\newcommand{\eps}{\epsilon}

\newcommand\ruone{{  \!\!\!\!\!\!\!\!\!\!\!\textrm{\normalsize \cyr \eoborotnoye}}}
\newcommand\rutwo{{  \!\!\!\!\!\!\!\!\!\!\!\textrm{\normalsize \cyr \cprime}}}
\newcommand\ruthree{{\!\!\!\!\!\!\!\!\!\!\!\!\textrm{\normalsize \cyr \cdprime}}}
\newcommand\rufour{{\!\!\!\!\!\!\!\!\!\!\!\!\textrm{\normalsize \cyr y}}}

\newcommand*\link[1]{\hspace*{0em plus 1fill}\makebox{#1}}

\definecolor{darkgreen}{rgb}{0, 0.6, 0}

\definecolor{energycolor}{RGB}{230,50,10}

\tikzset{
  energy/.style={->,
  energycolor,
  decoration={
      snake,
      amplitude=1pt,
      segment length=6pt,
      post length=1pt
    },
  decorate
  }
}

\makeatletter
\def\@fpheader{\ }
\makeatother

\title{Transverse spin in the light-ray OPE}
\author{Cyuan-Han Chang$^\ruone$, Murat Kolo\u{g}lu$^\rufour$, Petr Kravchuk$^\rutwo$, David Simmons-Duffin$^\ruone$, and Alexander Zhiboedov$^\ruthree$}
\affiliation{${}^\ruone$Walter Burke Institute for Theoretical Physics, Caltech, Pasadena, California 91125, USA \\
${}^\rufour$Mathematical Institute, University of Oxford, Oxford, OX2 6GG, UK\\
${}^\rutwo$School of Natural Sciences, Institute for Advanced Study, Princeton, New Jersey 08540, USA \\
${}^\ruthree$CERN, Theoretical Physics Department, 1211 Geneva 23, Switzerland
}

\date{}
\abstract{We study a product of null-integrated local operators $\mathcal{O}_1$ and $\mathcal{O}_2$ on the same null plane in a CFT. Such null-integrated operators transform like primaries in a fictitious $d-2$ dimensional CFT in the directions transverse to the null integrals. We give a complete description of the OPE in these transverse directions. The terms with low transverse spin are light-ray operators with spin $J_1+J_2-1$. The terms with higher transverse spin are primary descendants of light-ray operators with higher spins $J_1+J_2-1+n$, constructed using special conformally-invariant differential operators that appear precisely in the kinematics of the light-ray OPE. As an example, the OPE between average null energy operators contains light-ray operators with spin $3$ (as described by Hofman and Maldacena), but also novel terms with spin $5,7,9,$ etc.. These new terms are important for describing energy two-point correlators in non-rotationally-symmetric states, and for computing multi-point energy correlators. We check our formulas in a non-rotationally-symmetric energy correlator in $\cN=4$ SYM, finding perfect agreement.
}

\preprint{CALT-TH 2020-039 \\
\link{CERN-TH-2020-164}}

\begin{document}

\maketitle
\pagenumbering{roman}
\setcounter{page}{2}
\newpage
\pagenumbering{arabic}
\setcounter{page}{1}

\section{Introduction}

In Euclidean signature, the operator product expansion (OPE) gives a convergent expansion for correlation functions around coincident-point singularities. This expansion lets us formulate nonperturbative bootstrap conditions and perform myriad computations. Lorentzian correlators are in principle determined from Euclidean ones by analytic continuation. However, a given OPE may not commute with this continuation. Furthermore, Lorentzian signature allows for a much richer set of singularities than Euclidean signature \cite{Maldacena:2015iua}.  It is important to develop nonperturbative tools for understanding these singularities and efficiently computing Lorentzian observables.

The work \cite{Kologlu:2019mfz} introduced an intrinsically Lorentzian OPE for products of null-integrated operators on the same null plane.\footnote{See also \cite{Korchemsky:2019nzm} which derived the leading term in the expansion of a particular two-point event shape in $\cN=4$ SYM theory, and \cite{Dixon:2019uzg} which derived leading terms in collinear limits of energy-energy correlators in the setting of perturbative gauge theories.} This OPE can be applied to Lorentzian observables called ``event shapes,'' which measure the distribution of energy (and other quantities) in a collider-like experiment  \cite{Basham:1978zq,Basham:1978bw,Hofman:2008ar}:
\be\label{eq:firstee}
\<\Psi|\cE(\hat n_1)\cdots \cE(\hat n_k)|\Psi\>.
\ee
Here $|\Psi\>$ is a state, for example created by sending particles from past null infinity and letting them scatter. Each operator $\cE(\hat n_i)$ is a stress-tensor integrated along retarded time at future null infinity, at a fixed position $\hat n_i\in S^{d-2}$ on the celestial sphere. The $\cE(\hat n_i)$ act like calorimeters, measuring the energy flux at angle $\hat n_i$. We refer to the $\cE(\hat n_i)$ as ``detectors."

\begin{figure}[ht]
\begin{center}
\begin{overpic}[scale=0.166,unit=1mm]{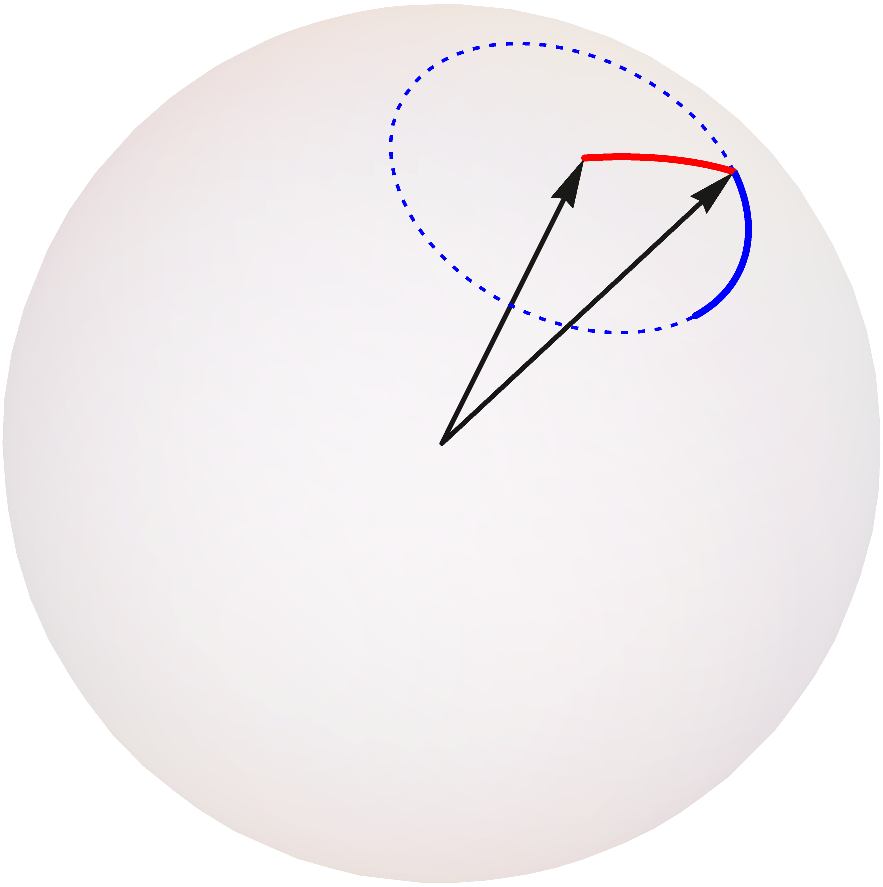}
    \put(44.5,35){{\color{blue} $\phi$}}
    \put(37.5,43.5){{\color{red} $\theta$}}
    \put(44.5,41.5){$\hat n_1$}
    \put(28,41.5){$\hat n_2$}
    \put(12,15){$S^{d-2}$}
\end{overpic}
\end{center}
\caption{The celestial sphere $S^{d-2}$ in a two-point event shape. The positions of the detectors are parametrized by $\hat n_1,\hat n_2\in S^{d-2}$. The light-ray OPE is an expansion in the angle $\theta$ between detectors  (solid red arc). Transverse spin $j$ is conjugate to the angle $\phi$ of one detector around the other on the celestial sphere (solid blue arc).}
\label{fig:celestialopeintro}
\end{figure}
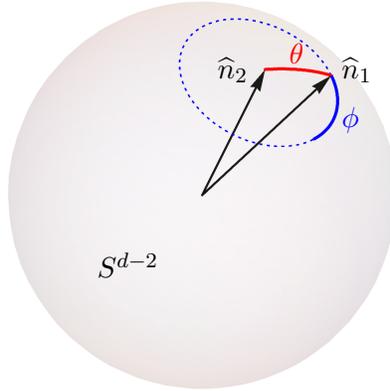

The OPE developed in \cite{Kologlu:2019mfz} gives a nonperturbative expansion for event shapes in the separation between a pair of detectors $1-\hat n_1 \. \hat n_2$. Specifically, \cite{Kologlu:2019mfz} gave a precise description of the low ``transverse spin" terms in this OPE.  Here, ``transverse spin" $j$ refers to spin on the celestial sphere: it is conjugate to rotation of the points $\hat n_1,\hat n_2$ around each other, see figure~\ref{fig:celestialopeintro}. The spin $J$, a different quantum number, is related to selection rules for the light-ray OPE.

The low transverse spin terms in the $\cE\x\cE$ OPE are given by spin $J=3$ light-ray operators \cite{Kologlu:2019mfz}, in accordance with an  earlier analysis of the light-ray OPE by Hofman and Maldacena \cite{Hofman:2008ar}. This is the complete OPE in 3d CFTs, where the transverse direction is 1-dimensional. Furthermore, low transverse spin terms are sufficient for studying two-point event shapes in rotationally-symmetric states in $d>3$ dimensions. This covers many of the cases studied in the literature, including simple energy two-point correlators in QCD \cite{Basham:1978bw,DelDuca:2016csb,Dixon:2018qgp,Luo:2019nig,Chen:2020vvp} and $\cN=4$ SYM \cite{Belitsky:2013xxa,Belitsky:2013ofa,Henn:2019gkr}.

In this work, we derive the remaining terms in the light-ray OPE, including arbitrary transverse spin. Higher transverse spin terms are important in $d>3$ dimensional theories when the initial state $|\Psi\>$ is not rotationally invariant or when other detectors are present. For example, in an energy three-point correlator $\<\Psi|\cE(\hat n_1)\cE(\hat n_2)\cE(\hat n_3)|\Psi\>$, transverse spin in the $\cE(\hat n_1)\x\cE(\hat n_2)$ OPE encodes dependence of the event shape on the direction of the tangent vector from $\hat n_1$ to $\hat n_2$ relative to the third direction $\hat n_3$.

\begin{figure}[t]
	\centering
	{\tikzset{tikzfig/.append style = {scale=1.5}}
	\begin{tikzpicture}
	\begin{pgfonlayer}{nodelayer}
		\node [style=none] (0) at (0, 5.3) {};
		\node [style=none] (1) at (0, -2) {};
		\node [style=none] (2) at (9, 0) {};
		\node [style=none] (3) at (-9, 0) {};
		\node [style=none] (4) at (0.5, 5.3) {$J$};
		\node [style=none] (5) at (9, -0.5) {$\Delta-\frac{d}{2}$};
		\node [style=none] at (0.3,1.2) {$3$};
		\node [style=none] at (0.3,2.0) {$5$};
		\node [style=none] at (0.3,2.8) {$7$};
		\node [style=none] at (0.3,3.6) {$9$};
		\node [style=none] at (0.4,4.4) {$11$};
		\node [style=none] at (5.8,5.3) {$j=0,$};
		\node [blue] at (6.82,5.3) {$2,$};
		\node [black!30!green] at (7.45,5.3) {$4,$};
		\node [style=none] at (8.2,5.3) {$0,$};
		\node [blue] at (8.82,5.3) {$2,$};
		\node [black!30!green] at (9.45,5.3) {$4\phantom{,}$};
		
	\end{pgfonlayer}
	\begin{pgfonlayer}{edgelayer}
		\draw [style=arrow] (3.center) to (2.center);
		\draw [style=arrow] (1.center) to (0.center);
		
		\draw [style={dashed_grey}] (-9,1.2) -- (9,1.2);
		\draw [style={dashed_grey}] (-9,2.0) -- (9,2.0);
		\draw [style={dashed_grey}] (-9,2.8) -- (9,2.8);
		\draw [style={dashed_grey}] (-9,3.6) -- (9,3.6);
		\draw [style={dashed_grey}] (-9,4.4) -- (9,4.4);

		\draw (0,1.2) -- (0.1,1.2);
		\draw (0,2.0) -- (0.1,2.0);
		\draw (0,2.8) -- (0.1,2.8);
		\draw (0,3.6) -- (0.1,3.6);
		\draw (0,4.4) -- (0.1,4.4);

		\draw  (6,5) to[out=-135,in=0,looseness=0.5] (0,0.5) to[out=180,in=-45,looseness=0.5] (-6,5);

		\draw [blue] (6.5,5) to[out=-135,in=30,looseness=0.5] (1.8,1);
		\draw [style={dashed_st},blue] (1.8,1) to[out=-150,in=20] (0.9,0.5);

		\draw [black!30!green] (7.2,5) to[out=-135,in=30,looseness=0.5] (2.6,1);
		\draw [style={dashed_st},black!30!green] (2.6,1) to[out=-150,in=20] (1.6,0.5);

		\draw (7.9,5) to[out=-135,in=30,looseness=0.5] (3.3,1);
		\draw [style={dashed_st}] (3.3,1) to[out=-150,in=20] (2.3,0.5);

		\draw [blue] (8.5,5) to[out=-135,in=30,looseness=0.5] (3.8,1);
		\draw [style={dashed_st},blue] (3.8,1) to[out=-150,in=20] (2.9,0.5);

		\draw [black!30!green] (9.2,5) to[out=-135,in=30,looseness=0.5] (4.6,1);
		\draw [style={dashed_st},black!30!green] (4.6,1) to[out=-150,in=20] (3.6,0.5);

		\draw [blue] (-6.5,5) to[out=-45,in=150,looseness=0.5] (-1.8,1);
		\draw [style={dashed_st},blue] (-1.8,1) to[out=-30,in=160] (-0.9,0.5);

		\draw [black!30!green] (-7.2,5) to[out=-45,in=150,looseness=0.5] (-2.6,1);
		\draw [style={dashed_st},black!30!green] (-2.6,1) to[out=-30,in=160] (-1.6,0.5);

		\draw (-7.9,5) to[out=-45,in=150,looseness=0.5] (-3.3,1);
		\draw [style={dashed_st}] (-3.3,1) to[out=-30,in=160] (-2.3,0.5);

		\draw [blue] (-8.5,5) to[out=-45,in=150,looseness=0.5] (-3.8,1);
		\draw [style={dashed_st},blue] (-3.8,1) to[out=-30,in=160] (-2.9,0.5);

		\draw [black!30!green] (-9.2,5) to[out=-45,in=150,looseness=0.5] (-4.6,1);
		\draw [style={dashed_st},black!30!green] (-4.6,1) to[out=-30,in=160] (-3.6,0.5);

		\node [style=point] (13) at (1.63, 1.2) {};
		\node [style=point,blue] (13) at (2.14, 1.2) {};
		\node [style=point,black!30!green] (13) at (2.91, 1.2) {};
		
		\node [style=point] (13) at (3.6, 1.2) {};
		\node [style=point,blue] (13) at (4.12, 1.2) {};
		\node [style=point,black!30!green] (13) at (4.91, 1.2) {};

		\node [style=point,black!30!green] (13) at (3.92, 2.0) {};
		\node [style=point,black!30!green] (13) at (5.92, 2.0) {};

		\node [style=point,black!30!green] (13) at (4.87, 2.8) {};
		\node [style=point,black!30!green] (13) at (6.87, 2.8) {};

		\node [style=point,black!30!green] (13) at (5.72, 3.6) {};
		\node [style=point,black!30!green] (13) at (7.72, 3.6) {};

		\node [style=point,black!30!green] (13) at (6.59, 4.4) {};
		\node [style=point,black!30!green] (13) at (8.59, 4.4) {};
		
	\end{pgfonlayer}
\end{tikzpicture}
	}
	\caption{Chew-Frautschi plot of light-ray operators in the $\cE \x \cE$ OPE in a hypothetical CFT. We show Regge trajectories of even signature operators with transverse spins $j=0$ (black curves), $j=2$ (blue curves), and $j=4$ (green curves). Light-ray operators that appear in the $\cE\x\cE$ OPE are marked with dots. They include operators with spin $J=3$ and transverse spins $j=0,2,4$. (These are the ``low transverse spin" terms described in \cite{Kologlu:2019mfz}.) In addition, there are primary descendants of light-ray operators with transverse spin $j=4$ and spin $J=5,7,9,\dots$.}
	\label{fig:CFplot}
\end{figure}
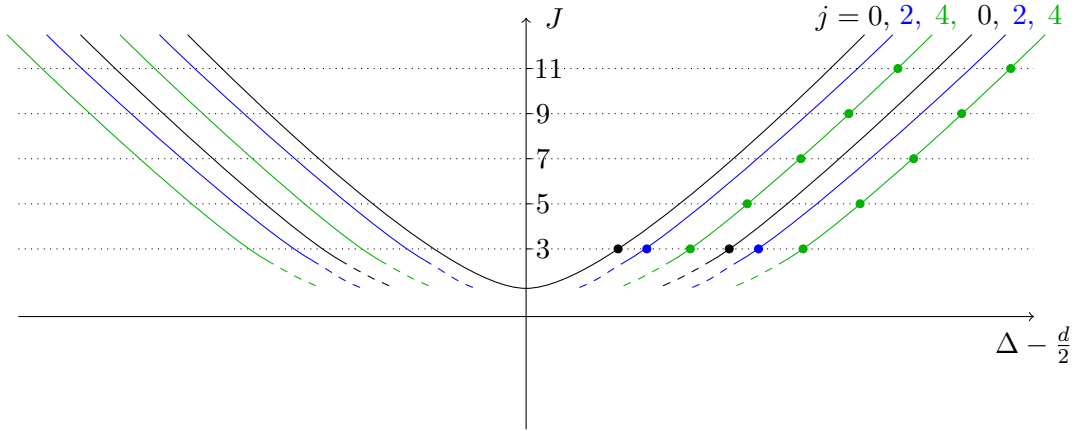

In seeking the higher transverse spin terms in the $\cE\x\cE$ OPE, we initially encounter a puzzle: there are no  spin $J=3$ primary light-ray operators of the type defined in \cite{Kravchuk:2018htv} that can do the job. The resolution is that higher transverse spin terms are ``primary descendants." They are given by the action of special conformally-invariant differential operators $\cD_{2n}$ on light-ray operators with higher spins $J=3+2n$.\footnote{In applications to event shapes such as~\eqref{eq:firstee}, where the light-ray operators appearing in the OPE are inserted at spatial infinity (and extend along the future null infinity), the operators $\cD_{2n}$ are actually polynomials in the special conformal generators $K_\mu$. This happens because an inversion which sends a finite point to infinity maps the translation generators $P_\mu$ (which act by derivatives) to $K_\mu$.} These conformally-invariant differential operators convert spin $J$ into transverse spin $j$. They are well-defined only when acting on objects with special quantum numbers --- precisely the quantum numbers that arise in the light-ray OPE. We find that the $\cE\x\cE$ OPE takes the schematic form 
\be
\label{eq:schematicEE}
	\cE\x\cE=\sum_i\p{\mathbb{O}^+_{i,J=3,j=0}+ \mathbb{O}^+_{i,J=3,j=2}+ \mathbb{O}^+_{i,J=3,j=4}}+\sum_{n,i} \cD_{2n}\mathbb{O}^{+}_{i,J=3+2n,j=4}.
\ee
Here, $\mathbb{O}_{i,J,j}^+$ are light-ray operators on the $i$-th Regge trajectory with spin $J$ and transverse spin $j$. The special differential operators $\cD_{2n}$ act on $\mathbb{O}^{+}_{i,J=3+2n,j=4}$ to give higher transverse spin terms. A Chew-Frautschi plot of the light-ray operators in (\ref{eq:schematicEE}) is depicted in figure~\ref{fig:CFplot}.

The higher transverse spin terms are new ingredients in the light-ray OPE. However, it is interesting that they do not require us to go outside the space of light-ray operators defined in \cite{Kravchuk:2018htv}. Instead, they are hidden in an interesting way inside the usual space of light-ray operators --- at higher values of $J$.  Along the way to understanding higher transverse spin terms in the light-ray OPE, we will also find a much simpler derivation of the original light-ray OPE from \cite{Kologlu:2019mfz}.

We begin in section~\ref{sec:eventshapeintro} with an introduction to the kinematics of transverse spin, focusing on the example of energy correlators in the process $e^+e^-\to \mathrm{hadrons}$. In section~\ref{sec:differential operators}, we provide a more detailed introduction to the underlying representation theory and the special conformally-invariant differential operators that raise transverse spin. In section~\ref{sec:scalar OPE}, we give a new derivation of the light-ray OPE, using null-integrated scalars as an example. In section~\ref{sec:general OPE}, we generalize this discussion to the OPE of null integrals of arbitrary local operators. In section~\ref{sec:examples}, we rederive the scalar light-ray OPE using the light-ray OPE formula for general operators, and give the light-ray operators that appear in the light-ray OPE of two charge detectors and the light-ray OPE of two energy detectors. In section~\ref{sec:N4example}, we check our formulas for an event shape in $\cN=4$ SYM with a non-rotationally-symmetric final state. We conclude in section~\ref{sec:discussion} with discussion and future directions.

\section{Event shapes, OPEs, and transverse spin}
\label{sec:eventshapeintro}

Event shapes describe patterns of excitations at future null infinity. Perhaps the most important examples are energy correlators
\be
\<\Psi|\cE(\hat n_1)\cdots \cE(\hat n_n)|\Psi\>,
\ee 
which measure the distribution of energy at future null infinity in the state $|\Psi\>$. Here, $\cE(\hat n_i)$ are calorimeters inserted at future null infinity in the direction $\hat n_i \in S^{d-2}$ on the celestial sphere. (We define them more precisely below.)

Suppose that $|\Psi\>$ is created by a linear combination of local operators $\cO_i$ acting on the vacuum,
\be
|\Psi\> &= \int d^d x \sum_i f_i(x) \cO_i(x)|\Omega\>.
\ee
The energy correlator is
\be
 \<\Psi|\cE(\hat n_1)\cdots \cE(\hat n_n)|\Psi\>
	& = \int d^dx d^dx' \sum_{i,j}f^*_i(x)f_j(x')\<\Omega|\cO^\dagger_i(x)\cE(\hat n_1)\cdots \cE(\hat n_n)\cO_j(x')|\Omega\>\nn\\
	& = \int \frac{d^dp}{(2\pi)^d} \sum_{i,j}\tl f_j^*(p)\tl f_i(p)\<\cO_i(p)|\cE(\hat n_1)\cdots \cE(\hat n_n)|\cO_j(p)\>,
	\label{eq:moreconcreteeventshape}
\ee
where
\be
	|\cO_i(p)\>=\int d^dx e^{ipx}\cO_i(x)|\Omega\>,
\ee
and $\tl f_i(p)$ is the Fourier transform of $f_i(x)$. The matrix element on the last line of (\ref{eq:moreconcreteeventshape}) is defined by stripping off a momentum-conserving delta function:
\be
	\<\cO_i(q)|\cE(\hat n_1)\cdots \cE(\hat n_n)|\cO_j(p)\>=\<\cO_i(p)|\cE(\hat n_1)\cdots \cE(\hat n_n)|\cO_j(p)\>(2\pi)^d\de^d(p-q).
\ee
The delta-function $\de^d(p-q)$ appears because the detectors $\cE(\hat n_i)$ are translation-invariant: translations do not alter the direction in which the excitations exit the system at future null infinity. Thus, we are naturally led to study expectation values of energy detectors in momentum eigenstates,
\be
	\<\cO_i(p)|\cE(\hat n_1)\cdots \cE(\hat n_n)|\cO_j(p)\>.
\ee

\subsection{Example: $e^+e^-\to \mathrm{hadrons}$}

As a concrete example, consider the Standard Model process $e^+e^- \to \g^* \to  \mathrm{hadrons}$, treated to leading order in the fine structure constant, but to all orders in the QCD coupling.\footnote{In practice, one considers the processes at high energies where hardonization corrections can be argued to be relatively small \cite{Webber:1994cp,Korchemsky:1999kt}, and computes the above event shape using perturbative QCD. Such calculations are one of the ways used to measure the strong coupling constant $\a_s$, see \cite{Verbytskyi:2019ikn} and references therein.} The relevant energy correlator is an expectation value in a state created by the electromagnetic current $J_\mu$,
\be
\label{eq:exptval}
\e^{*\nu} \< J_\nu(p)|\cE(\hat n_1)\cdots \cE(\hat n_n)|J_\mu(p)\>\e^\mu,
\ee
Here, $\e_\mu$ is a polarization vector for an off-shell photon that depends on the beam direction and helicities of the incoming $e^+e^-$ pair. (We imagine that the collision occurs at high energies, so the electrons can be treated as massless.) Suppose particle $1$ moves along the $\vec e_z$ direction and particle $2$ moves along the $-\vec e_z$ direction, with total momentum $p_1+p_2=p=(E,0,0,0)$. We henceforth set $E=1$. If the incoming helicities are $1^+2^-$, the corresponding polarization vector $\e^\mu$ is
\be
\e^\mu = \l_2 \s^\mu \bar \l_1 = (0,1,i,0).
\ee
More generally, for helicities $1^\pm 2^\mp$, we have
\be
\e^\mu = \e_\pm^\mu \equiv (0,1,\pm i, 0).
\ee

We can think of the ket and bra together in (\ref{eq:exptval}) as giving an (unnormalized) density matrix. For example, for helicities $1^+2^-$, we have the pure state
\be
\rho_{1^+2^-} &= |J_\mu(p)\>\<J_\nu(p)|\e_+^\mu \e_-^\nu,
\ee
where we used $\e_+^{*\mu} = \e_-^\mu$.
More generally, we may wish to study a mixed state, for example by averaging over incoming helicities
\be
\rho_{1 2}^\mathrm{av} &= |J_\mu(p)\>\<J_\nu(p)|\.\frac 1 2(\e_+^\mu \e_-^\nu+\e_-^\mu \e_+^\nu).
\ee
It is common to additionally average over the beam direction, replacing $\e_+^\mu \e_-^\nu \to \frac 2 3 (\eta^{\mu\nu}-\frac{p^\mu p^\nu}{p^2})$. However, this discards valuable information, as we explain below.

\subsubsection{Symmetries and the density matrix}

Let us understand how symmetries constrain energy correlators in the process $e^+e^-\to \mathrm{hadrons}$. Along the way, we will introduce the notion of light-ray operators with nonzero transverse spin and understand how they appear in event shapes.  Instead of QCD, we will work in a general 4-dimensional CFT. However, much of our analysis will not depend on conformal symmetry. We point out when conformal symmetry is used below.

We concentrate on a two-point energy correlator in a density matrix where we average over helicities (but not the beam direction)
\be
\Tr(\rho_{1 2}^\mathrm{av} \cE(\hat n_1) \cE(\hat n_2)) = \< J_\nu(p)|\cE(\hat n_1)\cE(\hat n_2)|J_\mu(p)\>\.\frac 1 2(\e_+^\mu \e_-^\nu+\e_-^\mu \e_+^\nu).
\ee
Let us separate the tensor $\frac 1 2(\e_+^\mu \e_-^\nu+\e_-^\mu \e_+^\nu)$ into irreducible components under the rotation group $\SO(3)$ that fixes $p$. Focusing on spatial components $i,j=1,2,3$, we have the traceless-symmetric and trace parts
\be
\label{eq:splitparts}
\frac 1 2(\e_+^i \e_-^j+\e_-^i \e_+^j) &= \p{\frac 1 2(\e_+^i \e_-^j+\e_-^i \e_+^j) - \frac{2}{3} \de^{ij}} + \frac 2 3 \de^{ij}.
\ee
The traceless-symmetric part can be written as a sum of products of null vectors $\vec q \in \C^3$:\footnote{This is a general fact about traceless symmetric tensors. An example decomposition in this case is
\be
\frac 1 2(\e_+^i \e_-^j+\e_-^i \e_+^j) - \frac{2}{3} \de^{ij} &= \sum_{\vec q \in Q} q^i q^j
\ee
where $Q$ contains four null vectors,
\be
Q &= \left\{\frac{1}{\sqrt 6}(1,0,\pm i),\frac{1}{\sqrt 6}(0,1,\pm i)\right\}.
\ee
}
\be
\label{eq:sumofnullstuff}
\frac 1 2(\e_+^i \e_-^j+\e_-^i \e_+^j) - \frac{2}{3} \de^{ij} &= \sum_{\vec q} q^i q^j . 
\ee
Plugging this in, we have
\be
\label{eq:sumoverqs}
\Tr(\rho_{1 2}^\mathrm{av} \cE(\hat n_1) \cE(\hat n_2)) &= \sum_{q} \< J_\nu(p)|\cE(\hat n_1)\cE(\hat n_2)|J_\mu(p)\> q^\mu q^\nu + \textrm{trace part}, 
\ee
where $q=(0,\vec q)$ and ``trace part" refers to the contribution of the second term in (\ref{eq:splitparts}). Note that each $q$ appearing in the sum is null and orthogonal to $p$:
\be
\label{eq:conditionsonq}
q^2 = q\. p = 0.
\ee
The form (\ref{eq:sumoverqs}) makes it easy to analyze the constraints of symmetries, since now we have only a single vector $q$ instead of a tensor. 

\subsubsection{Symmetries and detectors}

Next we need a more precise definition of the detector $\cE(\hat n)$. It can be expressed as an integral of the stress-tensor $T^{\mu\nu}$ over future null infinity. To state this more concretely, we can make a conformal transformation that maps future null infinity to the plane $x^-=0$. We then have
\be\label{eq:detector null integral}
	\cE(\hat n) \to 2\int dx^+ T_{++}(x^-=0, x^+, \vec x),
\ee
where $\vec x\in \R^{d-2}$ is a function of $\hat n$. As explained in detail in~\cite{Kravchuk:2018htv}, such null-integrated stress tensors can be interpreted in terms of the ``light-transform'' of $T$, denoted by
\be
	\wL[T](x,z),
\ee
where $x$ is a space-time position marking the starting point of the null integral, and $z$ is a future-directed null vector. The definition of $\wL$ is given in \eqref{eq:lighttransform} below. This description is useful because $\wL[T](x,z)$ transforms like a primary operator at $x$. If we send $x$ to past null infinity, $\wL[T](x,z)$ becomes the null integral in~\eqref{eq:detector null integral}. If we instead send $x$ to spatial infinity, then $\wL[T](x,z)$ becomes directly related to $\cE(\hat n)$, 
\be
	\cE(\hat n) = 2\wL[T](\oo, z)
\ee
where $z=(1,\hat n)$.

In this work, we derive a nonperturbative OPE between light-ray operators that takes the schematic form
\be
\label{eq:schematicope}
\wL[T](x,z_1) \wL[T](x,z_2) &= \sum_{j=0,2,\dots} \sum_i \cC_{\De_i-1,j}(z_1,z_2,\ptl_z,\ptl_w) \mathbb{D}_{\De_i,j}(x,z,w).
\ee
Here, $\cC_{\de,j}(z_1,z_2,\ptl_z,\ptl_w)$ is a differential operator that is fixed by Lorentz symmetry, and $\mathbb{D}_{\De_i,j}(x,z,w)$ are light-ray operators that we characterize in more detail shortly. For now, the only information about $\mathbb{D}_{\De_i,j}(x,z,w)$ that we need are its Lorentz transformation properties. It is a homogeneous function of null vectors $z \in \R^{d-1,1},w \in \C^{d-1,1}$ with homogeneities
\be
\mathbb{D}_{\De_i,j}(x,\a z, \b w) = \a^{1-\De_i} \b^j \mathbb{D}_{\De_i,j}(x, z, w).
\ee
Furthermore, $z,w$ are subject to the constraints
\be
\label{eq:constraintsonzandw}
z\.z=w\.w=w\.z = 0,
\ee
and the gauge redundancy
\be
\label{eq:gaugeredundnacy}
w &\sim w + \l z.
\ee

We can interpret $z_1,z_2$, and $z$ as embedding space coordinates~\cite{Costa:2011mg} on the celestial sphere $S^{d-2}$, and $w$ as an embedding space polarization vector on the celestial sphere. The quantum number $j$ labels spin on the celestial sphere, which we call ``transverse spin." Further, $\De_i-1$ is a dimension on the celestial sphere, and (\ref{eq:schematicope}) takes the form of an OPE in a fictitious $d{-}2$-dimensional Euclidean CFT.

The dependence of $\cE(\hat n)$,  $\mathbb{D}_{\De,j}$, and $\cC_{\de,j}$ on the coordinates $z_1,z_2,z,w$ relies only on Lorentz symmetry --- not full conformal symmetry.\footnote{Note that our use of $\De$ to denote one of the quantum numbers of $\mathbb{D}_{\De,j}$ might suggest that we are relying on conformal symmetry. We are not: here $1-\De$ is the Lorentz spin of $\mathbb{D}_{\De,j}$. The reason for this convention will become clear soon.}  We will use conformal symmetry later to derive (\ref{eq:schematicope}) --- in particular to constrain which $\mathbb{D}_{\De_i,j}$ can appear. However, it is possible that a similar OPE exists in non-conformal theories, and our analysis of the contributions of light-ray operators $\mathbb{D}_{\De_i,j}$ in this section will also apply in that case. 

\subsubsection{Symmetries and matrix elements}

Using the OPE in (\ref{eq:sumoverqs}), it suffices to compute matrix elements 
\be
\label{eq:jmatrixelements}
\< J_\nu(p)|\mathbb{D}_{\De_i,j}(\oo,z,w)|J_\mu(p)\> q^\mu q^\nu.
\ee
In fact, let us analyze a more general matrix element where the density matrix has $\SO(d-1)$ spin $l$:
\be
\cM(z,w;p,q) &= \< \cO_{\mu_1\cdots\mu_k} (p)|\mathbb{D}_{\De_i,j}(\oo,z,w)|\cO'_{\mu_{k+1}\cdots\mu_l}(p)\> q^{\mu_1} \cdots q^{\mu_l}.
\ee
The virtue of having classified the density matrix and light-ray operators into irreducible components is that $\cM$ is fixed by symmetry. The argument is as follows. Because of the gauge redundancy (\ref{eq:gaugeredundnacy}), $w$ can only appear in the gauge-invariant combination 
\be\label{eq:commutatordefinition}
[z,w]^{\mu\nu} &= z^\mu w^\nu - w^\mu z^\nu.
\ee
Because of (\ref{eq:constraintsonzandw}), the only antisymmetric tensor we can contract this with is $[p,q]$. Since $w$ must appear with homogeneity $j$, we have
\be
\cM(z,w;p,q) &= ([z,w]\.[p,q])^j \x \textrm{something}.
\ee
Finally, homogeneity in $q$ and $z$ fix the rest of the matrix element up to an overall coefficient:
\be
\label{eq:thematrixelement}
\cM(z,w;p,q) &\propto ([z,w]\.[p,q])^j (-z\.q)^{l-j} (-p\.z)^{1-\De_i-l}.
\ee

\subsubsection{A selection rule}

The result (\ref{eq:thematrixelement}) manifests a selection rule: Light-ray operators with transverse spin $j$ only have nonzero expectation values in density matrices with $\SO(d-1)$ spin $l$ at least $j$. In other words, for $\cM$ to be nonvanishing, we must have
\be
\label{eq:selectionrule}
j \leq l.
\ee
For example, if we average over the beam direction so that only density matrices with $l=0$ appear, we discard information about light-ray operators with nonzero transverse spin.

Because of the selection rule (\ref{eq:selectionrule}), the low transverse spin terms in the $\cE\x\cE$ OPE (\ref{eq:schematicEE}), which have $j=0,2,4$, are sufficient for computing two-point event shapes in density matrices with $\SO(d-1)$ spin $4$ or less. This includes the scalar density matrix studied in~\cite{Dixon:2018qgp,Luo:2019nig,Belitsky:2013ofa,Henn:2019gkr}. Higher transverse spin terms are important for density matrices with higher spin and in multi-point event shapes.

\subsubsection{The form of the light-ray OPE}

The differential operator $\cC_{\de,j}$ appearing in (\ref{eq:schematicope}) has an expansion in the angle $\th$ between detectors, or equivalently in small $-2z_1\. z_2\approx \th^2$.
To leading order in this expansion, $\cC_{\de,j}$ acts as
\be
\cC_{\de,j}(z_1,z_2,\ptl_z,\ptl_w) ([z,w]\.[p,q])^j f(z) &= (-2z_1\.z_2)^{\frac{\de-j-6}{2}} (-[z_1,z_2]\.[p,q])^j f(z_2) + \dots,
\ee
where $f(z)$ is any function of $z$ with the correct homogeneity, and ``$\dots$" indicates higher-order terms in $\th$. Applying this to (\ref{eq:thematrixelement}), we find
\be
\label{eq:expansionoftheform}
&\< \cO_{\mu_1\cdots\mu_k} (p)|\wL[T](\oo,z_1)\wL[T](\oo,z_2)|\cO'_{\mu_{k+1}\cdots\mu_l}(p)\> q^{\mu_1} \cdots q^{\mu_l}
\nn\\
&= \sum_{j,i}\l_{i,j}\p{(-2z_1\.z_2)^{\frac{\De_i-j-7}{2}}(-[z_1,z_2]\.[p,q])^j (-z_2\.q)^{l-j}(-z_2\.p)^{1-\De_i-l} + \dots},
\ee
where $\l_{i,j}$ are OPE coefficients that are not fixed by Lorentz symmetry. The ``$\dots$" on the right-hand side are fixed by symmetry and re-sum into a celestial block \cite{Kologlu:2019mfz}.\footnote{We give some example calculations of celestial blocks in section~\ref{sec:celestialblocks}.} We have written only the leading term of the celestial block for simplicity.

Let us specialize further to the kinematics of interest. Note that
\be
[z_1,z_2]\.[p,q] &= 2\vec n_{12}\.\vec q,
\ee
where $z_i=(1,\hat n_i)$, $\vec n_{12}=\hat n_1-\hat n_2$, and $\vec q$ are the spatial components of $q$.  From (\ref{eq:expansionoftheform}), we can compute the leading terms in the OPE for each possible value of $l$ and $j$:
\be
\label{eq:j2l2}
& |\vec n_{12}|^{\De_i-9} \p{(\vec \e_+\.\vec n_{12})(\vec \e_-\.\vec n_{12})-\frac 2 3 \vec n_{12}^2} + \dots && (j=2,l=2),\nn\\
&|\vec n_{12}|^{\De_i-7} \p{(\vec \e_+\.\hat n_2)(\vec \e_-\.\hat n_2) - \frac 2 3}+\dots && (j=0,l=2), \nn\\
&|\vec n_{12}|^{\De_i-7} + \dots && (j=0,l=0).
\ee
where we used (\ref{eq:sumofnullstuff}) to replace the sum over $q$.
In each case, the ``$\dots$" are fixed by symmetry and resum into a celestial block. The fact that light-ray operators with $j=0$ appear in two different ways reflects the fact that matrix elements of a given light-ray operator can admit multiple three-point structures, each of which comes with its own OPE coefficient. Light-ray operators with $j=1$ do not appear in the $\cE\x\cE$ OPE due to permutation symmetry under $\hat n_1\leftrightarrow \hat n_2$ (which follows from the fact that energy detectors commute).

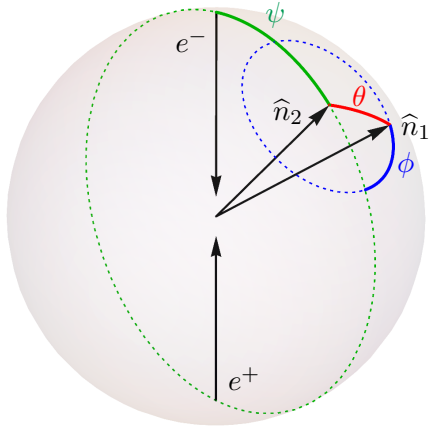
\begin{figure}
\begin{center}
\begin{overpic}[scale=0.166,unit=1mm]{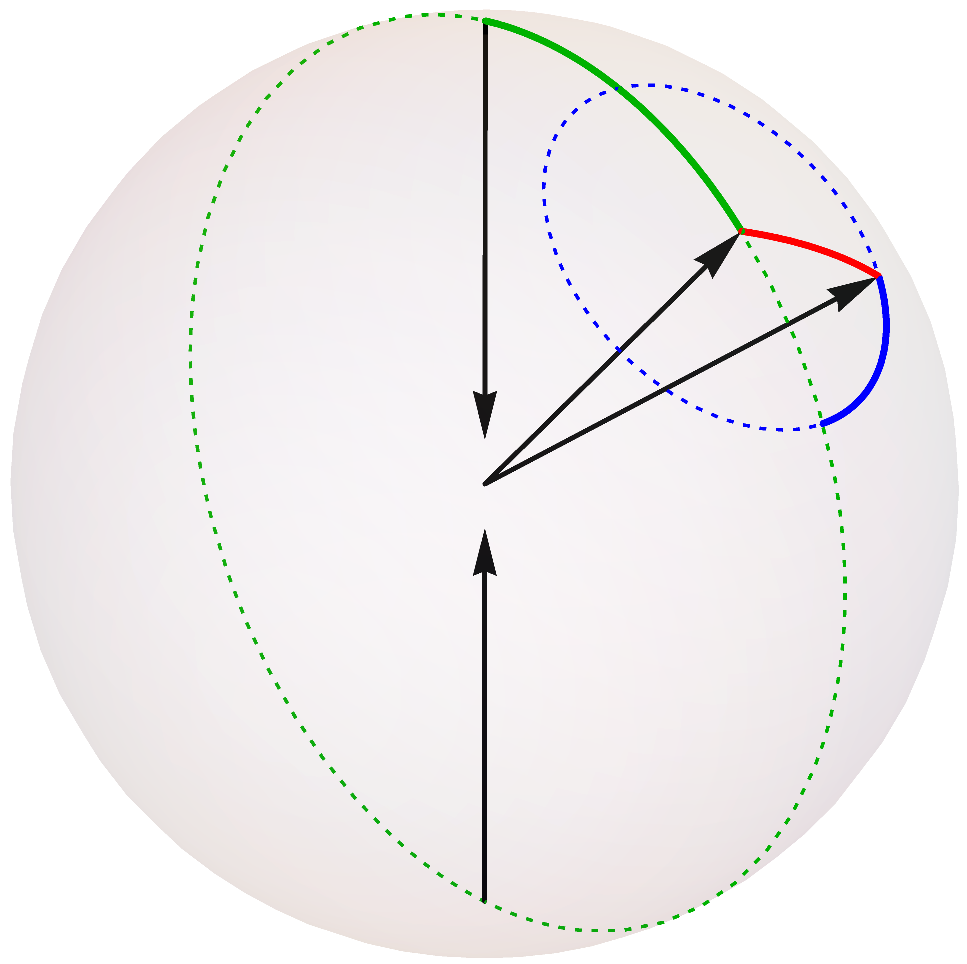}
    \put(35,54){{\color{ForestGreen} $\psi$}}
    \put(52.5,34){{\color{blue} $\phi$}}
    \put(46.5,42.8){{\color{red} $\theta$}}
    \put(30,5){$e^+$}
    \put(23,50){$e^-$}
    \put(53,39){$\hat n_1$}
    \put(36,41){$\hat n_2$}
\end{overpic}
\end{center}
\caption{Kinematics of a two-point energy correlator in the process $e^+e^-\to\mathrm{hadrons}$.  The picture shows the spatial geometry in the center of mass frame; time is suppressed. Particles $e^+$ and $e^-$ propagate in along the $\pm \vec e_z$ directions, and we measure energy flux in the directions $\hat n_1,\hat n_2$. (The correlator is invariant under swapping $e^+$ and $e^-$.) $\psi$ (solid green arc) is the angle between one of the detectors $\hat n_2$ and the beam direction, $\theta$ (solid red arc) is the angle between detectors, and $\f$ (solid blue arc) parametrizes the angle of $\hat n_1$ around $\hat n_2$ on the celestial sphere.}
\label{fig:celestialangles}
\end{figure}

To be completely explicit, let us parametrize the vectors as
\be
\hat n_2 &= R_{zx}(\psi)(0,0,1) \nn\\
\hat n_1 &= R_{zx}(\psi)(\sin\th\cos\f,\sin\th\sin\f,\cos\th).
\ee
where $R_{zx}(\psi)$ is a rotation by $\psi$ in the $zx$ plane. Here, $\psi$ is the angle between the (nearly coincident) detectors and the beam, $\th$ is the angle between detectors, and $\f$ represents a rotation of the two detectors around each other on the celestial sphere, see figure~\ref{fig:celestialangles}. The expressions (\ref{eq:j2l2}) become
\be
\label{eq:answerthingy}
&\th^{\De_i-7}\p{\frac 1 3 - \cos^2 \f \sin^2 \psi}  + \dots && (j=2,l=2),\nn\\
&\th^{\De_i-7}\p{\sin^2\psi - \frac 2 3} + \dots && (j=0,l=2), \nn\\
&\th^{\De_i-7} + \dots && (j=0,l=0).
\ee
where ``$\dots$" are higher order terms in $\th$. As a check on the first line of (\ref{eq:answerthingy}), note that if we set $\psi=0$ or $\pi$, so that the detectors are both approaching the beam direction, the dependence on $\f$ goes away, reflecting the fact that our density matrix is invariant under rotations around the beam direction.

Typically both theoretical and experimental analysis of the $e^+e^-\to \mathrm{hadrons}$ process focus on observables averaged over the beam direction. As explained above, this amounts to throwing
away the contribution of light-ray operators with non-zero transverse spin, which contain extra information and can provide further nontrivial tests of QCD.  Event shapes that are not averaged over the beam direction, so-called {\it oriented event shapes}, were studied experimentally at LEP by the DELPHI \cite{Abreu:2000ck} and OPAL \cite{Abbiendi:1998at} collaborations. For a recent discussion of oriented event shapes in QCD, see \cite{Mateu:2013gya}.

\subsection{A transverse spin puzzle}

So far, we have introduced the kinematics of the light-ray OPE inside a two-point event shape. The next question is: what are the operators $\mathbb{D}_{\De_i,j}$? In particular, what are the corresponding values of $\De_i$ and OPE coefficients?
In \cite{Kologlu:2019mfz}, we derived the low transverse spin terms in the light-ray OPE. For simplicity, consider a product of light-transformed scalars $\wL[\f_1]\wL[\f_2]$. The result of \cite{Kologlu:2019mfz} is\footnote{As explained in \cite{Kologlu:2019bco}, this product is only well-defined for a theory with a sufficiently low Regge intercept. Here we assume this is the case for the sake of illustration.}
\be\label{eq:scalarOPEfinalintro}
	\wL[\f_1](x,z_1)\wL[\f_2](x,z_2) &=\pi i\sum_i \cC_{\De_i-1,0}(z_1,z_2,\ptl_{z_2})\mathbb{O}^+_{\De_i,J=-1}(x,z_2)\nn\\
	&\quad+\textrm{higher transverse spin}.
\ee
The operators $\mathbb{O}_{\De_i,J=-1}^+(x,z)$ are analytic continuations of null-integrated operators $\wL[\cO_i]$ to spin $J=-1$, where $\cO_i$ appears in the $\f_1\x\f_2$ OPE. The quantum number $\De_i$ becomes the analytic continuation of scaling dimensions of $\cO_i$. The operators $\mathbb{O}_{\De_i,J=-1}^+(x,z)$ have transverse spin $j=0$ --- in particular they depend only on $x$ and $z$ and not on an additional polarization vector $w$. The reason is that only traceless symmetric tensors $\cO_i$ appear in the OPE of scalar operators. The light-transform $\wL[\cO_i]$ of a traceless symmetric tensor has vanishing transverse spin, and thus so do its analytic continuations. 

Note that in $d=3$ all (bosonic) operators are traceless-symmetric tensors and correspondingly there is no transverse spin in $d-2=1$ dimension. The higher transverse spin terms in~\eqref{eq:scalarOPEfinalintro} are absent in this case. However, in $d>3$ the transverse spin $j$ can be non-trivial and we expect infinitely many higher transverse spin terms to appear in~\eqref{eq:scalarOPEfinalintro}, since there is no reason to expect the event shapes to be independent of the angle $\phi$ discussed above.\footnote{The more accurate statement is that the first line of~\eqref{eq:scalarOPEfinalintro} already contains higher-transverse spin terms: they are generated by the operators $\cC_{\de,0}$. However, since the operators $\cC_{\de,0}$ are fixed by Lorentz symmetry, these contributions are determined in terms of $j=0$ contributions and therefore the event shapes would still be over-constrained if there were no additional contributions to~\eqref{eq:scalarOPEfinalintro}. For example, one would be able to write a differential equation in $z_1,z_2$ that the product $\wL[\f_1](x,z_1)\wL[\f_2](x,z_2)$ would have to satisfy. It can be checked that this differential equation is incompatible with the leading term of the $\wL[\f_2]\x\wL[\f_3]$ OPE in $\<\Psi|\wL[\f_1]\wL[\f_2]\wL[\f_3]|\Psi\>$.}

It is natural to expect that higher transverse-spin terms in (\ref{eq:scalarOPEfinalintro}) should also be related to the $\f_1\x\f_2$ OPE.  However, this presents a puzzle: All the primary light-ray operators $\mathbb{O}_{\De_i,J}(x,z)$ built from $\f_1\x\f_2$ using the construction of \cite{Kravchuk:2018htv} have vanishing transverse spin. How can we build light-ray operators with nonzero transverse spin to play the role of $\mathbb{D}_{\De_i,j}$ with $j>0$? It turns out that the $\mathbb{D}_{\De_i,j}$ are {\it primary descendants} of $\mathbb{O}_{\De_i,J}$ --- i.e.\ carefully chosen derivatives of $\mathbb{O}_{\De_i,J}$ that nonetheless transform like conformal primaries. In the next section, we explain how such primary descendants arise.

\section{Building transverse spin with differential operators}
\label{sec:differential operators}
\subsection{Local operators, light transforms, and shortening conditions}

Consider a local operator $\cO^{\mu_1\cdots\mu_J}(x)$ with dimension $\De$ and spin $J$. Throughout this work, we will use index-free notation, where we contract spin indices with an auxiliary null polarization vector $z\in \R^{d-1,1}$:
\be
\cO(x,z) &= \cO^{\mu_1\cdots\mu_J}(x)z_{\mu_1\cdots\mu_J},\qquad (z^2=0).
\ee
By construction, $\cO(x,z)$ is a homogeneous polynomial of degree $J$ in $z$.  Under a conformal transformation $U$, we have \be
\label{eq:conformaltransformationprop}
U\cO(x,z) U^{-1} &= \Omega(x')^\De \cO(x',R(x')z),
\ee
where $\Omega(x')$ and $R(x')$ are the local rescaling and rotation associated to $U$.

Index-free notation is more than a convenience. It allows us to describe a wider class of conformal representations than those associated to local operators. 
As an example, consider the light-transform 
\be
\label{eq:lighttransform}
\wL[\cO](x,z) &= \int_{-\oo}^\oo d\a (-\a)^{-\De-J} \cO\p{x-\frac{z}{\a},z},
\ee
which is an integral of $\cO$ in the direction of its polarization vector. Using index-free notation, we can interpret the light-transform as a conformally-invariant transform, changing the quantum numbers $(\De,J)$ by
\be
\wL:(\De,J) &\to (1-J,1-\De).
\ee
In other words, $\wL[\cO](x,z)$ satisfies the conformal transformation law (\ref{eq:conformaltransformationprop}), with $\De$ replaced by $1-J$ and $J$ replaced by $1-\De$. (This is clearest from the definition of the light-transform in the embedding space \cite{Kravchuk:2018htv}.) In general, we define the spin of an object as its homogeneity in the polarization vector $z$. Because $\wL[\cO](x,z)$ has non-integer spin $1-\De$, it cannot be written in terms of an underlying tensor with $1-\De$ indices.

Though index-free notation appears to treat the nonlocal operator $\wL[\cO]$ in the same way as the local operator $\cO$, there is still something special about the representations associated to local operators. Specifically, a local operator is a polynomial in its polarization vector. This can be phrased as a kind of shortening condition. Morally speaking,
$J+1$ derivatives of $\cO(x,z)$ with respect to $z$ must vanish:
\be
``\ptl_z^{\mu_1}\cdots \ptl_z^{\mu_{J+1}} \cO(x,z)" &= 0.
\ee
We must take care to write this condition correctly because $z$ is constrained, $z^2=0$. Let us parametrize $z$ by
\be\label{eq:zPoincare}
z &= (1,\vec y^2,\vec y),\qquad \vec y \in \R^{d-2},
\ee
where we use lightcone coordinates $z=(z^+,z^-,\vec z)$ with metric $dz^2=-dz^+dz^-+d\vec z\. d\vec z$.
A more proper shortening condition is
\be
\label{eq:morepropershortening}
\ptl_{\vec y}^{i_1}\cdots\ptl_{\vec y}^{i_{J+1}} \cO(x,z) - \textrm{traces} &= 0,
\ee
where we subtract traces using the metric on $\R^{d-2}$.
The argument for (\ref{eq:morepropershortening}) is as follows. $\cO(x,z)$ is a sum of products of $J$ factors of $1,\vec y^2$ and the components of $\vec y$. If $J+1$ $\vec y$-derivatives of some term is nonzero, at least two of those derivatives must act on the same $\vec y^2$ factor, resulting in a nonzero trace. By subtracting traces, we remove such terms.

The shortening condition (\ref{eq:morepropershortening}) is naturally a traceless symmetric tensor with spin $J+1$ in the ``transverse" space $\R^{d-2}$. We can write it more economically by introducing a null vector $\vec s\in \C^{d-2}$ such that $\vec s^2=0$: 
\be
\label{eq:moremorepropershortening}
(\vec s\.\ptl_{\vec y})^{J+1} \cO(x,z) &= 0.
\ee
We can make (\ref{eq:moremorepropershortening}) Lorentz-invariant by introducing a polarization vector $w=(0,2\vec y\.\vec s,\vec s)\in \C^{d-1,1}$. By construction, $w$ is null and transverse to $z$, i.e.\ $w\.z=w^2=0$. We would like $w$ to encode only the $d-2$ degrees of freedom in $\vec s$, so we must additionally impose a gauge-redundancy $w\sim w+\l z$. The condition (\ref{eq:moremorepropershortening}) finally becomes
\be
\label{eq:finalshorteningforO}
(w\.\ptl_z)^{J+1}\cO(x,z) &= 0.
\ee

One can check that the differential operator $(w\.\ptl_z)^{J+1}$ preserves the ideal generated by $z^2=w\.z=w^2=0$. More nontrivially, it is gauge invariant under $w\to w+\l z$ precisely when acting on functions with homogeneity $J$ in $z$ (we show this in~\eqref{eq:gaugeproof} below).

The shortening condition (\ref{eq:finalshorteningforO}) for $\cO$ implies a related shortening condition for $\wL[\cO]$. For simplicity, suppose $\cO$ is a scalar, i.e.\ $J=0$. By integrating by parts inside the light-transform (\ref{eq:lighttransform}), we find
\be
\wL (w\.\ptl_z) &= \frac{1}{2-\De} \Big( (z\.\ptl_x)(w\.\ptl_z) - (z\.\ptl_z)(w\.\ptl_x) \Big) \wL \qquad \textrm{(acting on scalars with dimension $\De$)}.
\ee
Consequently, $\wL[\cO]$ satisfies its own shortening condition
\be
\label{eq:funnyshortening}
\Big( (z\.\ptl_x)(w\.\ptl_z) - (z\.\ptl_z)(w\.\ptl_x) \Big) \wL[\cO](x,z) &= 0 \qquad \textrm{($\cO$ scalar)},
\ee
inherited from the shortening condition (\ref{eq:finalshorteningforO}) for $\cO$.
Just as $w\.\ptl_z$ is Lorentz-invariant (also conformally-invariant) only when acting on scalar representations, $(z\.\ptl_x)(w\.\ptl_z) - (z\.\ptl_z)(w\.\ptl_x)$ is conformally-invariant only when acting on representations with the quantum numbers of $\wL[\cO]$, i.e.\ with dimension $1-0=1$.

\subsection{Reducibility and primary descendants}

The operator $(z\.\ptl_x)(w\.\ptl_z) - (z\.\ptl_z)(w\.\ptl_x)$ can be compared to other conformally-invariant differential operators that exist for special quantum numbers. A well-known example is the operator that takes a current to its divergence 
\be
\cJ^\mu &\to \ptl_\mu \cJ^\mu=(\ptl_x\.\ptl_z)\cJ(x,z).
\ee
On the right-hand side, we have written the divergence in index-free notation.

The operator $\ptl_x\.\ptl_z$ changes quantum numbers $(\De,J)$ by
\be
\label{eq:conservationoperator}
\ptl_x\.\ptl_z : (d-1,1) \to (d,0).
\ee
It is conformally-invariant only when acting on operators with the correct dimension and spin. This has the following representation-theoretic interpretation. Let $V_{\De,J}$ be a long multiplet (i.e.\ a generalized Verma module, see e.g.\ \cite{Penedones:2015aga}) of the conformal group with dimension $\De$ and spin $J$. $V_{\De,1}$ is irreducible for generic $\De$. However it becomes reducible when $\De=d-1$:
\be
\label{eq:conservedcurrentsubspace}
V_{d-1,1} &\supset V_{d,0}.
\ee
The quotient $\tl V_{d-1,1}\equiv V_{d-1,1} / V_{d,0}$ is the short representation associated to a conserved current. If $|\cJ(z)\>=z_\mu|\cJ^\mu\>$ is the highest-weight state of $V_{d-1,1}$, then the highest-weight state of $V_{d,0}\subset V_{d-1,1}$ is
\be
\label{eq:primarydescendant}
P\.\ptl_z|\cJ(z)\> \qquad \textrm{(primary descendant)}.
\ee
We say that (\ref{eq:primarydescendant}) is a ``primary descendant."
More formally, there exists a homomorphism
\be
\Phi: V_{d,0} &\to V_{d-1,1},
\ee
sending the highest-weight state of $V_{d,0}$ to (\ref{eq:primarydescendant}) inside $V_{d-1,1}$. By replacing $P\to \ptl_x$ inside the expression for the primary descendant, we find the conformally-invariant differential operator (\ref{eq:conservationoperator}).

In summary, conformally-invariant differential operators are in correspondence with {\it reducible} generalized Verma modules --- i.e.\ generalized Verma modules that contain primary descendants. The differential operators arising in this way have been studied in the mathematics literature, and also in the physics literature due to their relation to poles in conformal blocks \cite{Kos:2013tga,Kos:2014bka,Penedones:2015aga,Erramilli:2019njx}. We refer the reader to \cite{Penedones:2015aga} for a lucid discussion and classification. In order to compare our operator (\ref{eq:funnyshortening}) to the classification in \cite{Penedones:2015aga}, we must understand the quantum numbers of the corresponding primary descendant.\footnote{Strictly speaking, we need to be more careful because the notion of generalized (aka parabolic) Verma module depends on a choice of a parabolic subalgebra of conformal algebra. This choice determines, in particular, which quantum numbers can be non-integer. In~\cite{Penedones:2015aga} the parabolic subalgebra is the maximal one for Euclidean conformal algebra, and it allows for non-integer $\De$. Here, we are interested in both $\De$ and $J$ being non-integer, which requires us to consider the maximal parabolic subalgebra of Lorentzian conformal algebra. Therefore, the classification of~\cite{Penedones:2015aga} is not, strictly speaking, applicable here. It would be interesting to have a full classification for Lorentzian conformal group. We take a more simplistic approach: our conformally-invariant differential operators can be identified with analytic continuations in $J$ of the differential operators classified in~\cite{Penedones:2015aga}, and this will be enough for our purposes.\label{footnote:parabolic}}

\subsection{Index-free notation for general tensor representations}

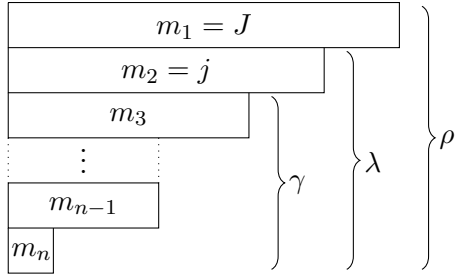
\begin{figure}[ht!]
	\centering 
	\begin{tikzpicture}
\draw[] (0,0) -- (5.2,0) -- (5.2,-\yheight) -- (0,-\yheight) -- (0,0);
\draw[] (0,-\yheight) -- (0,-2*\yheight) -- (4.2,-2*\yheight) -- (4.2,-\yheight);
\draw[] (0,-2*\yheight) -- (0,-3*\yheight) -- (3.2,-3*\yheight) -- (3.2,-2*\yheight);
\draw[dotted] (0,-3*\yheight) -- (0,-4*\yheight);
\draw[dotted] (2,-3*\yheight) -- (2,-4*\yheight);
\draw[] (0,-4*\yheight) -- (0,-5*\yheight) -- (2,-5*\yheight) -- (2,-4*\yheight) -- (0,-4*\yheight);
\draw[] (0,-5*\yheight) -- (0,-6*\yheight) -- (\yheight,-6*\yheight) -- (\yheight,-5*\yheight);
\node[] at (2.6,-0.5*\yheight-0.02) {$m_1=J$};
\node[] at (2.1,-1.5*\yheight-0.02) {$m_2=j$};
\node[] at (1.6,-2.5*\yheight-0.02) {$m_3$};
\node[] at (1,-3.5*\yheight+0.1) {$\vdots$};
\node[] at (1,-4.5*\yheight-0.02) {$m_{n-1}$};
\node[] at (0.5*\yheight+0.01,-5.5*\yheight-0.02) {$m_{n}$};

\draw[decorate,decoration={brace,amplitude=5pt}] (5.5,0-0.05) -- (5.5,-6*\yheight+0.05);
\draw[decorate,decoration={brace,amplitude=5pt}] (4.5,-\yheight-0.05) -- (4.5,-6*\yheight+0.05);
\draw[decorate,decoration={brace,amplitude=5pt}] (3.5,-2*\yheight-0.05) -- (3.5,-6*\yheight+0.05);
\node[right] at (5.6,-3*\yheight-0.02) {$\rho$};
\node[right] at (4.6,-3.5*\yheight-0.02) {$\l$};
\node[right] at (3.6,-4*\yheight-0.02) {$\g$};
	\end{tikzpicture}
	\caption{A Young diagram for an irreducible representation $\rho$ of $\SO(d-1,1)$. The rows have length $m_1,m_2,\dots,m_n$. We often write $m_1=J$ (spin) and $m_2=j$ (transverse spin). If we remove the first row of the Young diagram for $\rho$, the remaining rows $(m_2,\dots,m_n)$ make a Young diagram for an irreducible representation $\l$ of $\SO(d-2)$. If we remove another row, the remaining rows $(m_4,\dots,m_n)$ make a Young diagram for an irreducible representation $\g$ of $\SO(d-4)$.}
	\label{fig:youngdiagram}
\end{figure}

A general finite-dimensional tensor representation $\rho$ of $\SO(d-1,1)$ has a Young diagram with rows of length $(m_1,m_2,\dots,m_n)$, where $n=\lfloor \frac d 2 \rfloor$, see figure~\ref{fig:youngdiagram}. We define spin $J=m_1$ as the length of the first row of the Young diagram of $\rho$. The remaining rows $\l=(m_2,\dots,m_n)$ define a representation of $\SO(d-2)$. We define transverse spin $j=m_2$ as the length of the first row of the Young diagram of $\l$.

\tikzmath{
\yheight = 0.6;
}

An operator in the representation $\rho$ has indices $\cO^{\mu_1\cdots\mu_J;\nu_1\cdots \nu_{m_{2}};\cdots;\rho_{1}\cdots\rho_{m_{n}}}(x)$. Each index group delimited by semicolons is symmetric. Furthermore, $\cO$ is traceless in all its indices, and satisfies some additional symmetry conditions that we describe below.\footnote{For simplicity, we only consider the bosonic representations, i.e.\ the representations of $\SO(d-1,1)$ and not of $\mathrm{Spin}(d-1,1)$. Furthermore, we ignore the possible self-duality conditions in even $d$.}
To use index-free notation, we introduce null polarization vectors $z,w_1,\dots,w_{n-1}$ for each row: 
\be
\cO(x,z,\bw) &=
\cO(x,z,w_1,\dots,w_{n-1}) \nn\\
&= \cO^{\mu_1\cdots\mu_J;\nu_1\cdots \nu_{m_{2}};\cdots;\rho_{1}\cdots\rho_{m_{n}}}(x) z_{\mu_1}\cdots z_{\mu_J} w_{1,\nu_1}\cdots w_{1,\nu_{m_{2}}}\cdots w_{n-1,\r_1}\cdots w_{n-1,\r_{m_{n}}}.
\ee
Here we used the notation $\bw$ to denote the collection of vectors $w_i,\,i=1\cdots n-1$.
The polarization vectors satisfy the relations
\be
0 &= z^2  \nn\\
0 &= w_1^2 = w_1\.z \nn\\
0 &= w_2^2 = w_2\.w_1 = w_2\.z\nn\\
&\,\ \vdots \nn\\
0 &= w_{n-1}^2 = w_{n-1}\.w_{n-2} =\cdots = w_{n-1}\.z.
\ee
The additional symmetry properties of $\cO$ are equivalent to the statement that $\cO(x,z,\bw)$ is invariant under the gauge redundancies
\be
w_1 &\sim w_1 + \# z \nn\\
w_2 &\sim w_2 + \# w_1 + \# z \nn\\
&\,\ \vdots  \nn\\
w_{n-1} &\sim w_{n-1} + \# w_{n-2} + \cdots + \# z.
\ee
The object $\cO(x,z,\bw)$ is a homogeneous polynomial of the polarization vectors $z,w_1,\dots,w_{n-1}$ with degrees $(J,m_2,\dots,m_n)$. We often abbreviate $w_1=w$.

The index-free formalism we have just developed is essentially the embedding space formalism for the Lorentz group.\footnote{More generally, it comes from the Borel-Weil theorem, see \cite{Kologlu:2019mfz} for details.} In the embedding formalism, a traceless symmetric tensor operator in $d$-dimensions becomes a homogeneous function $\cO(X,Z)$ of variables $X,Z\in \R^{d+1,1}$ satisfying $X^2=X\.Z=Z^2=0$ and a gauge redundancy $Z\sim Z+\l X$ \cite{Costa:2011mg}. The Lorentz group $\SO(d-1,1)$ is of course the conformal group in $d-2$ dimensions. Thus the embedding formalism applies, with the $d+2$ dimensional $X$ and $Z$ replaced by the $d$-dimensional $z$ and $w$. In particular, $z$ can be interpreted as an embedding-space coordinate in $d-2$ dimensions.

\subsection{Raising transverse spin}

We can now recognize $w\.\ptl_z$ (with $w_1=w$) as a differential operator that raises transverse spin $j$, since it increases the degree in $w$ by $1$. Specifically, it changes the quantum numbers $(\De,J,j)$ by
\be
w\.\ptl_z : (\De,0,0) \to (\De,-1,1).
\ee
The operator $(z\.\ptl_x)(w\.\ptl_z) - (z\.\ptl_z)(w\.\ptl_x)$, obtained by commuting $w\.\ptl_z$ through the light-transform, changes (light-transformed) quantum numbers $(1-J,1-\De,j)$ by
\be
(z\.\ptl_x)(w\.\ptl_z) - (z\.\ptl_z)(w\.\ptl_x) : (1,1-\De,0) \to (2,1-\De,1).
\ee

More generally, let us define the operator
\be
\label{eq:dprimedef}
\cD'_n &= \frac{1}{n!}(w\.\ptl_z)^n,\\
\label{eq:dnprimequantumnumbers}
\cD_n'&:(\De,n+j-1,j)\to (\De,j-1,j+n).
\ee
The operator $\cD'_n$ provides the shortening condition for the Lorentz representation of local operators with spin $J=n+j-1$ and transverse spin $j$. Note that $\cD_n'$ preserves the ideal generated by $z^2=w\.z=w^2=0$. More nontrivially, it is gauge-invariant under $w\to w+\l z$. To see this, we act with the generator of a gauge transformation $z\.\ptl_w$ on $\cD_n'f(z,w)$, where $f(z,w)$ is gauge-invariant:
\be\label{eq:gaugeproof}
z\.\ptl_w (w\.\ptl_z)^n f(z,w) &= \sum_{j=0}^{n-1}(w\.\ptl_z)^{j}(z\.\ptl_z-w\.\ptl_w)(w\.\ptl_z)^{n-j-1} f(z,w) \nn\\
&= \sum_{j=0}^{n-1} (2j-n+1) (w\.\ptl_z)^{n-1} f(z,w) \nn\\
&= 0.
\ee
In the first line, we commute $z\.\ptl_w$ past other operators until it acts on $f(z,w)$, which it kills. In the second line, we use that $f(z,w)$ has homogeneity $n+j-1$ in $z$ and $j$ in $w$. This computation shows that $\cD_n'$ is Lorentz-invariant. Because local operators transform in irreducible representations of the Lorentz group, Schur's lemma implies that they must be killed by $\cD_n'$.

By commuting $\cD_n'$ through the light transform, 
\be\label{eq:LDcommutator}
\wL\cD_n' &= \cD_n \wL,
\ee
we find a new operator $\cD_n$ given by
\be
\label{eq:Dndef}
\cD_n &= \frac{(-1)^n\G(\De+j-2)}{\G(\De+j-2+n)\G(n+1)}(\ptl_x\.\cD_{z,w}^{0+})^n,\nn\\
\cD_{z,w;\mu}^{0+} &= w_\mu(w\.\ptl_w - z\.\ptl_z) + z_\mu w\.\ptl_z,
\ee
Here, $(\cD_{z,w}^{0+})^\mu$ is a weight-shifting operator for the Lorentz group $\SO(d-1,1)$ in the vector representation \cite{Karateev:2017jgd}. It can be obtained from the vector weight-shifting operators for the conformal group in \cite{Karateev:2017jgd} by using the analogy of $(z,w)$ with the embedding-space coordinates $(X,Z)$. The operator $\cD_n$ is conformally-invariant precisely when acting on the quantum numbers obtained by light-transforming (\ref{eq:dnprimequantumnumbers}):
\be
\cD_n &: (2-n-j,1-\De,j) \to (2-j,1-\De,j+n).
\ee
The operator $\cD_n$ falls into the classification of reducible Verma modules described in \cite{Penedones:2015aga}. In the notation of \cite{Penedones:2015aga}, it has type $\mathrm{I}_{2,n}$.\footnote{Similarly, we can think of $\cD'$ as an invariant differential operator with respect to $\SO(d-1,1)$ (the Lorentz group), thought of as the conformal group in $d-2$ dimensions. In the notation of \cite{Penedones:2015aga}, it has type $\mathrm{I}_{1,n}$.}${}^{,\ref{footnote:parabolic}}$

To summarize, by commuting $\cD_n'$ through the light transform, we obtain a special conformally-invariant differential operator $\cD_n$ that raises transverse spin. By construction, $\cD_n$ vanishes when acting on the light transform of a local operator with the appropriate quantum numbers, since
\be
\cD_n \wL[\cO] &= \wL[\cD_n'\cO] = 0.
\ee
Shortly, we will encounter light-ray operators $\mathbb{O}^\pm_{i,J,j}(x,z)$ that are conformal primaries with quantum numbers $(1-J,1-\De_i,j)$, but that are not light-transforms of local operators. When $J=n+j-1$ with $n\in \Z_{\geq 0}$, we can act on such operators with $\cD_{n}$ to obtain new primary descendants with higher transverse spin
\be
\cD_{n} \mathbb{O}^\pm_{i,J=n+j-1,j}(x,z) : (2-j,1-\De_i,j+n).
\ee
Primary descendants of this type will provide the higher transverse spin terms in the light-ray OPE.

\section{The complete OPE of scalar detectors}
\label{sec:scalar OPE}

In this section we will explicitly derive the OPE of the form~\eqref{eq:schematicope} for two scalar detectors $\wL[\f_1]\wL[\f_2]$. As discussed in~\cite{Kologlu:2019bco}, products of scalar detectors are not in general well-defined non-perturbatively. Nevertheless, they make sense kinematically, and give us a nice playground to demonstrate the key concepts, which we will then generalize to arbitrary detectors in section~\ref{sec:general OPE}. Furthermore, in section~\ref{sec:N4example} we will see an example where $\wL[\f_1]\wL[\f_2]$ is well-defined and the formulas from this section can be verified.

\subsection{Kinematics of the OPE}

Let us understand the general form that the OPE should take. First note that the product
\be
	\label{eq:Wz1z2def}
	\mathbb{W}(x,z_1,z_2)\equiv \wL[\f_1](x,z_1)\wL[\f_2](x,z_2),
\ee
is a conformal primary, if it is well-defined. This is because
\be
	[K_\mu,\mathbb{W}(0,z_1,z_2)] &= [K_\mu, \wL[\f_1](0,z_1)\wL[\f_2](0,z_2)]\nn\\
	&=[K_\mu, \wL[\f_1](0,z_1)]\wL[\f_2](0,z_2)+\wL[\f_1](0,z_1)[K_\mu,\wL[\f_2](0,z_2)]=0.
\ee
Similarly, the scaling dimension of $\mathbb{W}(x,z_1,z_2)$ is the sum of scaling dimensions of $\wL[\f_i]$,
\be
	\De_{\mathbb{W}}=(1-J_1)+(1-J_2)=2.
\ee
However, the operator $\mathbb{W}(x,z_1,z_2)$ does not transform irreducibly under the Lorentz group --- it depends on two polarization vectors, but they do not satisfy any of the relations described in section~\ref{sec:differential operators} among themselves. The OPE~\eqref{eq:schematicope} decomposes $\mathbb{W}(x,z_1,z_2)$ into irreducible components, since the detectors $\mathbb{D}(x,z,w)$ are irreducible under the Lorentz group.

In fact, the problem of decomposing $\mathbb{W}(x,z_1,z_2)$ into irreducible representations of the Lorentz group is a familiar one. The Lorentz group $\SO(d-1,1)$ is isomorphic to the Euclidean conformal group in $d-2$ dimensions. Under this isomorphism, $z_i$ become embedding-space coordinates~\cite{Weinberg:2010fx,Costa:2011mg} for a fictitious CFT${}_{d-2}$. By the definition of spin we have $\wL[\f_i](x,\l z_i)=\l^{1-\De_i}\wL[\f_i](x,z_i)$, where $\De_i$ is the scaling dimension of $\f_i$.\footnote{Recall that spin of $\wL[\phi_i]$ is $1-\De_i$.} At the same time, primary operators $\cP_\de$ in the embedding formalism for the fictitious CFT${}_{d-2}$ should satisfy $\cP_\de(\l z_1)=\l^{-\de}\cP_\de(z_1)$, where $\de$ is the $(d-2)$-dimensional scaling dimension. Thus, if we interpret the Lorentz group as a $(d-2)$-dimensional conformal group, the transformation properties of $\mathbb{W}(x,z_1,z_2)$ under Lorentz transformations can be described as\footnote{Provided that by Lorentz group we mean the group that fixes $x$. We take $x=0$ and use the standard Lorentz group for concreteness.}
\be
	\mathbb{W}(0,z_1,z_2)\sim \cP_{\de_1}(z_1)\cP_{\de_2}(z_2),
\ee
where $\de_i=\De_i-1$. In particular, if we parameterize $z_i$ as in~\eqref{eq:zPoincare},
\be\label{eq:zPoincare2}
	z_i^+=1,\quad z_i^-=\vec y^2,\quad z_i^\mu = y^\mu \,\,(\mu=2,\cdots, d-1),
\ee
for $\vec y\in \R^{d-2}$, then 
\be
	\mathbb{W}(0,\vec y_1,\vec y_2)\sim \cP_{\de_1}(\vec y_1)\cP_{\de_2}(\vec y_2)
\ee
transforms exactly like a pair of scalar primaries with dimensions $\de_1,\de_2$ at coordinates $\vec y_1,\vec y_2$ in the fictitious CFT${}_{d-2}$. 

We are familiar with taking the usual OPE between two scalar primaries,
\be\label{eq:CFTd-2 OPE}
	\cP_{\de_1}(\vec y_1)\cP_{\de_2}(\vec y_2)=\sum_{\de,j} \cC_{\de,j;\mu_1\cdots \mu_j}(\vec y_1,\vec y_2,\ptl_{\vec y_2})\cP_{\de,j}^{\mu_1\cdots \mu_j}(\vec y_2),
\ee
where $j$ is the $(d-2)$-dimensional traceless-symmetric spin.\footnote{Normally, we would have OPE coefficients in the right-hand side. Since here we are just discussing kinematics, we omit them. One can imagine that they have been absorbed into the definition of $\cP_{\de,j}$.} The differential operator $\cC_{\de,j;\mu_1\cdots \mu_j}(\vec y_1,\vec y_2,\ptl_{\vec y_2})$ is fixed by $(d-2)$-dimensional conformal symmetry and is the usual operator in Euclidean CFT${}_{d-2}$. It implicitly depends on $\de_1,\de_2$. For example, the leading term of this operator in the $\vec y_1\to \vec y_2$ limit is given by
\be
\label{eq:CFTd-2 C}
	\cC_{\de,j}^{\mu_1\cdots \mu_j}(\vec y_1,\vec y_2,\ptl_{\vec y_2}) = |\vec y_{12}|^{\de-\de_1-\de_2-j}(y_{12}^{\mu_1}\cdots y_{12}^{\mu_j}-\text{traces})+\cdots.
\ee

It will be convenient to work with~\eqref{eq:CFTd-2 OPE} written in embedding space coordinates. For this, note that traceless-symmetric primary operators $\cP_{\de,j}^{\mu_1\cdots \mu_j}(\vec y)$ are described in the embedding formalism as functions of $z,w$ satisfying $z^2=w^2=z\.w=0$ subject to
\be
	\cP_{\de,j}(z,w) &= \cP_{\de,j}(z,w+\a z),\\
	\cP_{\de,j}(z,\l w)&= \l^j \cP_{\de,j}(z,w).
\ee
In the notation of~\cite{Costa:2011mg} we have $P_\text{there} = z_\text{here}$, $Z_\text{there} = w_\text{here}$. So we can rewrite~\eqref{eq:CFTd-2 OPE} as
\be
	\cP_{\de_1}(z_1)\cP_{\de_2}(z_2)=\sum_{\de,j} \cC_{\de,j}(z_1,z_2,\ptl_{z_2},\ptl_{w_2})\cP_{\de,j}(z_2,w_2),
\ee
where $\cC_{\de,j}(z_1,z_2,\ptl_{z_2},\ptl_{w_2})$ is $\cC_{\de,j;\mu_1\cdots \mu_j}(\vec y_1,\vec y_2,\ptl_{\vec y_2})$, ``lifted'' to the embedding space. The embedding space expression for $\cC_{\de,j}$ has been studied in a number of papers (e.g.~\cite{Ferrara:1971vh,Ferrara:1972cq,Ferrara:1973yt,Fortin:2016lmf} to name a few), but we will not need its explicit form. All we need is that it is $\SO(d-1,1)$-invariant and upon restriction to the Poincar\'e section it becomes~\eqref{eq:CFTd-2 OPE}.

Translating this back to $\mathbb{W}(x,z_1,z_2)$, we see that the Lorentz symmetry requires the detector OPE to take the form
\be
	\mathbb{W}(x,\vec y_1,\vec y_2)=\sum_{i} \cC_{\de_i,j_i;\mu_1\cdots \mu_{j_i}}(\vec y_1,\vec y_2,\ptl_{\vec y_2})\mathbb{W}_{i}^{\mu_1\cdots \mu_{j_i}}(x,\vec y_2)
\ee
or
\be
	\label{eq:OPEansatz}
	\wL[\f_1](x,z_1)\wL[\f_2](x,z_2)=\mathbb{W}(x,z_1,z_2)=\sum_{i} \cC_{\de_i,j_i}(z_1,z_2,\ptl_{z_2},\ptl_{w_2})\mathbb{W}_{i}(x,z_2,w_2),
\ee
for some primary operators $\mathbb{W}_{i}(x,z,w)$ with $(d-2)$-dimensional quantum numbers $\de_i, j_i$.\footnote{Above, we had set $x=0$ but by translation invariance this is valid for any $x$.} Note that $w$ satisfies the same properties as the second-row polarization vector $w$ in section~\ref{sec:differential operators}, and so the operators $\mathbb{W}_{i}(x,z,w)$ transform in irreps of $\SO(d-1,1)$ with two-row Young diagrams, with the first row of length $-\de_i$ and the second row of length $j_i$. 

\subsection{Harmonic analysis on the celestial sphere}
\label{sec:harmonicanalysisscalar}

So far,~\eqref{eq:OPEansatz} is an ansatz based on kinematics. In this section we will show under which conditions~\eqref{eq:OPEansatz} holds, and provide a formula for $\mathbb{W}_{i}(x,z,w)$.

We will focus on a generic matrix element of $\mathbb{W}(x,z_1,z_2)$. Specifically, we will consider an expectation value of the form
\be
\label{eq:eventshapeweconsider}
	W(z_1,z_2)\equiv\<\cO_4(p)|\mathbb{W}(\oo,z_1,z_2)|\cO_3(p)\>,
\ee
for some local operators $\cO_3,\cO_4$.  
We allow $\cO_3$ and $\cO_4$ to carry arbitrary spin and do not require them to be primary.

In the language of the fictitious CFT${}_{d-2}$, $W(z_1,z_2)$ is a function of two points that at each point transforms as a scalar primary of dimension $\de_i$. The main idea is then to decompose $W(z_1,z_2)$ into $(d-2)$-dimensional conformal partial waves~\cite{Dobrev:1977qv,Karateev:2018oml}. This is equivalent to constructing the $\SO(d-1,1)$ Casimir operators from the Lorentz generators $M_{\mu\nu}$ defined by
\be
	M^{\mu\nu} &= M^{\mu\nu}_1+M^{\mu\nu}_2,\\
	M^{\mu\nu}_i&= z^\nu_i\tfrac{\ptl}{\ptl z_{\mu,i}}-z^\mu_i\tfrac{\ptl}{\ptl z_{\nu,i}},
\ee
and finding a basis of their common eigenfunctions.\footnote{These Casimir operators are only self-adjoint if $\de_i\in \frac{d-2}{2}+i\R$. The forthcoming discussion is also only strictly rigorous under this condition and for square-integrable $W(z_1,z_2)$. In practice, however, usually one can analytically continue to general $\de_i$ and work with non-square-integrable $W(z_1,z_2)$, provided certain care is taken in the process. See the discussions in~\cite{Caron-Huot:2017vep,Simmons-Duffin:2017nub,Kravchuk:2018htv} for details.\label{ft:harmoniccaveat}}
In fact, note that the standard conformally-invariant three-point functions 
\be\label{eq:standard3pt_celestial}
	\<\cP_{\de_1}(z_1)\cP_{\de_2}(z_2)\cP_{\de,j}(z,w)\>=\frac{
		(-4)^j(w\.z_2 z_1\.z-w\.z_1z_2\.z)^j
	}{
		(-2z_1\.z_2)^{\frac{\de_1+\de_2-\de+j}{2}}
		(-2z_1\.z)^{\frac{\de_1+\de-\de_2+j}{2}}
		(-2z_2\.z)^{\frac{\de_2+\de-\de_1+j}{2}}
	}
\ee
are eigenfunctions of all the Casimirs for any $z,w$ and $\de,j$.\footnote{Our conventions for tensor structures are summarized in appendix~\ref{app:structureconventions}.\label{foot:23convention}} This is because by conformal invariance, the action of the Casimirs on $z_1,z_2$ on (\ref{eq:standard3pt_celestial}) is equivalent to the action on $z,w$, and $\cP_{\de,j}$ is irreducible. Modulo technical details that we omit for simplicity (see~\cite{Karateev:2018oml,Kologlu:2019mfz}), these eigenfunctions form a complete basis and we have
\be\label{eq:W CPW pre}
	W(z_1,z_2)&=\sum_{j=0}^{\oo}\int_{\frac{d}{2}-i\oo}^{\frac{d}{2}+i\oo}\frac{d\de}{2\pi i}\int D^{d-2}z\,\<\cP_{\de_1}(z_1)\cP_{\de_2}(z_2)\cP_{\de,j}(z,D_w)\>W'_{\tl\de,j}(z,w)
\ee
for some $W'_{\tl\de,j}(z,w)$, $\tl \de = d-2-\de$. Here the integration measure in $z$ is the standard~\cite{SimmonsDuffin:2012uy}\footnote{Upon restriction to the Poincar\'{e} section~\eqref{eq:zPoincare2}, this measure becomes simply $d^{d-2}\vec y$.}
\be
	D^{d-2}z = 2\frac{d^dz \de(z^2)\theta(z^0)}{\vol \SO(1,1)},
\ee
and $D_w$ is roughly equivalent to $\ptl_w$~\cite{Costa:2011mg} and serves to contract the indices encoded by $w$ between $W'$ and $\cP_{\de,j}$.
This can be written in a form more similar to~\eqref{eq:OPEansatz} by using the fact
\be\label{eq:3pt=C 2pt}
	\<\cP_{\de_1}(z_1)\cP_{\de_2}(z_2)\cP_{\de,j}(z,w)\>=\cC_{\de,j}(z_1,z_2,\ptl_{z_2},\ptl_{w_2})\<\cP_{\de,j}(z_2,w_2)\cP_{\de,j}(z,w)\>,
\ee
for the normalization of $\cC_{\de,j}$ as in~\eqref{eq:CFTd-2 C} and for the standard two-point function given by${}^{\ref{foot:23convention}}$
\be\label{eq:standard2pt_celestial}
\<\cP_{\de,j}(z_1,w_1)\cP_{\de,j}(z_2,w_2)\>=\frac{(-2(w_1\.w_2)+2(w_1\.z_2)(w_2\.z_1)/(z_1\.z_2))^j}{(-2z_1\.z_2)^{\de}}.
\ee
Using (\ref{eq:3pt=C 2pt}), we find\footnote{A subtlety is that~\eqref{eq:3pt=C 2pt} does not converge for all values of the coordinates, for similar reasons that the usual Euclidean OPE does not converge in all Euclidean configurations. The proper justification of this step is the same as in the case when one passes from conformal partial waves to conformal blocks in partial conformal wave expansion of a four-point function~\cite{Dobrev:1975ru,Karateev:2018oml}.\label{foot:harmonicapology}
}
\be\label{eq:W CPW final}
	W(z_1,z_2)&=\sum_{j=0}^{\oo}\int_{\frac{d-2}{2}-i\oo}^{\frac{d-2}{2}+i\oo}\frac{d\de}{2\pi i}\,\cC_{\de,j}(z_1,z_2,\ptl_{z_2},\ptl_{w_2})W_{\de,j}(z_2,w_2),
\ee
where 
\be
	W_{\de,j}(z_2,w_2)=\int D^{d-2}z \<\cP_{\de,j}(z_2,w_2)\cP_{\de,j}(z,D_w)\>W'_{\tl\de,j}(z,w).
\ee

Using orthogonality of eigenfunctions, we can now find the inverse to~\eqref{eq:W CPW pre} to compute $W'$ and thus $W$. It is given by taking the inner product with the eigenfunctions,
\be\label{eq:scalarinverse}
	W_{\de,j}(z,w)&\equiv \a_{\de,j}\,\int D^{d-2}z_1D^{d-2}z_2 \<\tl\cP_{\de_1}(z_1)\tl\cP_{\de_2}(z_2)\cP_{\de,j}(z,w)\>W(z_1,z_2),
\ee
where the normalization coefficient has been computed, for example, in~\cite{Kologlu:2019mfz}
\be\label{eq:alphaexpression}
	\a_{\de,j} = \frac{(-1)^j\G(j+\frac{d-2}{2})\G(d-2+j-\de)\G(\de-1)}{2\pi^{d-2}\G(j+1)\G(\de-\frac{d-2}{2})\G(\de+j-1)}
	\frac{\G(\frac{\de+j+\de_1-\de_2}{2})\G(\frac{\de+j+\de_2-\de_1}{2})}
	{\G(\frac{d-\de+j+\de_1-\de_2}{2})\G(\frac{d-\de+j+\de_2-\de_1}{2})},
\ee
and $\tl\cP_{\de}\equiv \cP_{\tl\de}=\cP_{d-2-\de}$.\footnote{For $\de_i\in \frac{d-2}{2}+i\R$ we have $\tl\de_i=\de_i^*$ and~\eqref{eq:scalarinverse} just contains the complex conjugate of the wavefunction, as usual --- recall footnote~\ref{ft:harmoniccaveat}.}

What is the relationship between the OPE ansatz~\eqref{eq:OPEansatz} and the integral expression~\eqref{eq:W CPW final}? If $W_{\de,j}(z,w)$ is meromorphic in the right half-plane of $\de$, we can deform the integration contour in~\eqref{eq:W CPW final}.\footnote{In such deformations of the contour in the partial wave expansions one usually encounters ``spurious'' poles. Instead of representing physical contributions, they are purely kinematical and cancel among each other or against the possible discrete series contributions that we omitted in~\eqref{eq:W CPW pre}. See~\cite{Caron-Huot:2017vep} for an example of cancellation of these poles in usual four-point functions, and~\cite{Kologlu:2019mfz} for an example of cancellation in event shapes (for low transverse spin). To simplify the discussion, in this paper we assume that the spurious poles always cancel.\label{foot:spuriouspoles}} In this way, we would obtain
\be
	\<\cO_4(p)|\mathbb{W}_{i}(\oo,z,w)|\cO_3(p)\> = - \mathrm{res}_{\de=\de_i} W_{\de,j_i}(z,w).
\ee
Since $p$, $\cO_3$, and $\cO_4$ were arbitrary, this completely determines $\mathbb{W}_i(\oo,z,w)$ as an operator. 

\subsection{Relation to light-ray operators}

\subsubsection{Review of light-ray operators}
\label{sec:light-ray-ops}

We now show that $W_{\de,j}(z,w)$ is indeed expected to be meromorphic, with residues related to light-ray operators. Let us first review the definition and some basic properties of these operators.

Light-ray operators~\cite{Kravchuk:2018htv} are primary operators
\be
	\hat{\mathbb{O}}^\pm_{i,J,\l}(x,z,\bw),
\ee
parametrized by a spin $J$ (the length of the first row of the $\SO(d-1,1)$ Young diagram), an $\SO(d-2)$ representation $\l$ (the remaining rows of the $\SO(d-1,1)$ Young diagram), a Regge trajectory label $i$, and a signature $\pm$. Their defining property is that they are defined for generic complex spin $J$ and are related to local operators for certain integer values of $J$. Specifically, we have
\be\label{eq:LR-local-relation}
	\hat{\mathbb{O}}^\pm_{i,J,\l}(x,z,\bw) = \wL[\cO](x,z,\bw)\qquad (J\in \Z_{\geq 0}\textrm{ and }(-1)^J=\pm 1)
\ee
for some local operator $\cO$ with $\SO(d-1,1)$ representation $(J,\l)$ whenever $J$ is a non-negative integer such that $(-1)^J=\pm 1$ (where $\pm$ indicates the signature). That is, the operators $\hat{\mathbb{O}}^+_{i,J,\l}$ are related to even-spin local operators, while $\hat{\mathbb{O}}^-_{i,J,\l}$ are related to odd-spin local operators. The light-ray operators $\hat{\mathbb{O}}^\pm_{i,J,\l}$ thus organize the local operators into continuous families or ``Regge trajectories," and we label different trajectories by the index $i$.\footnote{The simple picture of operators organized into isolated Regge trajectories is not rigorously proven. We give a discussion of its correctness in appendix~\ref{app:light-ray-apology}. Even if it is not correct, the results of this paper are mostly unchanged, but become more awkward to phrase. For simplicity of presentation, we take it as an assumption in the main text, and delegate more nuanced discussion to appendix~\ref{app:light-ray-apology}.}

In practice we construct light-ray operators from a product of two primaries $\cO_1$ and $\cO_2$.\footnote{We expect that we get the same light-ray operators from any pair of local operators with the appropriate quantum numbers.} In this section, we are interested in light-ray operators that appear in the $\f_1,\f_2$ OPE, so we use $\cO_1=\f_1,\cO_2=\f_2$ to construct them. These operators all have $\l=0$ since only traceless-symmetric operators appear in the OPE of two scalars, and analytic continuation in spin does not affect the 2nd and higher rows of the Young diagram. We define, following~\cite{Kravchuk:2018htv}
\be\label{eq:LRdef scalars}
	\mathbb{O}^\pm_{\De,J}(x,z)\equiv &\pm\int'_{x_1\approx x^+\atop x_2\approx x} d^dx_1 d^dx_2 K^t_{\De,J}(x_1,x_2;x,z) \f_1(x_1)\f_2(x_2)\nn\\
	&+\int'_{x_1\approx x\atop x_2\approx x^+} d^dx_1 d^dx_2 K^u_{\De,J}(x_1,x_2;x,z)\f_2(x_2)\f_1(x_1),
\ee
where~\cite{Kravchuk:2018htv}
\be\label{eq:Kexpressions}
	K^t_{\De,J}(x_1,x_2;x,z)&=\b_{\De,J} \<0|\tl\f_1(x_1)\wL[\cO]\tl\f_2(x_2)|0\>,\nn\\
	K^u_{\De,J}(x_1,x_2;x,z)&=\b_{\De,J} \<0|\tl\f_2(x_2)\wL[\cO]\tl\f_1(x_1)|0\>,
\ee
where $\<0|\cdots|0\>$ denote the standard tensor structures for the Wightman three-point functions as defined in appendix~\ref{app:structureconventions}.\footnote{In particular, the structures appearing in $K^t$ and $K^u$ are not related to each other by a direct analytic continuation. Instead, they differ simply by the substitution $1\leftrightarrow 2$.} The primes on the integrals indicate that we should restrict $x_1$ and $x_2$ to an arbitrary small neighborhood of the null cone of $x$. The dependence on this arbitrary choice will go away momentarily. We also used the notation $a\approx b$ to indicate that points $a$ and $b$ are spacelike, and $x^+$ denotes the image of $x$ in the next Poincar\'{e} patch on the Lorentzian cylinder. The integrals are understood to be over the Lorentzian cylinder in order to preserve manifest conformal invariance. The coefficient $\b_{\De,J}$ has an expression in terms of the Plancherel measure of $\SO(d+1,1)$, shadow coefficients, and Euclidean three-point pairings~\cite{Kravchuk:2018htv}. Here we will only need the explicit expression
\be\label{eq:betaformula}
\b_{\De,J}=
\frac{
	\G(J+\frac{d}{2})\G(d+J-\De)\G(\De-1)
}{
	2\pi^d \G(J+1)\G(\De-\frac{d}{2})\G(\De+J-1)
}
\frac{
	\G(\frac{\De+J+\De_{12}}{2})\G(\frac{\De+J-\De_{12}}{2})
}{
	\G(\frac{d-\De+J+\De_{12}}{2})\G(\frac{d-\De+J-\De_{12}}{2})
},
\ee
where $\De_{12}=\De_1-\De_2$.

We then finally define
\be
	\mathbb{O}^\pm_{i,J}(x,z)=\mathrm{res}_{\De=\De_i(J)} \mathbb{O}^\pm_{\De,J}(x,z),
\ee
where $\De_i(J)$ are the scaling dimensions of the $i$-th Regge trajectory. To be more precise, from the Lorentzian inversion formula the matrix elements of $\mathbb{O}^\pm_{\De,J}(x,z)$ have poles at $\De=\De_i(J)$ for integer $J\geq 0, \,(-1)^J=\pm 1$, and these poles analytically continue to $J\in \C$ (modulo subtleties discussed in appendix~\ref{app:light-ray-apology}). The poles come from the integration region in~\eqref{eq:LRdef scalars} where $x_1$ and $x_2$ approach the null-cone of $x$, and thus the residues are independent of the precise choice of integration region for the primed integrals. The operators thus defined satisfy
\be
	\mathbb{O}^\pm_{i,J}(x,z)=f_{12\cO}\wL[\cO](x,z),\qquad \textrm{when}\ J\in \Z_{\geq 0},\  (-1)^J =\pm 1,
\ee
for some local operator $\cO$, where $f_{12\cO}$ is the OPE coefficient appearing in $\<\f_1\f_2\cO\>$. For general $J\in\C$ we have
\be
	\mathbb{O}^\pm_{i,J}(x,z)=f_{12i}(J) \hat{\mathbb{O}}^\pm_{i,J}(x,z),
\ee
where $f_{12i}(J)$ is the analytic continuation of $f_{12\cO}$ along the $i$-th Regge trajectory. Note that $\mathbb{O}^\pm_{i,J}(x,z)$ was constructed from a particular pair of local operators $\f_1,\f_2$. The above relation provides the connection between $\mathbb{O}^\pm_{i,J}(x,z)$ constructed from different pairs of local operators.

While the normalized operators $\hat{\mathbb{O}}^\pm_{i,J}(x,z)$ are more fundamental, for our purposes it will be more convenient to work with the operators $\mathbb{O}^\pm_{i,J}(x,z)$ constructed from $\f_1,\f_2$.

\subsubsection{The transverse spin puzzle}

We now relate $\mathbb{W}_{i}(x,z,w)$ in (\ref{eq:OPEansatz}) to the light-ray operators $\mathbb{O}^\pm_{i,J}(x,z)$. Consider the quantum numbers of $\mathbb{O}^\pm_{i,J}(x,z)$. By construction, they have scaling dimension $1-J$ and a one-row Lorentz representation with spin $1-\De_i(J)$. In particular, they have $j=0$. This is the analytic continuation of the fact that only traceless-symmetric operators appear in the OPE of two scalars.

On the other hand, the operators $\mathbb{W}_{i}(x,z,w)$ have scaling dimension $2$ and two-row Lorentz representations with spin $-\de_i$ and transverse spin $j_i$. If $j_i=0$, we can match quantum numbers with light-ray operators by setting $J=-1$ and $\de_i=\De_k(J=-1)-1$, leading us to expect
\be\label{eq:LTSconj}
	\mathbb{W}_{i}(x,z)\sim \mathbb{O}^+_{k,-1}(x,z)+\mathbb{O}^-_{k,-1}(x,z) \qquad (\textrm{for $\mathbb{W}_i$ with $j_i=0$}).
\ee
Here ``$\sim$" means that the left-hand side is a linear combination of objects on the right-hand side.
Note that the expressions such as
\be\label{eq:descendantfail}
	\mathbb{W}_{i}(x,z) \sim \ptl^2_x \mathbb{O}^+_{k',1}(x,z), \qquad (\textrm{wrong})
\ee
while allowed by dilatation and Lorentz symmetry, are forbidden by the full conformal symmetry, since $\mathbb{W}_{i}(x,z,w)$ must transform as primary operators.\footnote{Note that $\ptl_x^2$ can be conformally-invariant, i.e.\ produce a primary, but only if it acts on a scalar operator of dimension $\frac{d-2}{2}$, which is not the case here.}

A relation of the form~\eqref{eq:LTSconj} was proven in~\cite{Kologlu:2019mfz}, where $\mathbb{W}_{i}$ with $j_i=0$ were called  ``low transverse-spin'' terms. Having an expression for the low transverse-spin terms is sufficient for computing certain even shapes, as explored in~\cite{Kologlu:2019mfz}, but does not provide a complete OPE expansion. 

Finding an expression for $\mathbb{W}_{i}(x,z,w)$ with $j_i>0$ in terms of $\mathbb{O}_{i,J}^\pm(x,z)$ might seem hopeless due to the mismatch of quantum numbers. Luckily, our discussion in section~\ref{sec:differential operators} provides us with a family of conformally-invariant differential operators $\cD_n$ which have the property that
\be
(\cD_{n} \mathbb{O}^\pm_{i,J=n-1})(x,z,w)
\ee
transforms like a primary with scaling dimension $2$, spin $1-\De_i$ and transverse spin $n$. We can thus conjecture that for all $j_i$
\be
	\mathbb{W}_{i}(x,z,w)&\sim (\cD_{j_i} \mathbb{O}^+_{k,J=j_i-1})(x,z,w)+(\cD_{j_i} \mathbb{O}^-_{k,J=j_i-1})(x,z,w),
\ee
where $\de_i=\De_k(J=j_i-1)-1$. (We use the convention $\cD_0=1$.) This expression is similar in spirit to the wrong example~\eqref{eq:descendantfail} above, except that instead of $\ptl_x^2$ we use a carefully-constructed differential operator $\cD_n$ that makes the right-hand side transform as a primary. As discussed in section~\ref{sec:differential operators}, such operators are rare and it is an intriguing mathematical conspiracy that the quantum numbers of $\mathbb{W}_{i}(x,z,w)$ are precisely such that $\cD_n$ exist. In particular, one can verify that no other conformally-invariant differential operators exist in the classification of~\cite{Penedones:2015aga}${}^{\ref{footnote:parabolic}}$ that can be used in the ansatz for $\mathbb{W}_{i}(x,z,w)$. 

\subsubsection{Partial waves and light-ray operators}

As we discussed in section~\ref{sec:harmonicanalysisscalar}, the matrix elements of the operators $\mathbb{W}_{i}(x,z,w)$ are residues of $W_{\de,j}(x,z,w)$
defined by~\eqref{eq:scalarinverse}, which we can write as (restoring $x$-dependence)
\be\label{eq:WfromLL}
	&W_{\de,j}(x,z,w)=\nn\\
	&\a_{\de,j}\,\int D^{d-2}z_1D^{d-2}z_2 \<\tl\cP_{\de_1}(z_1)\tl\cP_{\de_2}(z_2)\cP_{\de,j}(z,w)\>
	\<\cO_4|\wL[\f_1](x,z_1)\wL[\f_2](x,z_2)|\cO_3\>.
\ee
By expanding the definition of $\wL$, this can be understood as a particular integral of
\be\label{eq:f1f2matrixelement}
	\<\cO_4|\f_1(x_1)\f_2(x_2)|\cO_3\>
\ee
over $x_1, x_2$. 

Similarly, the matrix elements of $(\cD_{j} \mathbb{O}^\pm_{i,J=j-1})(x,z,w)$ are residues of 
\be
	\<\cO_4|(\cD_{j} \mathbb{O}^\pm_{\De,J=j-1})(x,z,w)|\cO_3\>,
\ee
which itself is a $x_1,x_2$-integral of~\eqref{eq:f1f2matrixelement}. Therefore, we can prove the relation between $\mathbb{W}_{i}$ and $\cD_{j} \mathbb{O}^\pm_{i,J=j-1}$ by simply showing that these integrals are the same. Concretely, we will show that
\be\label{eq:WscalarResult}
	W_{\de,j}(x,z,w)=A\<\cO_4|(\cD_{j} \mathbb{O}^+_{\de+1,J=j-1})(x,z,w)-(\cD_{j} \mathbb{O}^-_{\de+1,J=j-1})(x,z,w)|\cO_3\>
\ee
for some constant $A$.

Let us start with rewriting~\eqref{eq:WfromLL} in terms of an integral of~\eqref{eq:f1f2matrixelement}:
\be
	W_{\de,j}(x,z,w)=\int d^dx_1 d^dx_2 \cL_{\de,j}(x_1,x_2;x,z,w) \<\cO_4|\f_1(x_1)\f_2(x_2)|\cO_3\>.
\ee
By the definition of the light-transform, the kernel $\cL_{\de,j}$ is given by
\be
\label{eq:cLkernel}
	\cL_{\de,j}(x_1,x_2;x,z,w)=\a_{\de,j}\int& d\a_1 d\a_2 D^{d-2}z_1D^{d-2}z_2\<\tl\cP_{\de_1}(z_1)\tl\cP_{\de_2}(z_2)\cP_{\de,j}(z,w)\>\times\nn\\
	&\times (-\a_1)^{-\de_1-1}(-\a_2)^{-\de_2-1}\de^d(x-z_1/\a_1-x_1)\de^d(x-z_2/\a_2-x_2).
\ee
To evaluate the $\a_i,z_i$-integrals, let us first consider the more general integral
\be
	\int d\a D^{d-2}z(-\a)^{-\de-1}\de^d(x+z/\a)f(z).
\ee
For the $z$-integration to make sense, we need $f(\l z) = \l^{-d+2+\de}f(z)$. Otherwise, we keep $f$ arbitrary. We restrict to $x$ belonging to the first Poincar\'{e} patch, in which case $\a$ must be negative in order to satisfy $x+z/\a=0$. We find
\be
	&\int_{\a<0} d\a D^{d-2}z(-\a)^{-\de-1}\de^d(x+z/\a)f(z)\nn\\
	&=\frac{2}{\vol\SO(1,1)}\int_{\a>0} d\a d^dz\de(z^2)\theta(z>0)\a^{-\de-1}\de^d(x-z/\a)f(z)\nn\\
	&=\frac{2}{\vol\SO(1,1)}\int_{\a>0} d\a\de(\a^2 x^2)\theta(\a x>0)\a^{d-\de-1}f(\a x)\nn\\
	&=\frac{2}{\vol\SO(1,1)}\int_{\a>0} d\a\de(x^2)\theta(x>0)\a^{-1}f(x)\nn\\
	&=2\de(x^2)\theta(x>0)f(x),
\ee
where we used that $\vol\SO(1,1)=\int_0^{\oo} d\a \a^{-1}$.\footnote{In a more formal derivation, one would fix the $\SO(1,1)$ freedom by Faddeev-Popov procedure.} Using this result in (\ref{eq:cLkernel}), we find
\be\label{eq:Lresult}
	\cL_{\de,j}(x_1,x_2;x_3,z,w)=4\a_{\de,j}\<\tl\cP_{\de_1}(x_{13})\tl\cP_{\de_2}(x_{23})\cP_{\de,j}(z,w)\>\de(x_{13}^2)\de(x_{23}^2)\theta(x_{13}>0,x_{23}>0).
\ee

Let us now perform the same exercise for the right-hand side of~\eqref{eq:WscalarResult}. First, it is clear why we want the difference of $\mathbb{O}^+$ and $\mathbb{O}^-$ operators with equal coefficients: by comparing to~\eqref{eq:LRdef scalars}, we see that this difference only involves the $K^t$ kernel and the ordering $\f_1\f_2$, which is precisely the ordering we need. We find
\be\label{eq:Ktintegral}
	&A\<\cO_4|(\cD_{j} \mathbb{O}^+_{\de+1,J=j-1})(x,z,w)-(\cD_{j} \mathbb{O}^-_{\de+1,J=j-1})(x,z,w)|\cO_3\>\nn\\
	&=2A\int' d^dx_1 d^dx_2 (\cD_{j}K^t_{\de+1,j-1}\theta(x_1\approx x^+, x_2\approx x))(x_1,x_2;x,z,w)\<\cO_4|\f_1(x_1)\f_2(x_2)|\cO_3\>.
\ee
Here we have taken into account the action of $\cD_j$ on $x$ in the integration limits in~\eqref{eq:LRdef scalars} by writing these limits as an explicit $\theta$-function.

We would like to relate the kernel in (\ref{eq:Ktintegral}) to the kernel  $\cL_{\de,j}$ in (\ref{eq:Lresult}). In doing so, there is an apparent problem: $K^t$ is non-zero for generic configurations of points $x_1,x_2,x$, while the kernel $\cL_{\de,j}$ only contains delta-functions. This problem is resolved by two mechanisms. Firstly, when $j>0$, the differential operator $\cD_j$ annihilates $K^t$ for generic configurations of points. Hence, its action in (\ref{eq:Ktintegral}) only has support on special loci --- precisely the loci where $\cL_{\de,j}$ has support. Meanwhile, when $j=0$, we have $\cD_0 = 1$ and this mechanism doesn't work. However, in this case the coefficient~$\b_{\de+1,j-1}$ in~\eqref{eq:Kexpressions} contains a 0 coming from $\G(J+1)$ in the denominator of~\eqref{eq:betaformula}. The zero is only cancelled on special loci, giving rise to delta-functions again. In the next section, we describe in more detail how delta-functions of~\eqref{eq:Lresult} emerge from (\ref{eq:Ktintegral}).

\subsection{Emergence of delta-functions from the light-ray kernel}

The light transform entering $K^t$ in~\eqref{eq:Kexpressions} is given by~\cite{Kravchuk:2018htv}
\be\label{eq:LtransformShadow3pt}
	&\<0|\tl\phi_1(x_1)\wL[\cO](x_3,z)\tl\phi_2(x_2)|0\>\nn\\
	&=
	L(\tl\f_1\tl\f_2[\cO])
	\frac{
		(2z\.x_{23} x_{13}^2 - 2z\.x_{13} x_{23}^2)^{1-\De}
	}{
		(x_{12}^2)^{\frac{\tl\De_1+\tl\De_2-(1-J)+(1-\De)}{2}}
		(-x_{13}^2)^{\frac{\tl\De_{12}+(1-J)+(1-\De)}{2}}
		(x_{23}^2)^{\frac{-\tl\De_{12}+(1-J)+(1-\De)}{2}}
	}
	\nn\\
	&\qquad(\textrm{in the configuration $x_1\approx x_2, x_1>x_3, x_2\approx x_3$}),
\ee
where
\be
	L(\tl\f_1\tl\f_2[\cO])=-2\pi i 
	\frac{
		\G(\De+J-1)
	}{
		\G(\frac{\De+\De_{12}+J}{2})
		\G(\frac{\De-\De_{12}+J}{2})
	}.
\ee
Acting with $\cD_j$ given in~\eqref{eq:Dndef} we find
\be\label{eq:DjLwightman}
	&\<0|\tl\phi_1(x_1)(\cD_j\wL[\cO])\tl\phi_2(x_2)|0\>\nn\\
	&=
	\frac{(-1)^j2^{2j}(-J)_j
	L(\tl\f_1\tl\f_2[\cO])(\De+j-2)}{(\De-2)\G(j+1)}(w\.x_{13}z\.x_{23}-w\.x_{23}z\.x_{13})^j\nn\\
	&\quad \times\frac{
		(2z\.x_{23} x_{13}^2 - 2z\.x_{13} x_{23}^2)^{1-\De-j}
	}{
		(x_{12}^2)^{\frac{\tl\De_1+\tl\De_2-(1-J)+(1-\De)}{2}}
		(-x_{13}^2)^{\frac{\tl\De_{12}+(1-J)+(1-\De)}{2}}
		(x_{23}^2)^{\frac{-\tl\De_{12}+(1-J)+(1-\De)}{2}}
	}.
\ee
If we set $J=j-1$ then for $j>0$ this indeed vanishes due to the $(-J)_j$ factor, and thus $\cD_j$ with $j>0$ annihilates $K^t$, as promised above. This can also be immediately concluded from the fact that for $J=j-1\geq 0$ we have above a light transform of an integer-spin three-point structure, and thus it must be annihilated by $\cD_j$ based on general properties of $\cD_j$ discussed in section~\ref{sec:differential operators}.

However, this calculation of the action of $\cD_j$ is not complete. This is because we have only computed the action of $\cD_j$ on $K^t$ in~\eqref{eq:Ktintegral}, but not on the $\theta$-function. Derivatives hitting the theta-function will produce delta-function contributions that are needed to match~\eqref{eq:Lresult}. To derive the explicit form of these delta-functions, we will take a slightly more general approach.

\subsubsection{Interlude: distributions and analytic continuation}
\label{sec:toymodeldist}

To illustrate the approach, it is helpful to consider a toy example
\be
\label{eq:derivofdelta}
\frac{\ptl}{\ptl x} \theta(x) &= \de(x).
\ee
While the right-hand side is obvious, let us re-derive it in a way that will be useful later.

We start with a new definition of the left-hand side. We can think of $\theta(x)$ as the case $a=0$ of the 1-parameter family of distributions $p_a(x) = x^a \theta(x)$. The derivative $p'_a(x)$ is represented by a locally-integrable function when $\Re a>0$:
\be
\label{eq:derivativecomputation}
p'_a(x) = a x^{a-1} \theta(x) + x^a \de(x) = a x^{a-1} \theta(x) \qquad (\Re a>0).
\ee
Specifically, when $\Re a>0$, we can ignore $\ptl/\ptl x$ acting on $\theta(x)$.
Let us define the left-hand side of (\ref{eq:derivofdelta}) as the analytic continuation of $p'_a(x)=a x^{a-1} \theta(x)$ from $\Re a>0$ to $a=0$.

As we explain in detail in appendix~\ref{app:deltapole}, the distribution $x^{a-1}\theta(x)$ has a pole at $a=0$ with residue $\de(x)$. It follows that $\lim_{a\to0} a x^{a-1} \theta(x)=\delta(x)$. To see this, consider the integral against a test function $f(x)$ with a regular Taylor series expansion at $x=0$,
\be
\int_0^\oo dx\, a x^{a-1} f(x) &= a \int_0^\oo dx\, x^{a-1} \p{f(0) + (f(x)-f(0))} \nn\\
&= a \p{\frac{f(0)}{a} + O(1)}\nn\\
&= f(0) + O(a) \qquad\qquad\qquad (\Re a>0).
\ee
In the second line, we used the fact that $\int dx\,x^{a-1}(f(x)-f(0))$ is $O(1)$ as $a\to 0$, since $f(x)-f(0)$ vanishes at least linearly there. Finally, taking $a\to 0$, we obtain $f(0)$, as claimed.

\subsubsection{Action of $\cD_j$ on $K^t$}

This strategy of starting in a region where the derivative of a kernel is simple (in particular where we can ignore derivatives acting on $\theta$-functions), and analytically continuing away works well in the case at hand.
First, let us define the expression
\be
\label{eq:theexpression}
	K^t_{\De,J}(x_1,x_2;x,z)\theta(x_1\approx x^+, x_2\approx x),
\ee
as a distribution. This is straightforward if we tune the parameters $\De_1,\De_2,\De,J$ to a region $U\in \C^4$ in which all the powers in~\eqref{eq:LtransformShadow3pt} are such that (\ref{eq:theexpression}) is locally bounded (in $x,x_i,z$). This is the analog of choosing $\Re a$ sufficiently large and positive in the toy example of section~\ref{sec:toymodeldist}. For $(\De_1,\De_2,\De,J) \in U$ we automatically get a distribution in $x,x_i,z$ that is analytic in $\De_1,\De_2,\De,J$. We can then define it for the values that we are interested in by analytic continuation. In the interior of $U$ we can ignore the theta-function for the purposes of acting with $\cD_j$: the theta function only jumps when some $x_{ij}^2$ in~\eqref{eq:LtransformShadow3pt} vanishes, but inside $U$ we have a product of all $x_{ij}^2$ appearing with positive powers, and so $K^t$ vanishes on the jump of the theta-function. This is the analog of the computation (\ref{eq:derivativecomputation}) in our toy example. Therefore, in order to compute the action of $\cD_j$ we can act in $U$ just on $K^t$ and then analytically continue to the desired values of $\De_1,\De_2,\De,J$.

Thus our goal is to analytically continue
\be\label{eq:tocontinute}
	&(\cD_jK^t_{\De,J})(x_1,x_2;x_3,z,w)\theta(x_1\approx x_3^+, x_2\approx x_3)\nn\\
	&=\b_{\De,J}\<0|\tl\phi_1(x_1)(\cD_j\wL[\cO])\tl\phi_2(x_2)|0\>\theta(x_1\approx x_3^+)\theta(x_2\approx x_3)
\ee
with the Wightman function given in~\eqref{eq:DjLwightman}. We introduce the following coordinates,
\be
	s=2z\.x_{23}x_{13}^2,\quad t=-2z\.x_{13}x_{23}^2.
\ee
Recall that the prime on the integral in~\eqref{eq:Ktintegral} means that we should only include the configurations where $x_2$ is near the future null cone of $x_3$ and so we can assume $z\.x_{23}<0$.\footnote{This is not true when $x_{23}$ is close to being proportional to $z$. However, here we focus on generic configurations.} We can then rewrite~\eqref{eq:tocontinute}, assuming for simplicity that all points are in the same Poincar\'{e} patch
\be\label{eq:intermediate1}
	&\b_{\De,J}	\frac{(-1)^j2^{2j}(-J)_j
		L(\tl\f_1\tl\f_2[\cO])(\De+j-2)}{(\De-2)\G(j+1)}(w\.x_{13}z\.x_{23}-w\.x_{23}z\.x_{13})^j\nn\\
	&\times\frac{
		(-2z\.x_{23})^{\frac{\tl\De_{12}+(1-J)+(1-\De)}{2}}
		(-2z\.x_{13})^{\frac{-\tl\De_{12}+(1-J)+(1-\De)}{2}}
	}{
		(x_{12}^2)^{\frac{\tl\De_1+\tl\De_2-(1-J)+(1-\De)}{2}}
	}
	\frac{
		(s+t)^{1-\De-j}\theta(s)\theta(t)
	}{
		s^{\frac{\tl\De_{12}+(1-J)+(1-\De)}{2}}
		t^{\frac{-\tl\De_{12}+(1-J)+(1-\De)}{2}}
	}.
\ee
The factors that are not written in terms of $s,t$ will not be important in our analytic continuation:
they are either analytic functions of the coordinates (the $w$-dependent factor), or can only become singular when
$x_1$ becomes almost proportional to $x_2$, or when $x_i$ becomes proportional to $z$, while at the moment we are interested
in more generic configurations.\footnote{Following the logic similar to below, one can check that no new contributions appear in these limiting configurations.}

To find the analytic continuation of the $s,t$-dependent factors, let us define
\be
	a=1-\De-j,\quad b = -\frac{\tl\De_{12}+(1-J)+(1-\De)}{2}, \quad c = -\frac{-\tl\De_{12}+(1-J)+(1-\De)}{2}.
\ee
Then the $s,t$-dependent part becomes
\be
(s+t)^a s^b t^c \theta(s)\theta(t).
\ee
This is well-defined as a distribution analytic in $a,b,c$ for $a,b,c>0$. In appendix~\ref{app:deltapole} we show that near $a+b+c+2=0$ its analytic continuation to general $a,b,c$ has a pole of the form
\be
\label{eq:deltasandgammas}
	(s+t)^a s^b t^c \theta(s)\theta(t)\sim \frac{1}{a+b+c+2}\frac{\G(b+1)\G(c+1)}{\G(b+c+2)}\de(s)\de(t).
\ee
With our definitions we have
\be
	a+b+c+2 = J-j+1,
\ee
which we want to set to $0$. The resulting divergence is canceled in~\eqref{eq:intermediate1} by $\b_{\De,J}$ for $j=0$ or by the $(-J)_j$ factor for $j>0$. Thus, plugging (\ref{eq:deltasandgammas}) into~\eqref{eq:intermediate1} and setting $J=j-1$, we find
\be
	&(\cD_{j}K^t_{\de+1,j-1}\theta(x_1\approx x_3^+, x_2\approx x_3))(x_1,x_2;x_3,z,w)\nn\\
	&=\frac{(-1)^j2^{1+2j}}{i\pi}\a_{\de,j}\frac{
		(w\.x_{13}z\.x_{23}-w\.x_{23}z\.x_{13})^j	
	}{
		(-2x_{13}\.x_{23})^{\frac{\tl\de_1+\tl\de_2+j-\de}{2}}
		(-2z\.x_{23})^{\frac{-\tl\de_{12}+j+\de}{2}}
		(-2z\.x_{13})^{\frac{\tl\de_{12}+j+\de}{2}}
	}\de(x_{13}^2)\de(x_{23}^2)
	,
\ee
Comparing this to~\eqref{eq:Lresult} and to the expression~\eqref{eq:standard3pt_celestial} for the 3-point structure, which we reproduce here with the relevant quantum numbers,
\be
	\<\tl\cP_{\de_1}(z_1)\tl\cP_{\de_2}(z_2)\cP_{\de,j}(z,w)\>=
	\frac{
		4^j(w\.z_1z_2\.z_3-w\.z_2 z_1\.z_3)^j
	}{
		(-2z_1\.z_2)^{\frac{\tl\de_1+\tl\de_2-\de+j}{2}}
		(-2z_1\.z_3)^{\frac{\tl\de_1+\de-\tl\de_2+j}{2}}
		(-2z_2\.z_3)^{\frac{\tl\de_2+\de-\tl\de_1+j}{2}}
	}
\ee
we find
\be
&(\cD_{j}K^t_{\de+1,j-1}\theta(x_1\approx x_3^+, x_2\approx x))(x_1,x_2;x_3,z,w)=
\frac{(-1)^j}{2\pi i}\cL_{\de,j}(x_1,x_2;x_3,z,w).
\ee
Thus~\eqref{eq:WscalarResult} indeed holds with
\be\label{eq:Aresult}
	A=(-1)^j i\pi.
\ee

\subsection{The final form of the scalar OPE}

Combining equations~\eqref{eq:W CPW final},~\eqref{eq:WscalarResult}, and~\eqref{eq:Aresult}, we find the OPE formula
\be\label{eq:scalarOPEfinal}
	&\wL[\f_1](x,z_1)\wL[\f_2](x,z_2)\nn\\
	&=\pi i\sum_{j=0}^{\oo}\int_{\frac{d-2}{2}-i\oo}^{\frac{d-2}{2}+i\oo}\frac{d\de}{2\pi i}\,\cC_{\de,j}(z_1,z_2,\ptl_{z_2},\ptl_{w_2})(-1)^j\p{(\cD_{j} \mathbb{O}^+_{\de+1,J=j-1})(x,z_2,w_2)-(\cD_{j} \mathbb{O}^-_{\de+1,J=j-1})(x,z_2,w_2)}.
\ee
Deforming the contour to the right and picking up the poles from light-ray operators,${}^{\ref{foot:spuriouspoles}}$ we can write
\be
	&\wL[\f_1](x,z_1)\wL[\f_2](x,z_2)\nn\\
	&=-\pi i\sum_{j=0}^{\oo}\sum_i \,\cC_{\De_i-1,j}(z_1,z_2,\ptl_{z_2},\ptl_{w_2})(-1)^j\p{(\cD_{j} \mathbb{O}^+_{i,J=j-1})(x,z_2,w_2)-(\cD_{j} \mathbb{O}^-_{i,J=j-1})(x,z_2,w_2)}.
\ee
Here the differential operator $\cC_{\de,j}$ is defined by~\eqref{eq:3pt=C 2pt}. For $j=0$, taking into account $\cD_0=1$, the result~\eqref{eq:scalarOPEfinal} agrees with~(3.96) in~\cite{Kologlu:2019mfz}, which has been tested in a number of examples~\cite{Kologlu:2019mfz}.\footnote{The minus in front of $\mathbb{O}^-$ has to do with a more explicit treatment of the analytic continuation of $(-1)^J$ factors in this paper.} For $j>0$ the result is new. We will explore an example in which the $j=1$ term is important in section~\ref{sec:N4example}.

Note that in the above expressions, the terms involving $\mathbb{O}^+$ for odd $j$ and terms involving $\mathbb{O}^-$ for even $j$ are related to light-transforms of local operators and are thus annihilated by $\cD_j$ for $j>0$. Similarly, for $j=0$ we have $\mathbb{O}^-_{\De,J=-1}=0$ due to the superconvergence sum rule~\cite{Kologlu:2019bco, Kologlu:2019mfz}, which holds whenever the leading Regge trajectory in $\phi_1\times\phi_2$ OPE has intercept below $J_1+J_2-1=-1$. We indeed have to assume that this is the case, in order for the left-hand side of~\eqref{eq:scalarOPEfinal} to be well-defined~\cite{Kologlu:2019bco}. We thus conclude that in fact the above expressions simplify to
\be\label{eq:scalarOPEfinal_simplified}
&\wL[\f_1](x,z_1)\wL[\f_2](x,z_2)\nn\\
&=\pi i\sum_{j=0}^{\oo}\int_{\frac{d-2}{2}-i\oo}^{\frac{d-2}{2}+i\oo}\frac{d\de}{2\pi i}\,\cC_{\de,j}(z_1,z_2,\ptl_{z_2},\ptl_{w_2})(\cD_{j} \mathbb{O}^{(-1)^{j}}_{\de+1,J=j-1})(x,z_2,w_2),
\ee
and
\be
&\wL[\f_1](x,z_1)\wL[\f_2](x,z_2)=-\pi i\sum_{j=0}^{\oo}\sum_i \,\cC_{\De_i-1,j}(z_1,z_2,\ptl_{z_2},\ptl_{w_2})(\cD_{j} \mathbb{O}^{(-1)^{j}}_{i,J=j-1})(x,z_2,w_2).
\ee

An interesting consequence of this simplification is vanishing of the commutator 
\be
[\wL[\f_1](x,z_1),\wL[\f_2](x,z_2)]=0.
\ee
Indeed, according to the discussion in~\cite{Kologlu:2019mfz}, this commutator has signature $(-1)^{J_1+J_2-1}=-1$. We claim that all the terms in~\eqref{eq:scalarOPEfinal_simplified} have signature~\cite{Kravchuk:2018htv,Kologlu:2019mfz} $+1$ and thus there are no contributions to the commutator. To see why 
\be
\cD_{j} \mathbb{O}^{(-1)^{j}}_{i,J=j-1}
\ee
has signature $+1$, we need to analyze two cases. Note that for $j=0$ we have
\be
\cD_{j} \mathbb{O}^{(-1)^{j}}_{i,J=j-1}=\mathbb{O}^{+}_{i,J=-1},
\ee
which manifestly has signature $+1$. For $j>0$ the operator $\cD_j$ is non-trivial, and we claim that it changes signature by $(-1)^j$, in which case we again see that
\be
\cD_{j} \mathbb{O}^{(-1)^{j}}_{i,J=j-1}
\ee
has signature $+1$. To understand how $\cD_j$ changes signature, it is convenient to use CRT symmetry $\cJ_0$ under which the light-ray operators have the property
\be
	\p{\cJ_0 \mathbb{O}^{\pm}_{i,J}(0,z,\bw) \cJ_0^{-1}}^\dagger=\pm\mathbb{O}^{\pm}_{i,J}(0,z,\bw)
\ee
for all $z,\bw$.\footnote{The CRT transformation $\cJ_0$ can be obtained from the CRT transformation $\cJ_\O$ described around~\eqref{eq:signatureop} below by conjugating with a conformal transformation which brings the point $(x^+=-\oo,x^-=0,x^i=0)$ to $x=0$ and the point $(x^+=0,x^-=-\oo,x^i=0)$ to spatial infinity.} Writing then
\be
	(\cD_j\mathbb{O}^{\pm}_{i,J})(0,z,\bw)\propto \cD_{z,w;\mu_1}^{0+}\cdots \cD_{z,w;\mu_j}^{0+}
	[P^{\mu_1},\cdots[P^{\mu_j},\mathbb{O}^{\pm}_{i,J}(0,z,\bw)]\cdots],
\ee
where $\cD_{z,w;\mu}^{0+}$ is a differential operator in $z,w$ defined in~\eqref{eq:Dndef}, and taking into account $\cJ_0 P^\mu \cJ_0^{-1}= -P^\mu$ and $P_\mu^\dagger=-P_\mu$, we find
\be
\p{\cJ_0(\cD_j\mathbb{O}^{\pm}_{i,J})(0,z,\bw)\cJ_0^{-1}}^\dagger=\pm (-1)^j(\cD_j\mathbb{O}^{\pm}_{i,J})(0,z,\bw),
\ee
and therefore $\cD_j$ indeed changes the signature by $(-1)^j$. Finally, note that the differential operator $\cC_{\de,j}$ does not affect signature because it acts in the transverse space (i.e.\ on $z,\bw$) which is not affected by $\cJ_0$.

\section{The complete OPE of general detectors}
\label{sec:general OPE}

In the previous section, we derived the form of the light-ray OPE for a product of light-transformed scalar operators $\wL[\f_1]\wL[\f_2]$. We now derive a generalization for light-transforms of operators in arbitrary Lorentz representations, of which the scalar formula~\eqref{eq:scalarOPEfinal} is a special case. The generalized light-ray OPE formula is
\be
\label{eq:lightrayope_generalops}
&\wL[\cO_1](x,z_1,\bw_1)\wL[\cO_2](x,z_2,\bw_2)\nn \\
&=\pi i(-1)^{J_1+J_2}\sum_{\l \in \L_{12}}\int_{\frac{d-2}{2}-i\oo}^{\frac{d-2}{2}+i\oo}\frac{d\de}{2\pi i}\cC_{\de,\l}^{(a)}(z_1,\bw_1,z_2,\bw_2,\ptl_{z_2},\ptl_{\bw_2})\mathbb{O}^{(-1)^{J_1+J_2}}_{\de+1,J_1+J_2-1,\l,(a)}(x,z_2,\bw_2)\nn \\
&+\pi i(-1)^{J_1+J_2}\sum_{n=1}^{\oo}\sum_{\g \in \G_{12}}\int_{\frac{d-2}{2}-i\oo}^{\frac{d-2}{2}+i\oo}\frac{d\de}{2\pi i}\cC_{\de,\l_\g(+n)}^{(a)}(z_1,\bw_1,z_2,\bw_2,\ptl_{z_2},\ptl_{\bw_2})(\cD_{n}\mathbb{O}^{(-1)^{J_1+J_2+n}}_{\de+1,J_1+J_2-1+n,\l_{\g},(a)})(x,z_2,\bw_2)
\ee
Owing to its generality, this expression is a bit unwieldy, so let us unpack it. (We also give several concrete examples in section~\ref{sec:examples}.) The left-hand side is a product of light-transforms of operators $\cO_1$ and $\cO_2$, which have the quantum numbers $(\De_i,J_i,\l_i)$.\footnote{For simplicity we focus on bosonic representations only.} The right-hand side contains two sets of terms. The first set (on line 2) contains what we refer to as the ``low transverse spin'' contributions, studied in detail in~\cite{Kologlu:2019mfz}. The second set (line 3) contains ``higher transverse spin'' contributions, which are new and require the use of the differential operators $\cD_n$ from section~\ref{sec:differential operators}. 

The operators $\cC^{(a)}_{\de,\l}$ are similar to the operators $\cC_{\de,j}$ from section~\ref{sec:scalar OPE}. They encode the kinematic structure of the OPE in a fictitious $(d-2)$-dimensional CFT on the celestial sphere. There are two important differences from section~\ref{sec:scalar OPE}. Firstly, in the general case considered here, there can be several OPE coefficients that enter a given OPE, each with its own $\cC^{(a)}_{\de,\l}$. The structure label $(a)$ labels these different OPE coefficients. A sum over $(a)$ is implicit in (\ref{eq:lightrayope_generalops}). As an example of the role of the structure label $(a)$, recall that deforming the $\de$-contour in~\eqref{eq:lightrayope_generalops} to the right and picking up the poles, we obtain a discrete sum of terms
\be\label{eq:DOcombination}
	\cC_{\de,\l}^{(a)}(z_1,\bw_1,z_2,\bw_2,\ptl_{z_2},\ptl_{\bw_2})\mathbb{O}^{\pm}_{i,J_1+J_2-1,\l(a)}(x,z_2,\bw_2).
\ee
Recall that $\mathbb{O}^{\pm}_{i,J_1+J_2-1,\l(a)}$ are built with reference to $\cO_1,\cO_2$ and are related to canonically-normalized light-ray operators as
\be
	\mathbb{O}^{\pm}_{i,J_1+J_2-1,\l(a)}=f_{12\cO^\dagger,(a)}\hat{\mathbb{O}}^{\pm}_{i,J_1+J_2-1,\l},
\ee
where $f_{12\cO^\dagger,(a)}$ are analytically-continued OPE coefficients in the $\cO_1\times\cO_2$ OPE. Equation~\eqref{eq:DOcombination} then becomes 
\be
f_{12\cO^\dagger,(a)}\cC_{\de,\l}^{(a)}(z_1,\bw_1,z_2,\bw_2,\ptl_{z_2},\ptl_{\bw_2})\hat{\mathbb{O}}^{\pm}_{i,J_1+J_2-1,\l}(x,z_2,\bw_2),
\ee
Thus, the index $(a)$ of $\cC_{\de,\l}^{(a)}$ is naturally contracted with OPE coefficients. Similar statements hold for higher transverse spin terms as well.

The second distinction has to do with the normalization of $\cC_{\de,\l}^{(a)}$. The operators $\cC_{\de,j}$ were normalized by equation~\eqref{eq:3pt=C 2pt}, which is formulated in terms of celestial three-point structures. On the other hand, the OPE coefficients $f_{12\cO^\dagger,(a)}$ are defined in terms of $d$-dimensional tensor structures, and thus the normalization condition~\eqref{eq:3pt=C 2pt} is only correct due to our specific choice of conventions for both $d$-dimensional as well as celestial tensor structures. The main result of this section will be the proof of~\eqref{eq:lightrayope_generalops} together with the simple, convention-independent, equations~\eqref{eq:ltsDefinitionSimple} and~\eqref{eq:htsDefinitionSimple} that determine the  $\cC_{\de,\l}^{(a)}$ in terms of $d$-dimensional data.\footnote{Note that up to an action by an invertible matrix on the index $(a)$, these operators are completely fixed by the $(d-2)$-dimensional conformal symmetry (equivalently, $d$-dimensional Lorentz symmetry). These equations thus simply determine a preferred basis of these operators.} We refer to these formulas as ``celestial map formulas'' because they map the $d$-dimensional three-point tensor structures (which naturally pair with the OPE coefficients $f_{12\cO^\dagger,(a)}$) to celestial sphere differential operators $\cC_{\de,\l}^{(a)}$.\footnote{As we explain below, it will sometimes happen that $\cC_{\de.\l}^{(a)}$ vanishes for some values of $a$, i.e.\ not all of the OPE coefficients actually appear in~\eqref{eq:DOcombination}.}

The remaining notation in~\eqref{eq:lightrayope_generalops} has to do with quantum numbers of exchanged operators and the respective selection rules. In the low transverse spin terms we are summing over transverse spins $\l\in\L_{12}$ which simply means all the transverse spins that appear in the usual local OPE $\cO_1\times \cO_2$ \textit{and} the celestial OPE of operators with $(d-2)$-dimensional spins $\l_1\times\l_2$.  These are the transverse spins for which the operators $\mathbb{O}^\pm_{i,J,\l,(a)}$ can be constructed from $\cO_1$ and $\cO_2$ and for which the operators $C^{(a)}_{\de,\l}$ make sense. 

As an example, when $\cO_1,\cO_2$ are scalars, $\Lambda_{12}$ contains only the trivial representation of $\SO(d-2)$ because only traceless-symmetric operators appear in OPE of $d$-dimensional scalars. In this case, the constraint that $\l\in \L_{12}$ should appear in the OPE of $(d-2)$ dimensional scalars is trivially satisfied. On the other hand, in the example of energy-energy OPE discussed in section~\ref{sec:examples} both constraints become non-trivial.

In the higher transverse spin terms, we sum over $\l_\g$, $\g\in \G_{12}$. These are the transverse spins for which the action of the operators $\cD_n$ is well-defined. Concretely, these are transverse spins with the first row of the $\SO(d-2)$ Young diagram (the second row of the $\SO(d-1,1)$ Young diagram) of length $J_1+J_2$. We write $\l_{\g}=(J_1+J_2,\g)$ with $\g$ an $\SO(d-4)$ irrep. Since the action of $\cD_n$ raises transverse spin, we use the OPE differential operator $\cC^{(a)}_{\de,\l_\g(+n)}$, where $\l_\g(+n)=(J_1+J_2+n,\g)$. The set $\G_{12}$ consists of $\g$ for which transverse spin $\l_\g$ appears in $d$-dimensional $\cO_1\times\cO_2$ OPE and at the same time $\l_\g(+n)$ appears in the $(d-2)$-dimensional celestial OPE. (As explained in appendix~\ref{app:tensorstructureselectionrule}, these two conditions are in fact equivalent.)

Before proceeding with the derivation, let us comment again on the relation of this section to~\cite{Kologlu:2019mfz}. In \cite{Kologlu:2019mfz}, the celestial map formula for lower transverse spin was derived by a rather non-trivial procedure using the Lorentzian inversion formula. In this section, we will give a much simpler derivation of the celestial map formula for both lower and higher transverse spin. The drawback of this simpler derivation is that it is based on the assumption that the light-ray kernel $K^{t}_{\De,J,\l(a)}(x_1,x_2;x,z)$ satisfying~\eqref{eq:lorentzianformulaforlightraykernel} localizes on the null cone of $x$ as we set $J$ to certain values and possibly act with $\cD_n$, analogously to the scalar case in section~\ref{sec:scalar OPE}. This assumption is plausible in the sense that it gives a natural generalization of the pattern observed in concrete examples, and is furthermore purely kinematical. It therefore appears to be a purely technical problem to prove it. (See sections~\ref{sec:lts} and~\ref{sec:hts} for the precise statement of our assumptions and the supporting evidence.)

For completeness, in appendix~\ref{app:fixingtheoldderivation}, we derive~\eqref{eq:lightrayope_generalops} by generalizing the derivation in \cite{Kologlu:2019mfz}. This derivation, although being more technical than the one in sections~\ref{sec:lts} and~\ref{sec:hts} (which is why it is relegated to an appendix), does not rely on the assumptions discussed above.

\subsection{A formula for the light-ray operator kernel}

Our starting point is the following intrinsically-Lorentzian description of light-ray operators. Recall that the
light-ray operators are defined by the integral~\cite{Kravchuk:2018htv}
\be
\label{eq:lightrayopgeneral}
	\mathbb{O}^\pm_{i,J,\l(a)}(x,z)=\mathrm{res}_{\De=\De_i}&\pm\int_{\substack{x\approx 2 \\ x\approx 1^-}} d^dx_1d^dx_2 K^{t}_{\De,J,\l(a)}(x_1,x_2;x,z)\cO_1\cO_2\nn\\
	&+\int_{\substack{x\approx 1 \\ x\approx 2^-}} d^dx_1d^dx_2 K^{u}_{\De,J,\l(a)}(x_2,x_1;x,z)\cO_2\cO_1.
\ee
We claim that the kernel $K^{t}$ is determined by the equation
\be
&\int_{\substack{2>x'>1^- \\ x\approx 2, 1^-}}\frac{d^dx_1d^dx_2}{\vol(\SO(1,1))^2}K^{t}_{\De,J,\l(a)}(x_1,x_2;x,z)\<0|\cO_2\wL[\cO^{\dag}](x',z')\cO_1|0\>^{(b)}_+ \nn \\
&=\frac{1}{2\pi i}\<\wL[\cO](x,z)\wL[\cO^{\dag}](x',z')\>\de^{(b)}_{(a)},
\label{eq:lorentzianformulaforlightraykernel}
\ee
together with the condition that it has the analyticity and conformal transformation properties of
\be
	\<0|\tl\cO_1^\dagger(x_1) \cO^L(x,z) \tl\cO_2^\dagger(x_2)|0\>,
\ee
where $\cO^L$ has the quantum numbers of $\wL[\cO]$.
The kernel $K^{u}$ is defined similarly and will be described below.\footnote{An intuitive picture behind~\eqref{eq:lorentzianformulaforlightraykernel} is as follows. The Wightman three-point structure in~\eqref{eq:lorentzianformulaforlightraykernel}, together with the condition $2>x'>1^-$, can be viewed as one of two parts of the light-transform of a time-ordered three-point structure~\cite{Kravchuk:2018htv}. The integration against $K^{t}$ is then similar to producing $\wL[\cO]$ from $\cO_1,\cO_2$ inside of this time-ordered three-point function, which should be equal to the two-point function in the right-hand side of~\eqref{eq:lorentzianformulaforlightraykernel}. It would be interesting if this intuitive reasoning could be made precise: our derivation of~\eqref{eq:lorentzianformulaforlightraykernel} is based on the generalized Lorentzian inversion formula of~\cite{Kravchuk:2018htv}, where the latter is derived from Euclidean harmonic analysis. It would be instructive to bypass the Euclidean argumentation altogether.} In~\eqref{eq:lightrayopgeneral} and in what follows we keep the transverse indices (encoded previously by polarization vectors $\bw$) implicit in order to avoid excessive clutter in the notation. The way these indices are contracted should in all cases be clear from the context.

Above, we use the notation that $\cO_i$ is at point $x_i$. The kernel $K^{t}$ carries Lorentz indices for points $x_1,x_2$ that are in the dual representations to $\cO_1,\cO_2$, and these indices are contracted with $\cO_1,\cO_2$ in (\ref{eq:lightrayopgeneral}) and (\ref{eq:lorentzianformulaforlightraykernel}).
We prove (\ref{eq:lorentzianformulaforlightraykernel}) in appendix~\ref{app:lorentziankernel}. An advantage of (\ref{eq:lorentzianformulaforlightraykernel}) relative to the definition in \cite{Kravchuk:2018htv} is that it makes reference only to Lorentzian objects. By contrast, the definition in \cite{Kravchuk:2018htv} includes the Plancherel measure for the Euclidean conformal group and Euclidean shadow coefficients, and is thus more awkward to use in a purely Lorentzian setting.

In order to use~\eqref{eq:lorentzianformulaforlightraykernel} for non-integer $J$, it remains to explain the meaning of the objects
\be
&\<\wL[\cO](x,z)\wL[\cO^{\dag}](x',z')\>,\label{eq:LLtwopt}\\
&\<0|\cO_2\wL[\cO^{\dag}](x',z')\cO_1|0\>^{(b)}_+\label{eq:Lthreept}
\ee
for such $J$.

\subsubsection{Analytic continuation in spin}
\label{sec:Jcontinuation}

We define~\eqref{eq:LLtwopt} and~\eqref{eq:Lthreept} by extending the definition of the underlying structures
\be
&\<\cO(x,z)\cO^{\dag}(x',z')\>,\label{eq:Jtwopt}\\
&\<0|\cO_2\cO^{\dag}(x',z')\cO_1|0\>^{(b)}_+\label{eq:Jthreept}
\ee
to non-integer $J$ and then taking the necessary light-transforms.

Note that fixing the normalization of local operators and setting the conventions for three-point structures of local operators involves
specifying the expressions for
\be
&\<\cO(x,z)\cO^{\dag}(x',z')\>,\\
&\<\cO_1(x_1,z_1)\cO_2(x_2,z_2)\cO^{\dag}(x,z)\>^{(b)}.
\ee
for integer spin $J$. We will now impose certain constraints on these choices which will simplify our general analysis with regard to factors of $(-1)^J$ which have ambiguous analytic continuation in $J$.

For the two-point function, note that for any fixed $n\in \Z$
\be
	\<\cO(x_1,z_1)\cO^{\dag}(x_2,z_2)\>(-2z_1\.I(x_{12})\.z_2)^{n-J}
\ee
is conformally-invariant with quantum numbers independent of $J$. Moreover, for sufficiently large $n$, the Lorentz weights with which it transforms are dominant\footnote{I.e.\ the first row of Young diagram is at least as long as the second one.} and we can take this structure to be equal to a fixed, $J$-independent, two-point function of local operators
\be
	\<\cO(x_1,z_1)\cO^{\dag}(x_2,z_2)\>(-2z_1\.I(x_{12})\.z_2)^{n-J}=f_0(x_1,z_1;x_2,z_2).
\ee
So we find that one can always choose 
\be\label{eq:twoptchoice}
	\<\cO(x_1,z_1)\cO^{\dag}(x_2,z_2)\>=f_0(x_1,z_1;x_2,z_2)(-z_1\.I(x_{12})\.z_2)^{J-n}.
\ee
We allow to modify this convention by exponential factors such as $2^J$ but not by $(-1)^J$.

Note that the expression~\eqref{eq:twoptchoice} can be rewritten as
\be
&f_0(x_1,z_1;x_2,z_2)(-z_1\.I(x_{12})\.z_2)^{J-n}\nn\\
&=f_0(x_1,z_1;x_2,z_2)(2(z_1\.x_{12})(z_2\.x_{12})-(z_1\.z_2) x_{12}^2)^{J-n}(x_{12}^2)^{n-J}.
\ee
Using the fact that $2(z_1\.x_{12})(z_2\.x_{12})-(z_1\.z_2) x_{12}^2>0$ for generic configurations,\footnote{To see this, note that one can always write $x_{12}=\a z_1+\b z_2+x_\perp$ for $x_\perp\. z_i=0$. We have then $2(z_1\.x_{12})(z_2\.x_{12})-(z_1\.z_2) x_{12}^2=(-z_1\.z_2)x_\perp^2>0$, where we used that $x_\perp$ is spacelike due to being orthogonal to the timelike vector $z_1+z_2$.} we see that the usual time-ordered $i\e$ prescription unambiguously defines the time-ordered two-point function~\eqref{eq:twoptchoice} for generic configurations of $x_i,z_i$. This is sufficient to apply the light-transforms in~\eqref{eq:LLtwopt}.

A similar argument for $\<\cO_1(x_1,z_1)\cO_2(x_2,z_2)\cO^{\dag}(x,z)\>^{(b)}$ shows that we can write, for sufficiently large integer $J$,
\be\label{eq:threeptchoice}
	\<\cO_1(x_1,z_1)\cO_2(x_2,z_2)\cO^{\dag}(x_3,z_3)\>^{(b)}=f_0^{(b)}(x_1,z_1;x_2,z_2;x_3,z_3)\frac{(2z\.x_{23}\, x_{13}^2-2z\.x_{13}\,x_{23}^2)^{J-n}}{x_{12}^{J-n}x_{23}^{J-n}x_{13}^{J-n}},
\ee
where $n$ is sufficiently large so that the Lorentz weights of $f_0$ are dominant, and the basis $f_0^{(b)}(x_1,z_1;x_2,z_2;x_3,z_3)$ can be chosen to be $J$-independent. There are two possible subtleties here. Firstly, the operators $\cO_1$ and $\cO_2$ can be identical, in which case permutation invariance will typically constrain the structures for even and odd $J$ differently. Secondly, some of $\cO_1,\cO_2$ can be conserved currents, in which case the conservation constraints will typically require some non-trivial polynomial dependence of $f_0^{(b)}$ on $J$. In both cases, our solution is to use the generic basis of structures, ignoring these constraints. The OPE coefficients computed in this basis may satisfy some linear equations, but this will not affect any of our arguments.

The problem of defining three-point structures for non-integer $J$ is complicated by the fact that $2z\.x_{23}\, x_{13}^2-2z\.x_{13}\,x_{23}^2$ is in general not sign-definite~\cite{Kravchuk:2018htv}. We will define the analytic continuation directly for
$\<0|\cO_2\cO^{\dag}(x_3,z_3)\cO_1|0\>^{(b)}$ in the configuration $2>3,1\approx 3,1\approx 2$, where we have
\be
2z\.x_{23}\, x_{13}^2-2z\.x_{13}\,x_{23}^2<0.
\ee
For integer $J$ we have
\be
	\<0|\cO_2\cO^{\dag}(x_3,z_3)\cO_1|0\>^{(b)}=(-1)^J(-2z\.x_{23}\, x_{13}^2+2z\.x_{13}\,x_{23}^2)^J\times (\cdots),
\ee
where the dots represent the standard analytic continuation of all the other factors. For general $J\in \C$ we define
\be
	\<0|\cO_2\cO^{\dag}(x_3,z_3)\cO_1|0\>^{(b)}_\pm=\pm(-2z\.x_{23}\, x_{13}^2+2z\.x_{13}\,x_{23}^2)^J\times (\cdots).
\ee
In all other configurations these Wightman functions are determined by the usual analytic continuation, which is unambiguous for this ordering even for $J\in \C$~\cite{Kravchuk:2018htv}

We will additionally use the ``time-ordered'' structures, which are defined to be equal to 
\be\label{eq:timeorederedcontinuation}
	\<\cO_1(x_1,z_1)\cO_2(x_2,z_2)\cO^{\dag}(x_3,z_3)\>^{(b)}_+&=f_0(x_1,z_1;x_2,z_2;x_3,z_3)\frac{(2z\.x_{23}\, x_{13}^2-2z\.x_{13}\,x_{23}^2)^{-n}}{x_{12}^{J-n}x_{23}^{J-n}x_{13}^{J-n}}\nn\\
	&\quad\times |2z\.x_{23}\, x_{13}^2-2z\.x_{13}\,x_{23}^2|^{J}
\ee
for spacelike-separated points and are defined in other configurations by usual time-ordered $i\e$-prescriptions applied to everything except $|\cdots|^J$. These structures are useful because
\be\label{eq:continuedLT}
	\<\cO_1\cO_2\wL[\cO^{\dag}]\>^{(b)}_+=
	\<0|\cO_2\wL[\cO^{\dag}]\cO_1|0\>^{(b)}_+ \theta(2>3>1^-)+(\cdots)\theta(1>3>2^-),
\ee
similarly to the relation between the light-transform of integer-spin time-ordered and Wightman correlators~\cite{Kravchuk:2018htv}. Here dots represent a structure that is a bit awkward to describe but which we won't need in what follows.

Our choice of conventions for traceless-symmetric operators described in appendix~\ref{app:structureconventions} satisfies the above constraints.

\subsubsection{Definition of $K^u$}

The kernel $K^{u}$ is defined by requiring that it has the analyticity and conformal transformation properties of
\be
\<0|\tl\cO_2^\dagger(x_2) \cO^L(x,z) \tl\cO_1^\dagger(x_1)|0\>,
\ee
and the requirement that when we set $x=-\oo z$ and $z^0=z^1=1, z^i=0$, the following equality holds
\be
	&\int_{\substack{x\approx 1 \\ x\approx 2^-}} d^dx_1d^dx_2 K^{u}_{\De,J,\l(a)}(x_2,x_1;x,z)\cO_2\cO_1
	=\overline{\int_{\substack{x\approx 2 \\ x\approx 1^-}} d^dx_1d^dx_2 K^{t}_{\De,J,\l(a)}(x_1,x_2;x,z)\cO_1\cO_2}.
\ee
Here, we have defined the linear operation
\be\label{eq:signatureop}
	\overline A = (\cJ_\O A \cJ_\O^{-1})^\dagger.
\ee
Here $\cJ_\O$ is the anti-unitary operator implementing the CRT symmetry, where the reflection acts as $x^1\to -x^1$. This relation should hold for all $J$. For integer $J$ it is equivalent to
\be
	&\int_{\substack{1>x'>2^- \\ x\approx 1, 2^-}}\frac{d^dx_1d^dx_2}{\vol(\SO(1,1))^2}K^{u}_{\De,J,\l(a)}(x_1,x_2;x,z)\<0|\cO_1\wL[\cO^{\dag}](x',z')\cO_2|0\>^{(b)} \nn \\
	&=\frac{1}{2\pi i}\<\wL[\cO](x,z)\wL[\cO^{\dag}](x',z')\>\de^{(b)}_{(a)}.
\ee

\subsection{Derivation of the OPE formula}\label{sec:celestialmap_lightransform}

\subsubsection{Harmonic analysis on celestial sphere}
\label{sec:harmonic_analysis_fact}

We have given a review of harmonic analysis on celestial sphere in section~\ref{sec:harmonicanalysisscalar}. In this section we  need a generalization which we state here without proof, and refer the reader instead to~\cite{Kologlu:2019mfz}.

Suppose we have a choice of two- and three-point structures on the celestial sphere
\be
\label{eq:choose2ptbasis}
	&\<\cP_{\de,\l}(z)\cP_{\de,\l}^\dagger(z')\>
	\\
	\label{eq:choose3ptbasis}
	&\<\cP_{\de_1,\l_1}(z_1)\cP_{\de_2,\l_2}(z_2)\cP^\dagger_{\de,\l}(z)\>^{(a)}
\ee
We assume that the three-point structures are linearly-independent and span the space of conformally-invariant tensor structures for the given quantum numbers. Given these structures, we can find OPE differential operators $\hat \cC_{\de,\l}^{(a)}$ satisfying
\be\label{eq:dcondition}
	\hat\cC^{(a)}_{\de,\l}(z_1,z_2;\ptl_{z_2})\<\cP_{\de,\l}(z_2)\cP_{\de,\l}^\dagger(z')\>=
	\<\cP_{\de_1,\l_1}(z_1)\cP_{\de_2,\l_2}(z_2)\cP^\dagger_{\de,\l}(z)\>^{(a)}.
\ee
Here and below, the $\cP_{\de,\l}$ carry $\SO(d-2)$ indices for $\l$, which we suppress for brevity. These indices are implicitly contracted between $\hat \cC_{\de,\l}^{(a)}$ and $\cP_{\de,\l}(z_2)$.

Now suppose that conformally-invariant kernels $k_{\de,\l,(a)}(z_1,z_2;z)$ solve the equation
\be\label{eq:kcondition}
	\int \frac{D^{d-2}z_1 D^{d-2}z_2}{\vol(\SO(1,1))} k_{\de,\l,(a)}(z_1,z_2;z)\<\cP_{\de_1,\l_1}(z_1)\cP_{\de_2,\l_2}(z_2)\cP^\dagger_{\de,\l}(z')\>^{(b)}=
	\<\cP_{\de,\l}(z)\cP_{\de,\l}^\dagger(z')\>\de^{(b)}_{(a)}.
\ee
Here, $k_{\de,\l,(a)}(z_1,z_2;z)$ carries $\SO(d-2)$ indices dual to $\cP_{\de_1,\l_1}$ and $\cP_{\de_2,\l_2}$ and these indices are implicitly contracted in (\ref{eq:kcondition}).
If the above conditions hold for all $\l$'s that can appear in the $(d-2)$-dimensional OPE $\cP_{\de_1,\l_1}\times\cP_{\de_2,\l_2}$, then 
we have
\be\label{eq:CPWresult}
	\wL[\cO_1](x,z_1)\wL[\cO_2](x,z_2)=\sum_\l \int^{\frac{d-2}{2}+i\oo}_{\frac{d-2}{2}-i\oo}
	\frac{d\de}{2\pi i}\hat\cC^{(a)}_{\de,\l}(z_1,z_2;\ptl_{z_2})\mathbb{W}_{\de,\l,(a)}(x,z_2),
\ee
where
\be
	\mathbb{W}_{\de,\l,(a)}(x,z)\equiv\int D^{d-2}z_1 D^{d-2}z_2 k_{\de,\l,(a)}(z_1,z_2;z)\wL[\cO_1](x,z_1)\wL[\cO_2](x,z_2).
\ee
For the validity of~\eqref{eq:CPWresult} the same caveats as in footnote~\ref{foot:harmonicapology} apply.
Note that it is not important which basis of structures one chooses in~\eqref{eq:choose3ptbasis} and~\eqref{eq:choose2ptbasis} --- the above statement is basis-independent. We utilize this freedom below, making
convenient choices when appropriate.

\subsubsection{Lower transverse spin}
\label{sec:lts}

For lower transverse spin, we will assume that when we set $J\to J_1+J_2-1$, the light-ray kernel $K^{t}_{\De,J,\l(a)}$ degenerates to
\be\label{eq:lowtransversespindegeneration}
	&\cN^{(a)}_{(c),\de,\l}\int_{x\approx 2\atop x\approx 1^-} d^dx_1d^dx_2K^{t}_{\de+1,J,\l,(a)}(x_1,x_2;x,z)\cO_1(x_1)\cO_2(x_2) \nn \\
	&\to \int D^{d-2}z_1D^{d-2}z_2k_{\de,\l,(c)}(z_1,z_2;z)\wL[\cO_1](x,z_1)\wL[\cO_2](x,z_2)
\ee
where $k_{\de,\l,(c)}$ is a basis of Lorentz-invariant kernels, and $\cN^{(a)}_{(c),\de,\l}$ is a rectangular matrix (implicit summation over $(a)$ is understood). The matrix $\cN$ is needed because not every $K^{t}_{\de+1,J,\l,(a)}$ degenerates in this way. We assume that all Lorentz-invariant kernels $k_{\de,\l,(c)}$ can be generated in this way i.e.\ $\cN$ is full rank.

The evidence for this assumption comes from several lines of reasoning. First of all, it agrees with the scalar example in section~\ref{sec:scalar OPE}. Moreover, we have additionally studied the example of $\wL[\f]\wL[V]$ OPE, where $V$ is an operator with spin 1, which has a non-trivial matrix $\cN$ and verified this statement. Furthermore, from the counting of structures in appendix~\ref{app:tensorstructureselectionrule} one can see that the dimensions of the spaces of structures match (specifically, the rank of $\cN$ can be predicted from~\eqref{eq:celetialmapkernel} below, and it matches the number of $k_{\de,\l,(c)}$). Finally,~\eqref{eq:lowtransversespindegeneration} yields the same results as in~\cite{Kologlu:2019mfz} and in appendix~\ref{app:fixingtheoldderivation}, where they have been obtained by an different, although more complicated, method. It would be interesting to find a direct derivation of~\eqref{eq:lowtransversespindegeneration}.

Plugging~\eqref{eq:lowtransversespindegeneration} into~\eqref{eq:lorentzianformulaforlightraykernel}, we obtain
\be\label{eq:kernelkdefinition_lowerj_pre}
	&\int D^{d-2}z_1D^{d-2}z_2k_{\de,\l(c)}(z_1,z_2;z)\frac{\<0|\wL^{+}[\cO_2](x,z_2)\wL[\cO^{\dag}](x',z')\wL^{-}[\cO_1](x,z_1)|0\>^{(a)}_+}{\vol(\SO(1,1))^2} \nn \\
	&=\frac{1}{2\pi i}\<\wL[\cO](x,z)\wL[\cO^{\dag}](x',z')\>\cN^{(a)}_{(c),\de,\l}.
\ee
where $\wL^{+}[\cO_2]$ indicates that the light transform contour is restricted to $2>x'$ and $\wL^{-}[\cO_1]$ indicates that the light transform contour is restricted to $x'>1^{-}$. On the left-hand side, one factor of $\vol(\SO(1,1))$ cancels against a zero-mode in the integral over $z_1,z_2$, and the other factor cancels against a zero-mode in the triple light-transform of the three-point structure. 

Let us define a matrix $\cM^{(c)}_{(a),\de,\l}$ with the property
\be\label{eq:Mdefn}
	\cM^{(c)}_{(a),\de,\l}\cN^{(a)}_{(c'),\de,\l}=\de^{(c)}_{(c')}.
\ee
This is always possible to do since $\cN$ is full rank. However, $\cM$ defined in this way is in general ambiguous because we can replace
\be\label{eq:Mambiguity}
	\cM^{(c)}_{(a),\de,\l}\to \cM^{(c)}_{(a),\de,\l}+m^{(c)}_{(a)}
\ee
for any $m^{(c)}_{(a)}$ such that $m^{(c)}_{(a)}\cN^{(a)}_{(c'),\de,\l}=0$. 
We make a choice of $\cM$ and rewrite~\eqref{eq:kernelkdefinition_lowerj_pre} as
\be
\label{eq:kernelkdefinition_lowerj}
&\int D^{d-2}z_1D^{d-2}z_2k_{\de,\l(c)}(z_1,z_2;z)\frac{\<0|\wL^{+}[\cO_2](x,z_2)\wL[\cO^{\dag}](x',z')\wL^{-}[\cO_1](x,z_1)|0\>^{(a)}_+}{\vol(\SO(1,1))^2}\cM^{(c')}_{(a),\de,\l} \nn \\
&=\frac{1}{2\pi i}\<\wL[\cO](x,z)\wL[\cO^{\dag}](x',z')\>\de^{(c')}_{(c)}.
\ee
In the left-hand side we have a celestial ``bubble integral'' of $k_{\de,\l(c)}(z_1,z_2;z)$ against
\be
\frac{\<0|\wL^{+}[\cO_2](x,z_2)\wL[\cO^{\dag}](x',z')\wL^{-}[\cO_1](x,z_1)|0\>^{(a)}_+}{\vol(\SO(1,1))}\cM^{(c')}_{(a),\de,\l}.
\ee
Since $k_{\de,\l(c)}(z_1,z_2;z)$ form a basis, it follows that this bubble integral is non-degenerate~\cite{Karateev:2018oml}, i.e.\ it vanishes for all $(c)$ if and only if the above expression vanishes. Due to~\eqref{eq:Mambiguity} it then follows that
\be\label{eq:celetialmapkernel}
\frac{\<0|\wL^{+}[\cO_2](x,z_2)\wL[\cO^{\dag}](x',z')\wL^{-}[\cO_1](x,z_1)|0\>^{(a)}_+}{\vol(\SO(1,1))}m^{(c')}_{(a)}=0
\ee
for any $m^{(c)}_{(a)}$ such that $m^{(c)}_{(a)}\cN^{(a)}_{(c'),\de,\l}=0$.

Equation~\eqref{eq:kernelkdefinition_lowerj} has almost the same structure as~\eqref{eq:kcondition} if we identify, for fixed $x, x'$,
\be
	\<\cP_{\de_1,\l_1}(z_1)\cP_{\de_2,\l_2}(z_2)\cP^\dagger_{\de,\l}(z)\>^{(c)} &\leftrightarrow 
	\frac{\<0|\wL^{+}[\cO_2](x,z_2)\wL[\cO^{\dag}](x',z')\wL^{-}[\cO_1](x,z_1)|0\>^{(a)}_+}{\vol(\SO(1,1))}\cM^{(c)}_{(a),\de,\l},\\
	\<\cP_{\de,\l}(z)\cP_{\de,\l}^\dagger(z')\>&\leftrightarrow \frac{1}{2\pi i}\<\wL[\cO](x,z)\wL[\cO^{\dag}](x',z')\>.
\ee
However, we cannot quite write equality signs in the relations above, because while $z_1,z_2$ and the transverse indices of $\wL^+[\cO_2], \wL^-[\cO_1]$ transform under the Lorentz group at $x$, the polarization $z'$ and transverse indices of $\wL[\cO^\dagger]$ transform under the Lorentz group at $x'$. Fortunately, we can perform two steps to rectify this problem. First, there is a subgroup $L_{xx'}\simeq\SO(d-1,1)$ of conformal transformations that fix both $x$ and $x'$, and act as Lorenz transformations locally at these points.\footnote{For example, we can set $x=0$ and $x'=\oo$, in which case the group in question is the standard Lorentz group.} Second, we can ``translate'' the indices and polarizations of $\wL[\cO^\dagger]$ from $x'$ to $x$, so that the action of $L_{xx'}$ is the same on all indices and polarizations. Specifically, we can write 
\be\label{eq:identificationLowSpin}
\<\cP_{\de_1,\l_1}(z_1)\cP_{\de_2,\l_2}(z_2)\cP^\dagger_{\de,\l}(z)\>^{(c)} &= 
\frac{\<0|\wL^{+}[\cO_2](x,z_2)(I_{xx'}\wL[\cO^{\dag}])(x',I_{xx'}z')\wL^{-}[\cO_1](x,z_1)|0\>^{(a)}_+}{\vol(\SO(1,1))}\cM^{(c)}_{(a),\de,\l},\\
\<\cP_{\de,\l}(z)\cP_{\de,\l}^\dagger(z')\>&= \frac{1}{2\pi i}\<\wL[\cO](x,z)(I_{xx'}\wL[\cO^{\dag}])(x',I_{xx'}z')\>.
\ee
where
\be
	(I_{xx'})^{\mu}{}_{\nu}=\de^\mu_\nu - 2x^\mu x_\nu/x^2,
\ee
and $I_{xx'}\wL[\cO^{\dag}]$ indicates the action of $I$ on the transverse spin indices. Importantly, the presence of $I_{xx'}$ means that while $z'$ and other indices now transform in the same way under $L_{xx'}$ as the indices and polarizations at $x$, the representation that they form is reflected relative to that of $\wL[\cO^\dagger]$. This reflection is crucial, because it is precisely the difference between the $d$-dimensional and $(d-2)$-dimensional $\dagger$ operation, and so the representations match in the above identifications.

Equation~\eqref{eq:kernelkdefinition_lowerj} implies that the set of three-point structures defined by~\eqref{eq:identificationLowSpin} is complete, otherwise we wouldn't be able to get the Kronecker delta in the right-hand side of~\eqref{eq:kernelkdefinition_lowerj}. We can therefore use the statement of section~\ref{sec:harmonic_analysis_fact}. Specifically, let us define operators $\hat \cC^{(a)}_{\de,\l}$ by~\eqref{eq:dcondition}, which now takes the form
\be
	&\frac{1}{2\pi i}\hat \cC^{(c)}_{\de,\l}(z_1,z_2,\ptl_{z_2})\<\wL[\cO](x,z_2)(I_{xx'}\wL[\cO^{\dag}])(x',I_{xx'}z')\>\nn\\
	&=
	\frac{\<0|\wL^{+}[\cO_2](x,z_2)(I_{xx'}\wL[\cO^{\dag}])(x',I_{xx'}z')\wL^{-}[\cO_1](x,z_1)|0\>^{(a)}_+}{\vol(\SO(1,1))}\cM^{(c)}_{(a),\de,\l}.
\ee
In this equation, we can cancel the $I_{xx'}$ matrices on both sides
\be\label{eq:ltsDdefinition_pre}
&\frac{1}{2\pi i}\hat\cC^{(c)}_{\de,\l}(z_1,z_2,\ptl_{z_2})\<\wL[\cO](x,z_2)\wL[\cO^{\dag}](x',z')\>
=
\frac{\<0|\wL^{+}[\cO_2](x,z_2)\wL[\cO^{\dag}](x',z')\wL^{-}[\cO_1](x,z_1)|0\>^{(a)}_+}{\vol(\SO(1,1))}\cM^{(c)}_{(a),\de,\l}.
\ee
Then we find
\be\label{eq:opeintermediate}
	\wL[\cO_1](x,z_1)\wL[\cO_2](x,z_2)=\sum_\l \int^{\frac{d-2}{2}+i\oo}_{\frac{d-2}{2}-i\oo}
	\frac{d\de}{2\pi i}\hat\cC^{(c)}_{\de,\l}(z_1,z_2;\ptl_{z_2})\mathbb{W}_{\de,\l,(c)}(x,z_2),
\ee
where
\be
	\mathbb{W}_{\de,\l,(c)}(x,z)\equiv\int D^{d-2}z_1 D^{d-2}z_2 k_{\de,\l,(c)}(z_1,z_2;z)\wL[\cO_1](x,z_1)\wL[\cO_2](x,z_2).
\ee
In our case, the kernel $k_{\de,\l,(c)}$ arose from $K^t$ through~\eqref{eq:lowtransversespindegeneration}. This means that we can alternatively rewrite $\mathbb{W}_{\de,\l,(c)}(x,z)$ as
\be
	\mathbb{W}_{\de,\l,(c)}(x,z)=\cN^{(a)}_{(c),\de,\l}\frac{1}{2}\p{
		\mathbb{O}^+_{\de+1,J_1+J_2-1,\l,(a)}(x,z)-
		\mathbb{O}^-_{\de+1,J_1+J_2-1,\l,(a)}(x,z)
	}.
\ee
Plugging this into~\eqref{eq:opeintermediate} we conclude
\be\label{eq:generalLTSresult_pre}
	&\wL[\cO_1](x,z_1)\wL[\cO_2](x,z_2)\nn\\
	&=\half\sum_{\l\in \L_{12}} \int^{\frac{d-2}{2}+i\oo}_{\frac{d-2}{2}-i\oo}
	\frac{d\de}{2\pi i}\cN^{(a)}_{(c),\de,\l}\hat\cC^{(c)}_{\de,\l}(z_1,z_2;\ptl_{z_2})\p{
 	\mathbb{O}^+_{\de+1,J_1+J_2-1,\l,(a)}(x,z)-
 	\mathbb{O}^-_{\de+1,J_1+J_2-1,\l,(a)}(x,z)
 }+\cdots.
\ee
In the last step we fixed a subtlety that we glossed over before: the identification of three-point structures in~\eqref{eq:identificationLowSpin} can only be performed for low transverse spin $\l$ since the right-hand side simply does not exist for higher transverse spin. Therefore, in~\eqref{eq:generalLTSresult_pre} we have only correctly identified the low transverse spin contributions, and $(\cdots)$ denotes the higher transverse spin contributions that we deal with in the next sections.

We can simplify equation~\eqref{eq:generalLTSresult_pre} further, by defining
\be
	\cC^{(a)}_{\de,\l}(z_1,z_2;\ptl_{z_2})=\frac{1}{2\pi i}\cN^{(a)}_{(c),\de,\l}\hat\cC^{(c)}_{\de,\l}(z_1,z_2;\ptl_{z_2}).
\ee
Note that while $\hat\cC^{(c)}_{\de,\l}$ are linearly-independent, $\cC^{(a)}_{\de,\l}$ are not. In terms of $\cC^{(a)}_{\de,\l}$ we get
\be\label{eq:generalLTSresult_pre2}
&\wL[\cO_1](x,z_1)\wL[\cO_2](x,z_2)\nn\\
&=\pi i\sum_{\l\in \L_{12}} \int^{\frac{d-2}{2}+i\oo}_{\frac{d-2}{2}-i\oo}
\frac{d\de}{2\pi i}\cC^{(a)}_{\de,\l}(z_1,z_2;\ptl_{z_2})\p{
	\mathbb{O}^+_{\de+1,J_1+J_2-1,\l,(a)}(x,z)-
	\mathbb{O}^-_{\de+1,J_1+J_2-1,\l,(a)}(x,z)
}+\cdots.
\ee
Using~\eqref{eq:ltsDdefinition_pre} and~\eqref{eq:Mdefn} we can give an equivalent characterization of $\cC^{(a)}_{\de,\l}$,
\be\label{eq:ltsDdefinition}
&\cC^{(a)}_{\de,\l}(z_1,z_2,\ptl_{z_2})\<\wL[\cO](x,z_2)\wL[\cO^{\dag}](x',z')\>
=
\frac{\<0|\wL^{+}[\cO_2](x,z_2)\wL[\cO^{\dag}](x',z')\wL^{-}[\cO_1](x,z_1)|0\>^{(a)}_+}{\vol(\SO(1,1))}.
\ee
Equations~\eqref{eq:ltsDdefinition} and~\eqref{eq:generalLTSresult_pre2} reproduce equations~(3.97) and~(3.98) of~\cite{Kologlu:2019mfz}, respectively.\footnote{There is a $(-)$ sign in~\eqref{eq:generalLTSresult_pre2} which is not in~(3.98) of~\cite{Kologlu:2019mfz}, which is due to the more explicit treatment of $(-1)^J$ signs in section~\ref{sec:Jcontinuation} of this paper.} We refer to~\eqref{eq:ltsDdefinition} as the ``celestial map'' formula because it maps the $d$-dimensional tensor structures appearing on the right-hand side to the $(d-2)$-dimensional OPE differential operators $\cC^{(a)}_{\de,\l}$. Note that~\eqref{eq:celetialmapkernel} implies that $m_{(a)}$ such that $\cN^{(a)}_{(c),\de,\l}m_{(a)}=0$ satisfy
\be
\frac{\<0|\wL^{+}[\cO_2](x,z_2)\wL[\cO^{\dag}](x',z')\wL^{-}[\cO_1](x,z_1)|0\>^{(a)}_+}{\vol(\SO(1,1))}m_{(a)}=0,
\ee
i.e.\ the kernel of $\cN^{(a)}_{(c),\de,\l}$ is the same as the kernel of the celestial map. 

Finally, to reproduce the low transverse spin terms in~\eqref{eq:lightrayope_generalops}, we note, analogously to the discussion around~\eqref{eq:scalarOPEfinal_simplified}, that in order for the left-hand side of~\eqref{eq:generalLTSresult_pre2} to be well-defined, we need the Regge intercept to satisfy $J_0<J_1+J_2-1$, in which case we can use a superconvergence sum rule~\cite{Kologlu:2019bco,Kologlu:2019mfz} which states that $\mathbb{O}^{(-1)^{J_1+J_2-1}}_{\de+1,J_1+J_2-1,\l,(a)}(x,z)=0$. This allows us to rewrite~\eqref{eq:generalLTSresult_pre2} as
\be\label{eq:generalLTSresult}
&\wL[\cO_1](x,z_1)\wL[\cO_2](x,z_2)\nn\\
&=\pi i(-1)^{J_1+J_2}\sum_{\l\in \L_{12}} \int^{\frac{d-2}{2}+i\oo}_{\frac{d-2}{2}-i\oo}
\frac{d\de}{2\pi i}\cC^{(a)}_{\de,\l}(z_1,z_2;\ptl_{z_2})
	\mathbb{O}^{(-1)^{J_1+J_2}}_{\de+1,J_1+J_2-1,\l,(a)}(x,z)+\cdots,
\ee
which reproduces the first sum in~\eqref{eq:lightrayope_generalops}.

\subsubsection{Higher transverse spin}
\label{sec:hts}

Now we would like to understand the higher transverse spin terms in the OPE~\eqref{eq:lightrayope_generalops}. The logic in this case is similar to the low transverse spin case, but with some extra complications. Similarly to~\eqref{eq:lowtransversespindegeneration}, we assume that
\be\label{eq:hightransversespindegeneration}
&\cN^{(a)}_{(c),\de,\l_\g,n}\cD_n\p{\int_{x\approx 2\atop x\approx 1^-} d^dx_1d^dx_2K^{t}_{\de+1,J,\l_\g,(a)}(x_1,x_2;x,z)\cO_1(x_1)\cO_2(x_2)} \nn \\
&\underset{J\to J_1+J_2-1+n}{\longrightarrow} \int D^{d-2}z_1D^{d-2}z_2k_{\de,\l_\g(+n),(c)}(z_1,z_2;z)\wL[\cO_1](x,z_1)\wL[\cO_2](x,z_2).
\ee
The evidence for this assumption is the same as for~\eqref{eq:lowtransversespindegeneration} and is discussed in section~\ref{sec:lts}.

Using this statement in~\eqref{eq:lorentzianformulaforlightraykernel} requires some care. Indeed, plugging in this relation (again defining $\cM$ to be the left inverse to $\cN$), we find
\be
\label{eq:kernelkdefinition_higherj}
&\int D^{d-2}z_1D^{d-2}z_2k_{\de,\l_\g(+n),(c)}(z_1,z_2;z)\frac{\<0|\wL^{+}[\cO_2](x,z_2)\wL[\cO^{\dag}](x',z')\wL^{-}[\cO_1](x,z_1)|0\>^{(a)}_+}{\vol(\SO(1,1))^2}\cM^{(c')}_{(a),\de,\l_\g,n} \nn \\
&=\frac{1}{2\pi i}\<(\cD_n\wL[\cO])(x,z)\wL[\cO^{\dag}](x',z')\>\de^{(c')}_{(c)}.
\ee
The equality holds if $J=J_n\equiv J_1+J_2-1+n$. But for this value of $J$, the action of $\cD_n$ is conformally-invariant, and the right-hand side of~\eqref{eq:kernelkdefinition_higherj} contains a conformally-invariant two-point function for operators $\cD_n\wL[\cO]$ and $\wL[\cO^\dagger]$. However, the quantum numbers of these operators are not Hermitian-conjugate to each other, and thus such two-point functions do not exist! This means that the right-hand side vanishes, and so the left-hand side must also vanish. Equation~\eqref{eq:kernelkdefinition_higherj} at $J=J_n$ is thus trivially satisfied.

To obtain nontrivial information, we must shift away from $J=J_n$. For this, it helps to act with $\cD_n$ on $\wL[\cO^\dagger]$ on both sides of~\eqref{eq:kernelkdefinition_higherj}. We obtain on the right-hand side
\be
&\frac{1}{2\pi i}\<(\cD_n\wL[\cO])(x,z)(\cD_n\wL[\cO^{\dag}])(x',z')\>\de^{(c')}_{(c)}\nn\\
&=(J-J_n)\lim_{J\to J_n}
	\frac{1}{2\pi i}\frac{\<(\cD_n\wL[\cO])(x,z)(\cD_n\wL[\cO^{\dag}])(x',z')\>}{J-J_n}\de^{(c')}_{(c)} + O((J-J_n)^2).
\ee
After acting with $\cD_n$ on the second operator, the limit
\be
\lim_{J\to J_n}
\frac{1}{2\pi i}\frac{\<(\cD_n\wL[\cO])(x,z)(\cD_n\wL[\cO^{\dag}])(x',z')\>}{J-J_n}
\ee
is a conformally-invariant two-point function~\cite{Erramilli:2019aa}. 

We now need to analyze the left-hand side of~\eqref{eq:lorentzianformulaforlightraykernel} away from $J=J_n$. One might worry that we would need to know the subleading term in~\eqref{eq:hightransversespindegeneration} in order to determine the leading non-zero piece in the left-hand side of~\eqref{eq:kernelkdefinition_higherj}. Fortunately, this is not required. To see this, let us first write for general $J$, from~\eqref{eq:lorentzianformulaforlightraykernel} and the above,
\be
&\int\frac{d^dx_1d^dx_2}{\vol(\SO(1,1))^2}\cD_n\p{K^{t,\pm}_{\De,J,\l(a)}(x_1,x_2;x,z)\theta(x\approx 2, 1^-)}\nn\\
&\hspace{2.5cm}\times \cD_n\p{\<0|\cO_2\wL[\cO^{\dag}](x',z')\cO_1|0\>^{(b)}_+\theta(2>x'>1^-)} \nn \\
&=(J-J_n)\lim_{J\to J_n}
\frac{1}{2\pi i}\frac{\<(\cD_n\wL[\cO])(x,z)(\cD_n\wL[\cO^{\dag}])(x',z')\>}{J-J_n}\de^{(b)}_{(a)} + O((J-J_n)^2)
\ee
Since the right-hand side is $O(J-J_n)$, this should be true for the left-hand side as well. We claim that in fact
\be\label{eq:vanishingintegral}
&\int\frac{d^dx_1d^dx_2}{\vol(\SO(1,1))^2}F(x_1,x_2;x,z) \cD_n\p{\<0|\cO_2\wL[\cO^{\dag}](x',z')\cO_1|0\>^{(b)}_+\theta(2>x'>1^-)}\in O(J-J_n)
\ee
for any conformally-invariant kernel $F$ that transforms at $x$ with the quantum numbers of $\cD_n\wL[\cO]$. This statement implies that we can use~\eqref{eq:kernelkdefinition_higherj} directly at $J=J_n=J_1+J_2-1+n$ for the purposes of determining the $O(J-J_n)$ term.

We actually need the following refined version of~\eqref{eq:vanishingintegral},
\be\label{eq:volSO11nonsense}
&\int\frac{d^dx_1d^dx_2}{\vol(\SO(1,1))^2}F(x_1,x_2;x,z) \cD_n\p{\<0|\cO_2\wL[\cO^{\dag}](x',z')\cO_1|0\>^{(b)}_+\theta(2>x'>1^-)}\nn\\
&=(J-J_n)\int\frac{d^dx_1d^dx_2}{\vol(\SO(1,1))}F(x_1,x_2;x,z) \cD_n\p{\<0|\cO_2\wL[\cO^{\dag}](x',z')\cO_1|0\>^{(b)}_+\theta(2>x'>1^-)}\Big\vert_{J=J_n}\nn\\
&\quad+O((J-J_n)^2).
\ee
Note that the expressions on the left-hand and right-hand sides differ by the power of the factor $\vol(\SO(1,1))$. The reason for this is that while at $J=J_n$ and $J\neq J_n$ both integrals have 2 reparameterization zero modes, in the integral on the right, one of the modes is similar to $x\to\a x$ in
\be
	\int dx\,\de(x)=1,
\ee
and doesn't require Faddeev-Popov fixing. A simple model for the above equation is provided by the integral
\be
\int_0^\oo \frac{dx}{\vol(\SO(1,1))}2\.x^{-(J-J_n)}\ptl_x (x^{J-J_n}\theta(x))=2(J-J_n).
\ee
Clearly, the coefficient $2$ in front of $(J-J_n)$ can be computed by setting $J=J_n$ and removing the $\vol(\SO(1,1))$ factor. The result~\eqref{eq:volSO11nonsense} can be shown by an explicit calculation in the case when $\cO_1,\cO_2$ are scalars,\footnote{It is easy to convince oneself that it suffices to ensure that the result of this section agrees with the result of section~\ref{sec:scalar OPE}.} and then noting that integrands in both sides have the same transformation properties under weight-shifting operators acting on $\cO_1,\cO_2,\cO$~\cite{Karateev:2017jgd,Karateev:2018oml}. 

Combining everything together, we find the following equation for the kernel $k_{\de,\l_\g(+n),(a)}$,
\be
&\int \frac{D^{d-2}z_1D^{d-2}z_2}{{\vol(\SO(1,1))}}k_{\de,\l_\g(+n),(c)}(z_1,z_2;z)\cD_n\p{\<0|\wL^{+}[\cO_2](x,z_2)\wL[\cO^{\dag}](x',z')\wL^{-}[\cO_1](x,z_1)|0\>^{(a)}_+}\cM^{(c')}_{(a),\de,\l_\g,n} \nn \\
&=\lim_{J\to J_n}
\frac{1}{2\pi i}\frac{\<(\cD_n\wL[\cO])(x,z)(\cD_n\wL[\cO^{\dag}])(x',z')\>}{J-J_n}\de^{(c')}_{(c)}.
\ee
From this equation, following precisely the same steps as in the derivation of low transverse spin terms in the previous section, we find that if the operators $\cC^{(a)}_{\de,\l_\g(+n)}$ are defined by
\be
\label{eq:htsDdefinition}
&\cC^{(a)}_{\de,\l_\g(+n)}(z_1,z_2,\ptl_{z_2})\lim_{J\to J_n}
\frac{\<(\cD_n\wL[\cO])(x,z_2)(\cD_n\wL[\cO^{\dag}])(x',z')\>}{J-J_n}\nn\\
&=(-1)^n\cD_n\p{\<0|\wL^{+}[\cO_2](x,z_2)\wL[\cO^{\dag}](x',z')\wL^{-}[\cO_1](x,z_1)|0\>^{(a)}_+},
\ee
then the contribution of higher transverse spin terms is given by the second sum in~\eqref{eq:lightrayope_generalops}.\footnote{Here we introduced a sign $(-1)^n$ by hand in order to simplify~\eqref{eq:lightrayope_generalops}.} In appendix~\ref{app:tensorstructureselectionrule} we consider which representations $\l_\g(+n)$ can be generated in this way, and show that there are enough $d$-dimensional structures that survive this celestial map to account for all the celestial OPE structures.

\subsection{Celestial map without light transforms}
In the previous section we derived the celestial map formulas~\eqref{eq:ltsDdefinition} and~\eqref{eq:htsDdefinition}
that determine the OPE differential operators $\cC^{(a)}_{\de,\l}$ appearing in~\eqref{eq:lightrayope_generalops}. These formulas
involve taking several light-transforms of tensor structures analytically continued to Wightman correlators, which in practice 
can be a difficult calculation. Fortunately, as first observed in~\cite{Kologlu:2019mfz} and proved for traceless-symmetric operators,
there exists a simpler version of the celestial map formulas that contains only simple algebraic manipulations.

We will show that the following is equivalent to~\eqref{eq:ltsDdefinition},
\be\label{eq:ltsDefinitionSimple}
&\cC_{\de,\l}^{(a)}(z_1,\bw_1, z_2,\bw_2,\ptl_{z_2}, \ptl_{\bw_2})
\p{
	(-2H_{20}){\<\cO(X_2,Z_2,\mathbf{W}_2)\cO^\dagger(X_0,Z_0,\mathbf{W}_0)\>}
}\Bigg\vert_\text{celestial}\nn\\
&=
X_{12}\left|-2V_{0,21}\right|
\<\cO_1(X_1,Z_1,\mathbf{W}_1)\cO^\dagger(X_0,Z_0,\mathbf{W}_0)\cO_2(X_2,Z_2,\mathbf{W}_2)\>^{(a)}_+
\Bigg\vert_\text{celestial},
\ee
and the following equivalent to~\eqref{eq:htsDdefinition},
\be\label{eq:htsDefinitionSimple}
	&\cC_{\de,\l_\g(+n)}^{(a)}(z_1,\bw_1, z_2,\bw_2,\ptl_{z_2}, \ptl_{\bw_2})
	\p{
		\lim_{J\to J_n} (-2H_{20})\frac{\<(\cD'_n \cO)(X_2,Z_2,\mathbf{W}_2)(\cD'_n \cO^\dagger)(X_0,Z_0,\mathbf{W}_0)\>}{J-J_n}
	}\Bigg\vert_\text{celestial}\nn\\
	&=
	\frac{(-1)^n X_{12}}{\de(-2V_{0,21})}\cD_n'\p{
		\theta(V_{0,12})\<\cO_1(X_1,Z_1,\mathbf{W}_1)\cO^\dagger(X_0,Z_0,\mathbf{W}_0)\cO_2(X_2,Z_2,\mathbf{W}_2)\>^{(a)}_+
	}\Bigg\vert_\text{celestial}.
\ee
(Note that we have used the $\<\cdots\>_+^{(a)}$ structures defined in section~\ref{sec:Jcontinuation}.) In the above equations, we have explicitly reintroduced polarization vectors $\bw_i$ for the second and higher rows of Young diagrams of $\cO_i$, as described in section~\ref{sec:differential operators}. Furthermore, we have used embedding space notation~\cite{Costa:2011mg} on the right-hand side, with the standard tensor structures 
\be
	X_{ij}&\equiv -2X_i\.X_j\\
	V_{i,jk}&\equiv \frac{Z_i\.X_j X_i\.X_k-Z_i\.X_k X_i\.X_j}{X_j\.X_k},   \label{eq:tensorstructure_V}\\
	H_{ij}&\equiv -2(Z_i\.Z_j X_i\.X_j-Z_i\.X_jZ_j\.X_i). \label{eq:tensorstructure_H}
\ee
Finally, the notation $(\cdots)\vert_\text{celestial}$ stands for substituting the following values for the embedding space coordinates,
\be\label{eq:celestiallocus}
&Z_0=-(1,0,0),\quad Z_1=-(0,1,0),\quad Z_2=-(0,1,0),\nn\\
&X_i=(0,0,z_i),\quad W_{i,j}=(0,0,w_{i,j}),
\ee
where we specify coordinates in the order $(Y^+,Y^-,Y^\mu)$, $\mu=0,\cdots,d-1$. Note that in~\eqref{eq:htsDefinitionSimple} we  divide by $\de(-2V_{0,21})$. By this we mean that the result of the action of $\cD'_n$ is proportional to $\de(-2V_{0,21})$, and we simply read off the coefficient of this delta-function. Note that this coefficient is only well-defined in configurations where $V_{0,21}=0$, which is indeed the case for the celestial locus~\eqref{eq:celestiallocus}.

\subsubsection{Factoring out the light-transforms}

We start by proving~\eqref{eq:htsDefinitionSimple}. The proof of~\eqref{eq:ltsDefinitionSimple} is only a simple modification that we comment on below. 

The proof proceeds with evaluation of light-transforms in both sides of~\eqref{eq:htsDdefinition}. We start with the right-hand side
\be\label{eq:3Ltransform}
\cD_n\p{\<0|\wL^{+}[\cO_2](x,z_2,\bw_2)\wL[\cO^{\dag}](x_0,z_0,\bw_0)\wL^{-}[\cO_1](x,z_1,\bw_1)|0\>^{(a)}_+}.
\ee
Here $\cD_n$ acts on $(x_0,z_0,\bw_0)$. We can rewrite this equivalently as
\be\label{eq:rewritingrhs}
&=\cD_n\wL_1\wL_2\p{\<0|\cO_2(x,z_2,\bw_2)\wL[\cO^{\dag}](x_0,z_0,\bw_0)\cO_1(x,z_1,\bw_1)|0\>^{(a)}_+\theta(2>0>1^-)}\nn\\
&=\cD_n\wL_1\wL_2\p{\<\cO_1(x,z_1,\bw_1)\cO_2(x,z_2,\bw_2)\wL[\cO^{\dag}](x_0,z_0,\bw_0)\>^{(a)}_+\theta(V_{0,12})},
\ee
where $\wL_i$ denotes light-transform acting on $(x_i,z_i,\bw_i)$. In the last equality we used the following fact. First of all, 
\be\label{eq:v012}
	V_{0,12}=\frac{z_0\.x_{10}x_{20}^2-z_0\.x_{20}x_{10}^2}{x_{12}^2}
\ee
is positive for $1\approx 2$ and $2>0>1^-$, and is negative for $1\approx 2$ and $1>0>2^-$. In~\eqref{eq:continuedLT} we have two terms, and multiplying by $\theta(V_{0,12})$ selects the first term, which is the one appearing in the first line of~\eqref{eq:rewritingrhs}.

Now we can use~\eqref{eq:LDcommutator} to rewrite this further as
\be\label{eq:3Ltransform'}
&=\wL_1\wL_2\wL_0\cD'_n\p{\<\cO_1(x,z_1,\bw_1)\cO_2(x,z_2,\bw_2)\cO^{\dag}(x_0,z_0,\bw_0)\>^{(a)}_+\theta(V_{0,12})},
\ee
where we took into account the easily verified fact that $V_{0,12}$ commutes with $\wL_0$. 

\subsubsection{Appearance of $\de(-2V_{0,12})$}
\label{sec:deltafunction}

Note that without the theta-function we would have
\be\label{eq:vanishD'}
\cD'_n\<\cO_1(x,z_1,\bw_1)\cO_2(x,z_2,\bw_2)\cO^{\dag}(x_0,z_0,\bw_0)\>^{(a)}=0.
\ee
This is because for $J=J_n$, all possible three-point tensor structures above are polynomial in $z_0$, and hence killed by $\cD'_n$. To see this, note that the way non-polynomial structures in $z_0$ appear is through the factors of the form
\be
	\<\cO_1(x,z_1,\bw_1)\cO_2(x,z_2,\bw_2)\cO^{\dag}(x_0,z_0,\bw_0)\>^{(a)} = (\cdots)V_{0,12}^{J-k}
\ee
when $J<k$, where $k>0$ is the degree of $z_0$ in $(\cdots)$. Non-polynomiality cannot appear in any other way, because we require that all polarizations except $z_0$ enter polynomially, and $V_{0,12}$ is the only invariant that involves only $z_0$. We see that for sufficiently large integer $J$ all structures are therefore polynomial, and the appearance of non-polynomial structures is indicated by the reduction in the number of polynomial structures. The number of polynomial structures can be computed using group-theoretic counting rules~\cite{Kravchuk:2016qvl}, and a simple calculation shows that the polynomial structures start disappearing for $J=J_1+J_2-1$ (see appendix~\ref{app:tensorstructureselectionrule}). This means that for $J=J_1+J_2-1$ the maximal non-polynomiality is $V_{0,12}^{-1}$, and for $J=J_n=J_1+J_2-1+n$ with $n>0$ there are no non-polynomial structures. More generally, the smallest possible exponent of $V_{0,12}$ is 
\be
V_{0,12}^{J-J_1-J_2}.
\ee
Since all structures are polynomial for $J=J_n$, the properties of $\cD_n'$ discussed in section~\ref{sec:differential operators} ensure~\eqref{eq:vanishD'}.

However, we are interested in
\be\label{eq:thingwithdeltafunction}
\cD'_n\p{\<\cO_1(x,z_1,\bw_1)\cO_2(x,z_2,\bw_2)\cO^{\dag}(x_0,z_0,\bw_0)\>^{(a)}_+\theta(V_{0,12})}.
\ee
We can obtain derivatives of delta functions $\de^{(m)}(-2V_{0,12})$ from $\cD'_n$ hitting the theta-function. We claim that we get a result which is proportional to $\de(-2V_{0,12})$. To see this, we regularize by analytic continuation in $J$. Note that in~\eqref{eq:thingwithdeltafunction} the most negative power of $V_{0,12}$ is 
\be
	V_{0,12}^{J-J_1-J_2-n}=V_{0,12}^{J-J_n-1},
\ee
because $\cD'_n$ has $n$ derivatives which will therefore hit $V_{0,12}^{J-J_1-J_2}$, which is the most negative power of $V_{0,12}$ before acting with $\cD'_n$, at most $n$ times. In other words, we have for each $(a)$
\be
&\cD'_n\p{\<\cO_1(x,z_1,\bw_1)\cO_2(x,z_2,\bw_2)\cO^{\dag}(x_0,z_0,\bw_0)\>^{(a)}_+\theta(V_{0,12})}\nn\\
&=(J-J_n)(\cdots)V_{0,12}^{J-J_n-1}\theta(V_{0,12})+R,
\ee
where $R$ involves higher powers of $V_{0,12}$ or $(J-J_n)$. As we send $J\to J_n$, we then find $R\to 0$ and 
\be
(J-J_n)V_{0,12}^{J-J_n-1}\theta(V_{0,12})\to \de(V_{0,12}).
\ee
We therefore conclude that
\be
\cD'_n\p{\<\cO_1(x,z_1,\bw_1)\cO_2(x,z_2,\bw_2)\cO^{\dag}(x_0,z_0,\bw_0)\>^{(a)}_+\theta(V_{0,12})}=\de(-2V_{0,21})\times\text{finite}.
\ee

\subsubsection{Symmetries of the integrand}
Defining
\be
	&f(X_1,Z_1,\mathbf{W}_1;X_2,Z_2,\mathbf{W}_2;X_0,Z_0,\mathbf{W}_0)\nn\\
	&\equiv
	\cD'_n\p{\<\cO_1(X_1,Z_1,\mathbf{W}_1)\cO_2(X_2,Z_2,\mathbf{W}_2)\cO^\dagger(X_0,Z_0,\mathbf{W}_0)\>^{(a)}_+\theta(V_{0,12})},
\ee
and
\be\label{eq:faaadefn}
	f(\a_1,\a_2,\a_0)=f(Z_1-\a_1 X_\oo, -X_\oo,\mathbf{W}_1; Z_2-\a_2 X_\oo,-X_\oo,\mathbf{W}_2; Z_0-\a_0X_0,-X_0,\mathbf{W}_0),
\ee
where
\be
	&X_\oo=(0,1,0),\quad X_0=(1,0,0),\\
	&Z_i=(0,0,z_i),\quad W_{i,j}=(0,0,w_{i,j}),
\ee
we can rewrite~\eqref{eq:3Ltransform'} and thus~\eqref{eq:3Ltransform} for $x=\oo,x'=0$ as
\be\label{eq:tocompute}
\cD_n\<0|\wL^{+}[\cO_2](\oo,z_2,\bw_2)\wL[\cO^{\dag}](0,z_0,\bw_0)\wL^{-}[\cO_1](\oo,z_1,\bw_1)|0\>^{(a)}_+=
	\int_{-\oo}^{+\oo}d\a_1d\a_2d\a_0 f(\a_1,\a_2,\a_0).
\ee

Our goal here will be to find an expression for $f(\a_1,\a_2,\a_0)$ based solely on symmetries. First, we check that the boost in the embedding space which sends
\be
	X_\oo\to \l X_\oo,\quad X_0\to \l^{-1}X_0
\ee
implies that for $\l>0$
\be
	f(\a_1,\a_2,\a_0)=\l^{J_1+J_2-J'}f(\l\a_1,\l\a_2,\l^{-1}\a_0),
\ee
where $J'=J_n-n=J_1+J_2-1$ is the spin we get after the action of $\cD_n'$. That is, we have
\be
	f(\a_1,\a_2,\a_0)=\l f(\l\a_1,\l\a_2,\l^{-1}\a_0).
\ee

Second, the symmetry that in Minkowski space is represented by translation along $z_0$ acts on embedding space coordinates as
\be
	X_0\to X_0+\l Z_0,\quad Z_i\to Z_i+2(z_i\.z_0)\l X_{\oo}\quad (i=1,2),
\ee
leaving all other coordinates invariant.\footnote{To be more precise, we have
\be
W_{i,j}\to W_{i,j}+2(w_{i,j}\.z_0)\l X_\oo\quad (i=1,2),
\ee
but since in~\eqref{eq:faaadefn} $W_{i,j}$ are inserted with first-row polarization $-X_\oo$, the shift by $X_\oo$ has no effect on the value of $f$.
}
We can check that it implies the following equation for $f$,
\be
	f(\a_1,\a_2,\a_0)=(1-\a_0 \l)^{-\De-J'}f(\a_1-2(z_1\.z_0)\l, \a_2-2(z_2,z_0)\l,(\a_0^{-1}-\l)^{-1})
\ee
for $1-\a_0\l>0$.

Now recall from the previous discussion that $f(\a_1,\a_2,\a_0)$ is proportional to $\de(-2V_{0,21})$, and so we can write
\be
	f(\a_1,\a_2,\a_0)=\de\p{\frac{\a_1(z_0\.z_2)-\a_2(z_0\.z_2)}{(z_1\.z_2)}}g(\a_2,\a_0),
\ee
for some $g$. Here we eliminated the $\a_1$-dependence using the delta-function. In terms of $g$, the symmetries discussed above read
\be
	g(\a_2,\a_0)&=g(\l\a_2,\l^{-1}\a_0),\\
	g(\a_2,\a_0)&=(1-\a_0\l)^{-\De-J'}g(\a_2-2(z_2,z_0)\l,(\a_0^{-1}-\l)^{-1}),
\ee
for $\l>0$ and $1-\a_0\l>0$ respectively. Let us consider $g_0$ defined by
\be
	g_0(\a_2,\a_0)=|\a_2\a_0-2(z_2\.z_0)|^{-\De-J'}.
\ee
It is easy to check that $g_0$ satisfies the same symmetries as $g$, and thus $g/g_0$ is simply invariant under the above transformations of $\a_2,\a_0$. Note that we have two continuous families of transformations, and it is easy to verify that the 2 vector fields by which they are generated are generically linearly-independent. Since we have only two coordinates $\a_2,\a_0$, we find that $g/g_0$ should be locally constant and can only change value where these vector fields degenerate. This only happens when $\a_2\a_0-2(z_2\.z_0)=0$ or $\a_0=0$.

Note that $g(\a_2,\a_0)$ comes from a three-point structure, and the three-point structures are analytic away from $X_{ij}=0$, which are given by
\be
	X_{12}&=-2(z_1\.z_2),\\
	X_{10}&=\a_1\a_0-2(z_1\.z_0),\\
	X_{20}&=\a_2\a_0-2(z_2\.z_0),
\ee
so $g/g_0$ cannot have a discontinuity at $\a_0=0$. We thus only need to determine how $g/g_0$ changes when crossing $X_{20}=0$. Since $g$ comes from a time-ordered three-point structure, a non-analyticity near $X_{20}=0$ must have the form
\be
	(X_{20}+i\e)^{\#},
\ee
where we use the usual $i\e$-prescription for time-ordered correlators. It follows that 
\be
	g(\a_2,\a_0)=A(\a_2\a_0-2(z_2\.z_0)+i\e)^{-\De-J'}
\ee
for some $A$, for all real values of $\a_2,\a_0$. It is convenient to express $A$ in terms of $g(0,0)$,
\be
	g(\a_2,\a_0)=(-2 z_2\.z_0)^{\De+J'}g(0,0)(\a_2\a_0-2(z_2\.z_0)+i\e)^{-\De-J'}.
\ee

\subsubsection{Computing the light-transforms}

We can now compute the integrals in~\eqref{eq:tocompute},
\be\label{eq:computed}
	&\int d\a_1 d\a_2 d\a_0f(\a_1,\a_2,\a_0)\nn\\
	&=\int d\a_1 d\a_2 \de\p{\frac{\a_1(z_0\.z_2)-\a_2(z_0\.z_2)}{(z_1\.z_2)}}
	\int d\a_0(-2 z_2\.z_0)^{\De+J'}g(0,0)(\a_2\a_0-2(z_2\.z_0)+i\e)^{-\De-J'}\nn\\
	&=g(0,0)(-2 z_2\.z_0)\int d\a_1 d\a_2 \de\p{\frac{\a_1(z_0\.z_2)-\a_2(z_0\.z_2)}{(z_1\.z_2)}}
	\int d\a_0(\a_2\a_0+1+i\e)^{-\De-J'}\nn\\
	&=\frac{-2\pi i}{\De+J'-1}g(0,0)(-2 z_2\.z_0)\int d\a_1 d\a_2 \de\p{\frac{\a_1(z_0\.z_2)-\a_2(z_0\.z_2)}{(z_1\.z_2)}}
	\de(\a_2)\nn\\
	&=\frac{-2\pi i}{\De+J'-1}(-2 z_1\.z_2)g(0,0),
\ee
where we used the equation
\be\label{eq:funnyintegral}
	\int_{-\oo}^{+\oo} \frac{dx}{(xy+1+i\e)^a}=-\frac{2\pi i}{a-1}\de(y)
\ee
which we prove in appendix~\ref{sec:distributionidentity}.
Unwinding the definitions, we check that
\be\label{eq:copmutedtocelestial}
&(-2 z_1\.z_2)g(0,0)\nn\\
&=\frac{X_{12}}{\de(-2V_{0,21})}\cD_n'\p{
	\theta(V_{0,12})\<\cO_1(X_1,Z_1,\mathbf{W}_1)\cO^\dagger(X_0,Z_0,\mathbf{W}_0)\cO_2(X_2,Z_2,\mathbf{W}_2)\>^{(a)}_+
}\Bigg\vert_\text{celestial}.
\ee

It only remains to compute the light-transforms in the left-hand side of~\eqref{eq:htsDdefinition}.
For this, note that 
\be\label{eq:rewrite2pt}
&\lim_{J\to J_n}
\frac{\<(\cD_n\wL[\cO])(x_2,z_2,\bw_2)(\cD_n\wL[\cO^{\dag}])(x_0,z_0,\bw_0)\>}{J-J_n}\nn\\
&=\wL_2\wL_0\lim_{J\to J_n}\frac{\<(\cD'_n \cO)(X_2,Z_2,\mathbf{W}_2)(\cD'_n \cO^\dagger)(X_0,Z_0,\mathbf{W}_0)\>}{J-J_n}
\ee
and defining
\be
	g'(X_2,Z_2,\mathbf{W}_2;X_0,Z_0,\mathbf{W}_0)&\equiv \lim_{J\to J_n}\frac{\<(\cD'_n \cO)(X_2,Z_2,\mathbf{W}_2)(\cD'_n \cO^\dagger)(X_0,Z_0,\mathbf{W}_0)\>}{J-J_n},\\
	g'(\a_2,\a_0)&\equiv g'(Z_2-\a_2 X_\oo,-X_\oo,\mathbf{W}_2; Z_0-\a_0X_0,-X_0,\mathbf{W}_0),
\ee
we find that $g'$ satisfies the same properties as $g$ above, and the same arguments lead to
\be
	g'(\a_2,\a_0)=(-2 z_2\.z_0)^{\De+J'}g'(0,0)(\a_2\a_0-2(z_2\.z_0)+i\e)^{-\De-J'}.
\ee
We can then similarly compute 
\be\label{eq:computed2pt}
	&\wL_2\wL_0\lim_{J\to J_n}\frac{\<(\cD'_n \cO)(X_2,Z_2,\mathbf{W}_2)(\cD'_n \cO^\dagger)(X_0,Z_0,\mathbf{W}_0)\>}{J-J_n}=\int d\a_2 d\a_0 g'(\a_2,\a_0)\nn\\
	&=\frac{-2\pi i}{\De+J'-1}(-2 z_2\.z_0) g'(0,0).
\ee
Since $(-2z_2\.z_0)=-2H_{20}|_\text{celestial}$, we find
\be\label{eq:computed2pttocelestial}
	(-2 z_2\.z_0) g'(0,0)=\p{
		\lim_{J\to J_n} (-2H_{20})\frac{\<(\cD'_n \cO)(X_2,Z_2,\mathbf{W}_2)(\cD'_n \cO^\dagger)(X_0,Z_0,\mathbf{W}_0)\>}{J-J_n}
	}\Bigg\vert_\text{celestial}.
\ee
Combining with~\eqref{eq:htsDdefinition},~\eqref{eq:tocompute},~\eqref{eq:computed},~\eqref{eq:rewrite2pt},~\eqref{eq:computed2pt}, and~\eqref{eq:copmutedtocelestial}, we arrive at~\eqref{eq:htsDefinitionSimple}.

\subsubsection{Low transverse spin case}

The only modification to the above proof required for the case of low transverse spin ---~\eqref{eq:ltsDdefinition} and~\eqref{eq:ltsDefinitionSimple} --- concerns the appearance of the delta-function in the analogue of~\eqref{eq:thingwithdeltafunction}.
\be
(\<\cO_1(x,z_1,\bw_1)\cO_2(x,z_2,\bw_2)\cO^{\dag}(x_0,z_0,\bw_0)\>^{(a)}_+\theta(V_{0,12})).
\ee
In principle, there is no delta-function here. Instead, from the discussion in section~\ref{sec:deltafunction} we know that for $J=J_1+J_2-1$ there are structures which contain the most negative power of $V_{0,21}$ which is
\be
	(\cdots)(-2V_{0,21})^{-1}\theta(V_{0,12}).
\ee
This inverse power of $V_{0,21}$ leads to the divergence in the triple light-transform in~\eqref{eq:ltsDdefinition} which is canceled by the $\SO(1,1)$ factor. Similarly to the discussion around~\eqref{eq:volSO11nonsense} we can remove the $\vol(\SO(1,1))$ factor and replace (in fact, this is exactly the same replacement as in~\eqref{eq:volSO11nonsense})
\be
	(-2V_{0,21})^{-1}\theta(V_{0,12})\to \de(-2V_{0,21})
\ee
and then the above proof can be applied to~\eqref{eq:ltsDefinitionSimple}. The only remaining difference is that in the final formula we prefer to factor out $(-2V_{0,21})^{-1}$ before the above substitution instead of factoring out $\de(-2V_{0,21})$ after the substitution.

\section{Examples}
\label{sec:examples}

\subsection{Re-deriving the $\wL[\f_1]\x\wL[\f_2]$ OPE from general formulas}

In this section we rederive the scalar detector OPE~\eqref{eq:scalarOPEfinal_simplified} from the general result~\eqref{eq:lightrayope_generalops}.

The first step is to work out the low-transverse spin terms in~\eqref{eq:scalarOPEfinal_simplified}. The set of transverse spins
$\L_{12}$ is constrained to consist of traceless-symmetric representations $\l$ of spin $j$ because $\wL[\f_1]$ and $\wL[\f_2]$ both
transform as scalars on the celestial sphere. In $d$-dimensional language, the transverse spin $j$ is the length of the second row of the Young diagram of a $\SO(d-1,1)$ irrep, and is therefore constrained to be $j=0$ since only traceless-symmetric light-ray operators appear in $\f_1\x\f_2$ OPE. The low-transverse spin contributions are then given by
\be
	\wL[\f_1](x,z_1)\wL[\f_2](x,z_2)=\pi i\int^{\frac{d-2}{2}+i\oo}_{\frac{d-2}{2}-i\oo}
	\frac{d\de}{2\pi i}\cC_{\de,j=0}(z_1,z_2,\ptl_{z_2})\mathbb{O}^+_{\de+1,J=-1}(x,z_2)+\cdots,
\ee
where we also substituted $J_1=J_2=0$ and removed dependence on the transverse polarizations $\bw_i$ since all operators are traceless-symmetric. In order for this to agree with the $j=0$ term of~\eqref{eq:scalarOPEfinal_simplified}, we need to verify that $C_{\de,j=0}$ normalized as in~\eqref{eq:3pt=C 2pt} also satisfies the celestial map~\eqref{eq:ltsDefinitionSimple}. We do this by computing the structures entering in~\eqref{eq:ltsDefinitionSimple} and comparing them to the structures in~\eqref{eq:3pt=C 2pt}.

First of all, we need to determine the expression for the three-point structure entering~\eqref{eq:ltsDefinitionSimple}. Our $d$-dimensional structures for integer $J$ are defined in appendix~\ref{app:structureconventions}, and the analytic continuation should be performed following the conventions of section~\ref{sec:Jcontinuation}. Comparing~\eqref{eq:ffOconventionPoincare} and~\eqref{eq:threeptchoice} we see that $n=0$ in~\eqref{eq:threeptchoice} and $f_0$ is given by some product of distances $x_{ij}$ which is positive for space-like separated points. Looking at~\eqref{eq:timeorederedcontinuation} we see that in our case the analytically-continued three-point structures appearing in~\eqref{eq:ltsDefinitionSimple}, when all points are spacelike-separated, are equal to the absolute value of~\eqref{eq:ffOconventionPoincare}.

We can therefore use~\eqref{eq:ffOconvetionEF} and substitute the celestial locus values~\eqref{eq:celestiallocus} into it. We find for $J=-1$,
\be
	&X_{12}|\!-\!2V_{3,12}|\<\f_1(X_1)\f_2(X_2)\cO(X_3,Z_3)\>\vert_{\text{celestial}}=\frac{1}
	{X_{12}^{\frac{\De_1+\De_2-\De-1}{2}}
	X_{13}^{\frac{\De_1+\De-\De_2-1}{2}}
	X_{23}^{\frac{\De_2+\De-\De_1-1}{2}}}\Bigg\vert_{\text{celestial}} \nn \\
	&=\frac{1}
	{(-2z_1\.z_2)^{\frac{\de_1+\de_2-\de}{2}}
	(-2z_1\.z_3)^{\frac{\de_1+\de-\de_2}{2}}
	(-2z_2\.z_3)^{\frac{\de_2+\de-\de_1}{2}}},
\ee
and since this is positive and all $X_i$ are space-like separated in~\eqref{eq:celestiallocus}, it follows, according to the discussion above, that this is equal to the right-hand side of~\eqref{eq:ltsDefinitionSimple} after substitution $3\to 0$.

The left-hand side of~\eqref{eq:ltsDefinitionSimple} is easily computed from~\eqref{eq:OOconvetionEF} to be equal to
\be
	(-2H_{23})\<\cO(X_2,Z_2)\cO(X_3,Z_3)\>\vert_{\text{celestial}}=(-2z_2\.z_3)^{-\de},
\ee
after substitution $3\to 0$. Using these results in~\eqref{eq:ltsDefinitionSimple} we see immediately that it gives the same normalization of $\cC_{\de,j=0}$ as~\eqref{eq:3pt=C 2pt}.

A similar logic works for the higher transverse spin terms in~\eqref{eq:lightrayope_generalops}. We have $\l_\g(+n)=(n, \g)$. The set $\G_{12}$ in~\eqref{eq:lightrayope_generalops} then consists of just the trivial representation, because only traceless-symmetric representations appear in the $\f_1\x\f_2$ OPE. We thus find that the contribution of higher transverse spins is
\be
&\wL[\cO_1](x,z_1,\bw_1)\wL[\cO_2](x,z_2,\bw_2)\nn \\
&=\pi i\sum_{n=1}^{\oo}\int_{\frac{d-2}{2}-i\oo}^{\frac{d-2}{2}+i\oo}\frac{d\de}{2\pi i}\cC_{\de,j=n}(z_1,z_2,\ptl_{z_2},\ptl_{\bw_2})(\cD_{n}\mathbb{O}^{(-1)^{n}}_{\de+1,n-1})(x,z_2,\bw_2)+\cdots.
\ee
Therefore, in order to verify that~\eqref{eq:lightrayope_generalops} reproduces~\eqref{eq:scalarOPEfinal_simplified}, we need to check that the normalizations of $\cC_{\de,j}$ defined by~\eqref{eq:3pt=C 2pt} and~\eqref{eq:htsDefinitionSimple} are consistent.

Reasoning analogously to the lower transverse spin case, we find that we need to compute the action of $\cD'_n$ defined by~\eqref{eq:dprimedef} on the absolute value of~\eqref{eq:ffOconventionPoincare}. Since in~\eqref{eq:htsDefinitionSimple} we have $\theta(V_{0,12})$, we need to restrict to the region where $V_{0,12}$ is positive. In this region, we have 
\be
	&\cD_n'\<\f_1(x_1)\f_2(x_2)\cO(x_3,z)\>_+=	\cD_n'\frac{(-2z\.x_{23}\, x_{13}^2+2z\.x_{13}\,x_{23}^2)^J}
{x_{12}^{\De_1+\De_2-\De+J}x_{13}^{\De_1+\De-\De_2+J}x_{23}^{\De_2+\De-\De_1+J}}\nn\\
&=\frac{J(J-1)\cdots (J-n+1)}{n!}(-2w\.x_{23}\, x_{13}^2+2w\.x_{13}\,x_{23}^2)^n\frac{(-2z\.x_{23}\, x_{13}^2+2z\.x_{13}\,x_{23}^2)^{J-n}}
{x_{12}^{\De_1+\De_2-\De+J}x_{13}^{\De_1+\De-\De_2+J}x_{23}^{\De_2+\De-\De_1+J}}.
\ee
Multiplying by~$\theta(V_{0,12})$ and taking limit $J\to n-1$, we find
\be
\cD_n'\p{\<\f_1(x_1)\f_2(x_2)\cO(x_3,z)\>_+\theta(V_{0,12})}=\frac{1}{n}\frac{(-2w\.x_{23}\, x_{13}^2+2w\.x_{13}\,x_{23}^2)^n}
{x_{12}^{\De_1+\De_2-\De+n+1}x_{13}^{\De_1+\De-\De_2+n-1}x_{23}^{\De_2+\De-\De_1+n-1}}\de(-2V_{0,12}).
\ee
Lifting this to embedding space and evaluating at the celestial locus~\eqref{eq:celestiallocus}, we find that the right-hand side of~\eqref{eq:htsDefinitionSimple} is
\be
	&\frac{(-1)^n X_{12}}{\de(-2V_{0,21})}\cD_n'\p{
	\theta(V_{0,12})\<\f_1(X_1)\cO(X_0,Z_0)\f_2(X_2)\>^{(a)}_+
}\Bigg\vert_\text{celestial}\nn\\
&=\frac{1}{n}\frac{(-4W\.X_{2}\, X_1\.X_3+4W\.X_{1}\,X_2\.X_3)^n}
{X_{12}^{\frac{\De_1+\De_2-\De+n-1}{2}}X_{13}^{\frac{\De_1+\De-\De_2+n-1}{2}}X_{23}^{\frac{\De_2+\De-\De_1+n-1}{2}}}\Bigg\vert_\text{celestial}\nn\\
&=\frac{1}{n}\frac{(-4w\.z_{2}\, z_1\.z_3+4w\.z_{1}\,z_2\.z_3)^n}
{(-2z_1\.z_2)^{\frac{\de_1+\de_2-\de+n}{2}}(-2z_1\.z_3)^{\frac{\de_1+\de-\de_2+n}{2}}(-2z_2\.z_3)^{\frac{\de_2+\de-\de_1+n}{2}}}
\ee
which up to a factor of $1/n$ agrees with the standard three-point structure~\eqref{eq:standard3pt_celestial} which appears in~\eqref{eq:3pt=C 2pt}.

We now need to compute the left-hand side of~\eqref{eq:htsDefinitionSimple}. We have
\be
	&\<(\cD'_n \cO)(x_2,z_2)(\cD'_n \cO)(x_0,z_0)\>=\cD'_{n,2}\cD'_{n,0}\frac{(-2z_2\.I(x_{20})\.z_0)^J}{x_{20}^{2\De}}\nn\\
	&=
	\frac{(J)_{(n)}}{n!}\cD'_{n,2} (-2z_2\.I(x_{20})\.w_0)^{n}\frac{(-2z_2\.I(x_{20})\.z_0)^{J-n}}{x_{20}^{2\De}}\nn\\
	&=\sum_{k=0}^n \binom{n}{k}\frac{(J)_{(n)}}{n!}\frac{(n)_{(k)}(J-n)_{(n-k)}}{n!} (-2w_2\.I(x_{20})\.w_0)^k(-2z_2\.I(x_{20})\.w_0)^{n-k}\nn\\
	&\qquad\times (-2w_2\.I(x_{20})\.z_0)^{n-k}\frac{(-2z_2\.I(x_{20})\.z_0)^{J-2n+k}}{x_{20}^{2\De}},
\ee
where we have defined $(a)_{(b)}\equiv a(a-1)\cdots (a-b+1)$. We now send $J\to n-1$ to find
\be
&\lim_{J\to n-1}\frac{1}{J-n+1}\<(\cD'_n \cO)(x_2,z_2)(\cD'_n \cO)(x_0,z_0)\>\nn\\
&=\sum_{k=0}^n \binom{n}{k}\frac{1}{n}(-1)^{n-k}(-2w_2\.I(x_{20})\.w_0)^k(-2z_2\.I(x_{20})\.w_0)^{n-k}\nn\\
&\qquad\times (-2w_2\.I(x_{20})\.z_0)^{n-k}\frac{(-2z_2\.I(x_{20})\.z_0)^{-1-n+k}}{x_{20}^{2\De}},\nn\\
&=\frac{1}{n}\p{(-2w_2\.I(x_{20})\.w_0)(-2z_2\.I(x_{20})\.z_0)-(-2w_2\.I(x_{20})\.z_0)(-2z_2\.I(x_{20})\.w_0)}^n\nn\\
&\qquad\frac{(-2z_2\.I(x_{20})\.z_0)^{-1-n}}{x_{20}^{2\De}}\nn\\
&=\frac{1}{n}\p{(-2H_{20}^{WW})(-2H_{20})-(-2H_{20}^{WZ})(-2H_{20}^{ZW})}^n\frac{(-2H_{20})^{-1-n}}{X_{20}^{\De+n-1}}
\ee
where $H^{AB}_{ij}$ is defined by replacing $Z_i$ by $A_i$ and $Z_j$ by $B_j$ in $H_{ij}$. This can now be evaluated in the configuration~\eqref{eq:celestiallocus} which yields, after multiplying by $-2H_{02}$, for the structure in the left-hand side of~\eqref{eq:htsDefinitionSimple}
\be
&\p{
	\lim_{J\to n-1} (-2H_{20})\frac{\<(\cD'_n \cO)(X_2,Z_2,\mathbf{W}_2)(\cD'_n \cO^\dagger)(X_0,Z_0,\mathbf{W}_0)\>}{J-J_n}
}\Bigg\vert_{\text{celestial}}\nn\\
&=\frac{1}{n}\frac{\p{(4z_2\.z_0 w_2\.w_0-4z_2\.w_0 z_0\.w_2)}^n}{(-2 z_2\.z_0)^{\de+n}}.
\ee
This agrees up to a factor of $1/n$ with the standard two-point structure~\eqref{eq:standard2pt_celestial} which appears in~\eqref{eq:3pt=C 2pt}. We thus find that both sides of~\eqref{eq:htsDefinitionSimple} differ from~\eqref{eq:3pt=C 2pt} by a factor of $1/n$, and therefore the two equations are equivalent.

\subsection{Selection rules in the $\wL[\cJ]\x\wL[\cJ]$ OPE}

In this section we consider the light-ray operators that contribute to the light-ray OPE~\eqref{eq:lightrayope_generalops} of two identical charge detectors, i.e.\ to the two light-transforms of identical $U(1)$ currents $\wL[\cJ]$. (The analysis for the non-abelian case is similar.\footnote{Although note the discussion of contact terms in~\cite{Kologlu:2019mfz}.}) For concreteness we focus on $d=4$ and we do not assume parity symmetry. However, the results we find will be valid in any dimension $d\geq 4$.

In $d=4$ the $\SO(d-1,1)=\SO(3,1)$ representations are parametrized by two-row Young diagrams, which can be supplemented with self- or anti-self duality constraints. However, since we are considering a non-chiral setup, it is convenient to use real tensor representations of $\SO(d-1,1)$ which do not have self-duality constraints. We thus parametrize these representations by pairs $(J,j)$. Local operators always have $J\geq j$.

First, we consider the local OPE of a $U(1)$ current $\cJ$ with itself. Using the counting rules of~\cite{Kravchuk:2016qvl} it is easy to see that for sufficiently large $J$ we have operators in the $\cJ\x \cJ$ OPE in irreps $(J,0),(J,1)$ with even and odd $J$, and in irreps $(J,2)$ with even $J$. This generic-$J$ behavior determines the light-ray operators that appear in the $\cJ\x\cJ$ OPE. To see this explicitly, recall that for even $J+j$ the number of structures is given by the dimension of~\cite{Kravchuk:2016qvl}
\be
	\p{S^2\myng{(1)}\otimes \mathrm{Res}^{\SO(3,1)}_{\SO(3)} (J,j)}^{\SO(3)},
\ee
where $\myng{(1)}$ is the $\SO(3)$ vector irrep\footnote{Using an $\SO(3)$ irrep instead of an $\SO(4)$ irrep takes into account the conservation of $\cJ$.} and for odd $J+j$ we need instead
\be
	\p{\wedge^2\myng{(1)}\otimes \mathrm{Res}^{\SO(3,1)}_{\SO(3)} (J,j)}^{\SO(3)}.
\ee
We have
\be\label{eq:SAvec}
	S^2\myng{(1)}=\myng{(2)}\oplus \bullet,\quad \wedge^2\myng{(1)}=\myng{(1)},
\ee
where $\bullet$ is the trivial representation, and the restriction of $(J,j)$ to $\SO(3)$ is
\be\label{eq:dimred}
	\mathrm{Res}^{\SO(3,1)}_{\SO(3)} (J,j)=\sum_{l=j}^J (l)
\ee
where $(l)$ is the spin-$l$ irrep of $\SO(3)$. We get tensor structures by matching $\SO(3)$ irreps between~\eqref{eq:SAvec} and~\eqref{eq:dimred}. We immediately see that there are no structures with $j>2$, and that for $j=2$ the spin $J$ must be even. For $j=0,1$ $J$ can be of any parity.

The transverse spins that appear in the $\cJ\x \cJ$ OPE are thus $0,1,2$. All these spins are traceless-symmetric in $d-2$ dimensions and thus are allowed to appear in the celestial OPE.\footnote{Note in $d>5$ three-row Young diagrams would appear in the $\cJ\x\cJ$ OPE, but the corresponding transverse spins are not allowed to appear in the celestial OPE of two scalars.} So, the set $\L_{12}$ in~\eqref{eq:lightrayope_generalops} is given by $\L_{12}=\{0,1,2\}$. We then have the low transverse-spin contributions, schematically
\be\label{eq:JJltsnaive}
	\wL[\cJ]\x\wL[\cJ]=\sum_i \mathbb{O}^+_{i,J=1,j=0}+ \mathbb{O}^+_{i,J=1,j=1}+ \mathbb{O}^+_{i,J=1,j=2}+\cdots.
\ee

A subtlety here is that the celestial map~\eqref{eq:ltsDefinitionSimple} typically maps multiple three-point tensor structures to zero, see appendix~\ref{app:tensorstructureselectionrule} for details. As we discuss there, the only structures that survive are those that contain $V_{0,12}^{-1}$, which are precisely those that are polynomial for $J>J_1+J_2-1$ but stop being polynomial exactly at $J=J_1+J_2-1$. 
In our case we are interested in even-spin structures which are polynomial for $J=2$ but are not polynomial for $J=1$, i.e.\ those which disappear from the counting above as we change $J=2$ to $J=1$.\footnote{Here we need to detach the notion of signature (even-spin or odd-spin) from parity of $J$ since we are analytically continuing in $J$. That is, for $J=1$ we still use $S^2\myng{(1)}$ for $j=0,2$ and $\wedge^2 \myng{(1)}$ for $j=1$.} We see that for $j=2$ the number of structures changes from $1$ to $0$ and for $j=0$ from $2$ to $1$. This happens because the $\SO(3)$ content of $(J,j)$ changes: the spin-2 irrep disappears and cannot be paired with the spin-2 irrep in $S^2\myng{(1)}$. However, for $j=1$ the number of structures doesn't change because the only structure comes from pairing with the spin-1 irrep in $\wedge^2\myng{(1)}$, and thus all $j=1$ structures are annihilated by the celestial map.

The final form of the low transverse spin contributions is therefore
\be
	\wL[\cJ]\x\wL[\cJ]=\sum_i \mathbb{O}^+_{i,J=1,j=0}+ \mathbb{O}^+_{i,J=1,j=2}+\cdots.
\ee
This is of course consistent with the fact that on the celestial sphere we have two identical scalars, and thus only even $j$ should be allowed.\footnote{Here we assume, as usual, that the product $\wL[\cJ]\x\wL[\cJ]$ is well-defined and thus the two light-transforms commute. This requires the Regge intercept $J_0$ to satisfy $J_0<J_1+J_2-1=1$~\cite{Kologlu:2019bco}.}

For higher transverse spin we find that the label~$\g$ in~\eqref{eq:lightrayope_generalops} is trivial because $\SO(d-4)$ is trivial in our case. We have $\l(+n)=2+n$ and thus the higher transverse spin terms take the schematic form
\be
	\wL[\cJ]\x\wL[\cJ]=\sum_{n,i} \cD_n\mathbb{O}^{(-1)^n}_{i,J=1+n,j=2}+\cdots.
\ee
However, since $\mathbb{O}^{-}_{i,J,j=2}$ do not appear in the $\cJ\x\cJ$ OPE, we find
\be
	\wL[\cJ]\x\wL[\cJ]=\sum_{n,i} \cD_{2n}\mathbb{O}^{+}_{i,J=1+2n,j=2}+\cdots.
\ee
Note that these contributions have even transverse spin $j=2+2n$ and thus this expansion is again consistent with permutation symmetry on the celestial sphere.

Summarizing, we have the following schematic contributions to the OPE of two charge detectors in $d=4$,
\be\label{eq:JJschematic}
	\wL[\cJ]\x\wL[\cJ]=\sum_i\p{ \mathbb{O}^+_{i,J=1,j=0}+ \mathbb{O}^+_{i,J=1,j=2}}+\sum_{n,i} \cD_{2n}\mathbb{O}^{+}_{i,J=1+2n,j=2}.
\ee

\subsection{Selection rules in the $\wL[T]\x\wL[T]$ OPE}

We now discuss the case of the OPE of two energy detectors, i.e.\ two light-transforms of $T$. We use the same setup as in the previous section, i.e.\ we work in $d=4$ and in terms of real tensor irreps of~$\SO(3,1)$.

The transverse spins appearing in the $T\x T$ OPE are analogous to the $\cJ\x\cJ$ case. We have spins $j=0,1,2,3$ for both even and odd $J$ and spin $j=4$ for even $J$. This translates to the following analogue of~\eqref{eq:JJltsnaive}
\be
\wL[T]\x\wL[T]=\sum_i \mathbb{O}^+_{i,J=3,j=0}+ \mathbb{O}^+_{i,J=3,j=1}+ \mathbb{O}^+_{i,J=3,j=2}+ \mathbb{O}^+_{i,J=3,j=3}+ \mathbb{O}^+_{i,J=3,j=4}+\cdots.
\ee
However, we again must take care of the fact that the celestial map~\eqref{eq:ltsDefinitionSimple} annihilates some tensor structures. In this case, using the same logic as before, we find that only $j=0,2,4$ operators have tensor structures that are non-vanishing under the celestial map. Therefore, the low-transverse spin contribution is actually
\be
\wL[T]\x\wL[T]=\sum_i \mathbb{O}^+_{i,J=3,j=0}+ \mathbb{O}^+_{i,J=3,j=2}+  \mathbb{O}^+_{i,J=3,j=4}+\cdots.
\ee
Again, this is consistent with the permutation symmetry on the celestial sphere that only allows even $j$.

The analysis of higher transverse spin contributions is also the same as in the $\cJ\x\cJ$ case. We have from~\eqref{eq:lightrayope_generalops}
\be
	\wL[T]\x\wL[T]=\sum_{n,i} \cD_n\mathbb{O}^{(-1)^n}_{i,J=3+n,j=4}+\cdots.
\ee
Taking into account that only even-spin $j=4$ operators appear in $T\x T$ and combining with the low transverse spin terms, we find
\be
\label{eq:formofee}
	\wL[T]\x\wL[T]=\sum_i\p{\mathbb{O}^+_{i,J=3,j=0}+ \mathbb{O}^+_{i,J=3,j=2}+ \mathbb{O}^+_{i,J=3,j=4}}+\sum_{n,i} \cD_{2n}\mathbb{O}^{+}_{i,J=3+2n,j=4}.
\ee
As mentioned above, although we have derived (\ref{eq:formofee}) in $d=4$, the result is valid for any $d\geq 4$.

It is interesting to ask what are the leading operators appearing at various transverse spins in the above expansion in a weakly-coupled gauge theory. At $j=0$ it is well-known that the leading twist is $\tau_0=2$ operators which can take the schematic form,
\be
	\bar\phi\ptl_{\b_1\dot\a_1}\cdots\ptl_{\b_{J}\dot\a_J}\phi,\quad 
	\bar\psi_{\dot\a_1}\ptl_{\b_2\dot\a_2}\cdots\ptl_{\b_{J}\dot\a_J}\psi_{\b_1},\quad 
	\bar F_{\dot\a_1\dot\a_2}\ptl_{\b_3\dot\a_3}\cdots\ptl_{\b_{J}\dot\a_J}F_{\b_1\b_2},
\ee
where we assume that the gauge indices are implicitly contracted, and the dotted and undotted indices are implicitly symmetrized. At $j=2$ the leading twist\footnote{For all transverse spins we define twist as $\De-J$.} is also $\tau_2= 2$, for the operators
\be
	F_{\b_1\b_2}\ptl_{\b_5\dot\a_5}\cdots\ptl_{\b_{J+2}\dot\a_{J+2}}F_{\b_3\b_4}.
\ee
To see that this is the minimal possible twist, note that $\tau\geq 2$ is the unitarity bound in $d=4$ for generic-$J$ operators. To see that there are no other operators, note that the twist of a product of symbols is bounded from below by the sum of constituent twists, and the classical twist of all fundamental fields is $1$, while the twist of a derivative is $0$. Therefore only products of two fundamental fields and any number of derivatives can have twist $\approx 2$, provided no indices are contracted. Fixing the value of transverse spin then leaves us with the above options.

From this argument it is clear that no $j=4$ operators with twist $\tau=2$ exist. To build the lowest-twist $j=4$ operators we need to use more than $2$ fundamental fields but as few as possible. Since $F$ carries the most transverse spin among all fundamental fields, we find that the lowest-twist operators $j=4$ have twist $\tau_4=4$ and are schematically given by,
\be
	F\ptl^{J_1}F\ptl^{J_2}F\ptl^{J_3}F,
\ee
where we keep spinor indices uncontracted and symmetrized, with $J_1+J_2+J_3+4=J$. Note that there are multiple ways in which the gauge indices can be contracted. The fact that $j=4$ operators have higher twist than required by the unitarity bound is in general a consequence of the improved unitarity bounds of~\cite{Cordova:2017dhq}, which state
\be
	\tau \geq \max\{2,j\}.
\ee

In terms of the celestial quantum numbers, we get the following contributions. At $j=0$ and $j=2$ we have the leading contributions with dimension
\be
	\de=\De-1\approx \tau_j+J-1=\tau_j+2=4.
\ee
Note that this dimension corresponds to the singularity of the form $\theta^{\de-\de_1-\de_2}=\theta^{-2}$.
For $j=4+2n\geq 4$ we get 
\be
	\de=\De-1\approx\tau_4+J-1=\tau_4+2n+2=6+2n,
\ee
which corresponds to the leading short-angle dependence of the form $\theta^{2n}$. Since the leading classical twists $\tau_j$ are all at the (improved) unitarity bounds, the anomalous dimensions should be positive. Therefore, the leading short-angle asymptotics from these contributions in the interacting theory should be softer than the ones given above.

\section{Example: event shape  in $\cN=4$ SYM}
\label{sec:N4example}

In this section we consider an example of an event shape that includes the transverse spin structures discussed above. More precisely, we consider the following event shape in $\cN=4$ SYM:
\be \label{eq:JSSO event shape}
\<\cO_{20'}(p)|\wL[\cO_{20'}](\oo,z_1)\wL[\cO_{20'}](\oo,z_2)|J(p,z_3)\>,
\ee
where $J$ is the R-symmetry current, which is in the same multiplet as the half-BPS operator $\cO_{20'}$. We will first compute~\eqref{eq:JSSO event shape} directly by performing the light transform of the relevant four-point function. Next, we compute it using the light-ray OPE formula. In both cases, we will derive a Ward identity that relates the event shape~\eqref{eq:JSSO event shape} to the energy-energy correlator calculated in~\cite{Kologlu:2019mfz}. The result is given by~\eqref{eq:relationtoEEC} and~\eqref{eq:relationtoEECfromlightray}. Despite the simplicity of our result, the fact that the Ward identity can be derived in two independent ways still provides a nontrivial check of our formulas.

\subsection{Direct computation}
\label{sec:directcomputation}

The computation takes a few steps, summarized as follows. We start with the expression for the correlator $\< \cO_{20'} \cO_{20'} \cO_{20'} J \> $ in terms of  the scalar correlator $\< \cO_{20'} \cO_{20'} \cO_{20'} \cO_{20'} \> $. The two are related via the superconformal Ward identities \cite{Belitsky:2014zha}. Then, we go to the Mellin space representation of the correlator, and perform the light transforms there. Finally, we Fourier transform the separation of the in and out states to obtain the desired event shape.

The relevant four-point function is given by the following expression, see (3.15) in  \cite{Belitsky:2014zha},
\be\label{eq:JOOO4pt}
\<   \cO_{20'}(x_4) \cO_{20'}(x_1)  \cO_{20'}(x_2)  J_{\alpha \dot \alpha, a a'}(x_3)  \> &= {1 \over 4} (\partial_{x_3} )_{\ \dot \alpha}^{\beta}  (y_{12}^2 y_{14}^2 Y_{324} - v y_{12}^2 y_{24}^2 Y_{314} - u y_{24}^2 y_{14}^2 Y_{321})_{a a'} \nn \\
&\quad\times \< X_{324}, X_{314} \>_{(\alpha \beta)} {\Phi(u,v) \over x_{12}^2 x_{24}^2 x_{14}^2} ,
\ee
where our spinor conventions can be found in appendix ~\ref{sec:spinor conventions}. The function $\Phi(u,v)$ is the part of the four-point function of $\cO_{20'}$s that contains the nontrivial dynamical data, see e.g.\ section 7.2 in \cite{Kologlu:2019mfz}. It satisfies $\Phi(u,v) = \Phi(v,u) = {1 \over v} \Phi({u \over v}, {1 \over v})$. The rest of the ingredients are various kinematical factors that require some unpacking. The $\a,\dot \a=1,2$ are spinor indices for the Lorentz group $SU(2)_L\times SU(2)_R$. The $y_i$ are auxiliary variables keeping track of the $SU(4)$ R-symmetry, see \cite{Belitsky:2014zha} for details. The structures $X_{ijk}$ and $Y_{ijk}$ are defined as follows:
\be
(X_{i j k})_{\alpha \dot \alpha} = {(x_{ij})_{\alpha \dot \beta} \over x_{ij}^2} (x_{j k})^{\dot \beta \gamma}  {(x_{ki})_{\gamma \dot \alpha} \over x_{ik}^2} \ , \quad (Y_{ijk})_{a a'} = (y_{ij})_{ab'} (y_{jk})^{b'b} (y_{ki})_{b a'}.
\ee The commutator bracket $\<~,~\>$ for spinor indices\footnote{We use angular brackets to denote the commutator in spinor indices to avoid a clash with the commutator in vector indices denoted by the traditional square brackets.} is defined as
\be
\< a, b \>_{(\alpha \beta)} \equiv a_{\alpha \dot \alpha} b^{\dot \alpha \gamma} \eps_{\gamma \beta} - b_{\alpha \dot \alpha} a^{\dot \alpha \gamma} \eps_{\gamma \beta}\, .
\ee
Note that $\< a, b\>_{(\alpha \beta)}$ is automatically symmetric under permutation of $\alpha$ and $\beta$. 

To connect with the event shape, we set $x_4=0$, and $x_3$ will eventually be Fourier transformed with momentum $p$. For now, we can replace the derivative $(\partial_{x_3} )_{\ \dot \alpha}^{\beta} $ with ${i \over 2} p_{\ \dot \alpha}^{\beta}$.
We pass to index-free notation by introducing the polarization vector ${1 \over 2} z_3^{\dot \alpha \alpha}$, such that $z_3^2 = 0$, and contracting ${1 \over 2} z_3^{\dot \alpha \alpha} J_{\alpha \dot \alpha} = z_3^\mu J_{\mu}$. Contracting with the polarization vector produces a term
\be \label{eq:z dot p}
p_{\ \dot \alpha}^{\beta} z_3^{\dot \alpha \alpha} = {1 \over 2} \< z_3,p \>^{(\beta \alpha)} + (z_3 \cdot p) \eps^{\beta \alpha} \ .
\ee
In terms of the polarization vector, the correlator becomes
\be
\<  \cO_{20'}(x_4) \cO_{20'}(x_1)  \cO_{20'}(x_2)  J_{a a'}(x_3, z_3) \> &= {i \over 32} (y_{12}^2 y_{14}^2 Y_{324} - v y_{12}^2 y_{24}^2 Y_{314} - u y_{24}^2 y_{14}^2 Y_{321})_{a a'} \nn \\
&\quad \times \< z_3,p \>^{(\alpha \beta)} \< X_{324}, X_{314}\>_{(\alpha \beta)} {\Phi(u,v) \over x_{12}^2 x_{24}^2 x_{14}^2} .
\ee
In this formula we only Fourier transformed the external derivative while keeping the rest in coordinate space. 
The contraction of the brackets can be performed by the identity
\be
\< a,b \>^{(\a\b)}\< c,d\>_{(\a\b)} = 8(g^{\mu \s}g^{\nu\rho} -g^{\mu\rho}g^{\nu\s} +i \e^{\mu\nu\rho\s}) a_\mu b_\nu c_\rho d_\s \,.
\ee

To compute the event shape, we place the detectors at embedding space coordinates
\be
X_i = X_\infty = (0,1,0), \quad Z_i=(0,0,z_i), \quad \text{for}~i=1,2,
\ee 
and take the light-transforms 
\be
\label{eq:defdet}
\wL[\cO_{20'}](\oo,z_i)= \int_{-\infty}^\infty d\a_i \, \cO_{20'}(Z_i-\a_i X_{\oo}) ,
\ee 
while the external states are placed at $X_3 = (1,x_3^2,x_3)$ and $X_4 = (1,0,0)$. 

We find it convenient to work in the $x$ Poincar\'{e} patch and approach the spatial infinity insertion as follows
\be
\label{eq:approach}
\wL[\cO_{20'}](\oo,z_i)=  \int_{-\infty}^\infty d\a_i \lim_{r_i \to \infty} r_i^{2} \cO_{20'}( r_i z_i + \alpha_i  \bar z_i ) ,
\ee
where $\bar z_i^2 = 0$, $(- z_i \cdot \bar z_i) = {1 \over 2}$, and $\bar z_i$ is arbitrary otherwise. In the embedding space, (\ref{eq:approach}) corresponds to choosing
\be
Z_i &= \lim_{r \to \infty} Z_i^r = \p{{1 \over r_i} , 0 , z_i} , \nn \\
X_{\oo} &= \lim_{r \to \infty} X_\oo^{r_i} = \p{0,1,{\bar z_i \over r_i}} .
\ee
Note that $(Z_i^r)^2 = (X_\oo^{r_i})^2 = (Z_i^r \cdot X_\oo^{r_i}) = 0$. It is clear from the definition (\ref{eq:defdet}) that the final result does not depend on the particular choice of $\bar z_i$.
For convenience, we also define the null coordinates
\be
x_{i-} &\equiv (- x_i \cdot z_i)  =  \a_i/2 \ ,\nn \\
x_{3i-} &\equiv (- x_3\cdot z_i) - x_{i-} \ , 
\ee for $i=1,2$. 

Next we evaluate the integrand in the kinematics above. The cross ratios $u$ and $v$ take the form
\be
u = \frac{x_3^2 z_1\cdot z_2}{2 x_{2 -} x_{31 -}}  \ , \qquad v= \frac{x_{1 -} x_{32 -}}{x_{2 -} x_{31 -}}\ ,
\ee and the commutator becomes 
\be
\label{eq:celcomm}
\<X_{324},X_{314}\>_{(\alpha \beta)} = {1 \over 2} {1 \over x_3^2 x_{32-} x_{31-}} \left( -x_{32-} \< z_1, x_3\>_{(\alpha \beta)} + x_{31-} \< z_2, x_3\>_{(\alpha \beta)} + {1 \over 2} x_3^2 \<z_2, z_1\>_{(\alpha \beta)} \right).
\ee 
With these expressions, the light transforms are evaluated by the integral
\be
\label{eq:integralFN4}
&\<  \cO_{20'}(0) | \wL [\cO_{20'}](\infty, z_1) \wL[\cO_{20'}](\infty, z_2)  | J_{a a'}(x_3, z_3) \> \nn
\\ &\qquad =
\int_{-\infty}^{\infty} d x_{1-} d x_{2-} {i \over 32} (y_{12}^2 y_{14}^2 Y_{324} - v y_{12}^2 y_{24}^2 Y_{314} - u y_{24}^2 y_{14}^2 Y_{321})_{a a'} \nn \\
&\qquad \qquad\times \< z_3,p\>^{(\alpha \beta)} \<X_{324}, X_{314}\>_{(\alpha \beta)} {\Phi(u,v) \over (-2z_1 \cdot z_2) x_{1-} x_{2-}} .
\ee

To perform the light transform integrals above, it is very convenient to use the Mellin representation for $\Phi(u,v)$, see e.g.\ \cite{Belitsky:2013bja},
\be
\label{eq:Mellinform}
\Phi(u,v) &={v \over u} \int_{{\cal C}_0} {d \gamma_{12} d \gamma_{14}  \over (2 \pi i)^2} \Gamma(\gamma_{12})^2 \Gamma(\gamma_{14})^2 \Gamma(2- \gamma_{12}-\gamma_{14})^2 M(\gamma_{12}, \gamma_{14}) u^{- \gamma_{12}} v^{- \gamma_{14}} \ , \nn \\
{\cal C}_0 &: ~~~ {\rm Re}[\gamma_{12}]>-1,~~~ {\rm Re}[\gamma_{14}],{\rm Re}[\gamma_{13}]>1 ,
\ee
where $\gamma_{12} + \gamma_{13} + \gamma_{14} = 2$. The condition ${\rm Re}[\gamma_{13}]>1$ thus becomes ${\rm Re}[\gamma_{12} + \gamma_{14}]<1$. The weak and strong coupling results take the form
\be
\label{eq:Mellin amplitudes at weak and strong coupling}
M^{\text{weak}}(\gamma_{12}, \gamma_{14}) &= - {a \over 4} {\gamma_{12}^2 \over (\gamma_{14}-1)^2 (\gamma_{13}-1)^2} , \nn \\
M^{\text{strong}}(\gamma_{12}, \gamma_{14}) &= - {1 \over 2} {\gamma_{12}^2 (1+\gamma_{12}) \over (\gamma_{14} - 1) (\gamma_{13} - 1)} \ .
\ee

We can plug the Mellin representation~(\ref{eq:Mellinform}) above into (\ref{eq:integralFN4}), and perform the light-transform integrals using the formula
\be
\int_{- \infty}^{\infty} d x_- \ (x_- + i \eps)^{-a} ( (-z\cdot x) - x_- + i \eps)^{- b} = -2 \pi i (-z \cdot x)^{1-a-b} {\Gamma(a+b-1) \over \Gamma(a) \Gamma(b)} \ .
\ee
The integral converges for ${\rm Re}[a+b]>1$. 

At this point, we observe that the terms in the correlator proportional to $ y_{12}^2$ produce a divergent result. This means that the event shapes for the corresponding R-symmetry structures are not well-defined. To obtain a well-defined event shape, we set $y_{12}^2=0$, so that only the $- u y_{24}^2 y_{14}^2 Y_{321}$ term survives. This term produces a finite result for the integral, which converges for ${\rm Re}[\gamma_{12}]<0$. Note that the condition $y_{12}^2 = 0$ keeps representations ${\bf 84}$, ${\bf 105}$, ${\bf 175}$ in the OPE of scalars $\cO_{ 20'}$ \cite{Belitsky:2013bja}, whereas in the $J\times \cO_{ 20'}$ OPE we have the representations
\be
{\bf 15} \times {\bf 20}' = {\bf 15} + {\bf 20}' + {\bf 45} + \bar {\bf 45} + {\bf 175} \ .
\ee Therefore, the only representation that appears in the event shape determined by $y_{12}^2 = 0$ is ${\bf 175}$. From now on, we will focus on this finite event shape in the ${\bf 175}$ R-symmetry channel:
\be\label{eq:175Rsymmetrychannel}
\< \cO_{20'} | \wL[\cO_{20'}]\wL[\cO_{20'}] |J_{aa'} \>|_{y_{12}^2 =0} = (y_{24}^2 y_{14}^2 Y_{321})_{aa'} \< \cO_{20'} | \wL[\cO_{20'}]\wL[\cO_{20'}] |J \>_{{\bf 175}} .
\ee

Performing the light-transforms for the event shape in the ${\bf 175}$ channel, we arrive at the expression
\be
&\< \cO_{20'}(p) | \wL[\cO_{20'}](\oo,z_1)\wL[\cO_{20'}](\oo,z_2) |J(p,z_3) \>_{{\bf 175}} \nn \\
&\qquad=- {i \over 32}  {1 \over (z_1 \cdot z_2)^2}  \int_{{\cal C}_0} {d \gamma_{12} d \gamma_{14}  \over (2 \pi i)^2} {M(\gamma_{12}, \gamma_{14}) \over \gamma_{12}} {2 \pi^4 \over (\sin \pi \gamma_{12})^2}  \cD_p \int d^4 x {e^{i p x} \over x^4} \gamma^{\gamma_{12} - 1}  , 
\label{eq:event shape as differential eq acting on fourier}
\ee
where $\gamma =2 {(-x \cdot z_1) (-x \cdot z_2) \over x^2 (z_1 \cdot z_2)}$. The differential operator $\cD_p$ is given by
\be
\cD_p&\equiv\< z_3,p\>^{(\alpha \beta)}\left( (\gamma_{14}-1) \<z_1, \partial_p\>_{(\alpha \beta)}   z_2 \cdot \partial_p  -  (\gamma_{13}-1) \<z_2, \partial_p\>_{(\alpha \beta)}  z_1 \cdot \partial_p    - {1 \over 2}  \g_{12} \<z_2, z_1\>_{(\alpha \beta)} \partial_p^2\right) .
\ee
Finally, the Fourier transform can be performed using the following master formula:
\be
\int d^d x {e^{i p \cdot x}  \over (x^{2} - i \eps x^0)^a} \gamma^b = \theta(p) \times  2^{1-2a+d} \pi^{1+{d \over 2}} (-p^2)^{a - {d \over 2}}{ \z^{-b}  \ _2 F_1 (-b,-b,1+a-b-{d \over 2}, \z) \over \Gamma(a+b) \Gamma(1+a-b-{d \over 2})} \ ,
\ee where we introduced $\theta(p) \equiv \theta(p^0) \theta(-p^2) $ and the cross ratio
\be
\label{eq:event shape cross ratio}
\zeta = {(-2z_1 \cdot z_2) (-p^2 ) \over  (-2p \cdot z_1) (-2p \cdot z_2)}.
\ee 
Setting $a = 2$, $b = \gamma_{12} - 1$ and $d=4$ we obtain
\be
\label{eq:JSSO fourier result}
\int d^4 x {e^{i p x} \over x^4} \gamma^{\gamma_{12} - 1} = { 2 \pi^2 \sin \pi \gamma_{12} \over \gamma_{12}} \int_0^\zeta d w w^{-\gamma_{12}} (1-w)^{\gamma_{12} - 1} .
\ee
Acting with $\cD_p$ on the result of the Fourier transform~(\ref{eq:JSSO fourier result}) and using crossing symmetry of the Mellin amplitude $M(\gamma_{12}, \gamma_{14})=M(\gamma_{12}, \gamma_{13})$, we get
\be
\label{eq:resultf1}
&\< \cO_{20'}(p) | \wL[\cO_{20'}](\oo,z_1)\wL[\cO_{20'}](\oo,z_2)  | J(p,z_3)\>_{{\bf 175}} \nn \\
&= i {[z_3,p]\cdot[z_1, z_2] \theta(p)  \over (-2z_1 \cdot z_2)^2} 
\int_{{\cal C}_0} {d \gamma_{12} d \gamma_{14}  \over (2 \pi i)^2} M(\gamma_{12}, \gamma_{14}) {\pi^6 \over \sin \pi \gamma_{12}} {4 \over p^2} \left( {\zeta \over 1- \zeta} \right)^{1 -  \gamma_{12}} ,
\ee
where the commutator $[a,b]^{\mu\nu}$ is defined in the same way as~\eqref{eq:commutatordefinition}, and contracting a pair gives
\be
\label{eq:prod}
[a,b]\cdot[c, d] =  2 \left[  (a \cdot c) (b \cdot d) - (a \cdot d) (b \cdot c) \right].
\ee
We can rewrite the result above as 
\be
&\< \cO_{20'}(p) | \wL[\cO_{20'}](\oo,z_1)\wL[\cO_{20'}](\oo,z_2)  | J
(p,z_3) \>_{{\bf 175}}  =16 i \pi^5  
{ [z_3,p]\cdot[z_1, z_2]   (-p^2) \theta(p) \over (-2 p \cdot z_1)^2 (-2 p \cdot z_2)^2} {\cal F}_{JO}(\zeta) \ ,
\ee
where
\be
\label{eq:finalresultB}
{\cal F}_{JO}(\zeta)=-{1 \over 4 \z^2} \int_{{\cal C}_0} {d \gamma_{12} d  \gamma_{14}  \over (2 \pi i)^2}  M(\gamma_{12},  \gamma_{14})  {\pi \over \sin \pi \gamma_{12}}  \left( {\zeta \over 1- \zeta} \right)^{1 - \gamma_{12}}   \ . 
\ee
The result (\ref{eq:finalresultB}), see e.g.\ \cite{Belitsky:2013xxa}, immediately implies that
\be
\label{eq:relationtoEEC}
{\cal F}_{JO}(\zeta)= - {\z \over 2} {\cal F}_{\cE}(\zeta) ,
\ee
where ${\cal F}_{\cE}(\zeta)$ was defined in \cite{Kologlu:2019mfz} and computes the energy-energy correlator.

Using (\ref{eq:relationtoEEC}) and the formulas \cite{Kologlu:2019mfz}, or directly computing the Mellin integral in~(\ref{eq:finalresultB}) with the Mellin amplitudes~(\ref{eq:Mellin amplitudes at weak and strong coupling}), we get the results at weak and at strong coupling
\be
{\cal F}_{JO}^{{\rm weak}}(\zeta) &= {a \over 8} {\log 1 - \z \over \z (1-\z)} , \quad \quad \quad {\cal F}_{JO}^{\text{strong}}(\zeta) = - {\z \over 4} .
\ee

\subsection{Computation using the light-ray OPE}

Now we compute the event shape~\eqref{eq:JSSO event shape} using the light-ray OPE formula. The complete light-ray OPE formula for two scalars including higher transverse spin terms is given by~\eqref{eq:scalarOPEfinal_simplified}. Plugging the formula into~\eqref{eq:JSSO event shape}, we have
\be\label{eq:N=4SYMeventshapeexpression1}
&\<\cO_{20'}(p)|\wL[\cO_{20'}](\oo,z_1)\wL[\cO_{20'}](\oo,z_2)|J(p,z_3)\>\nn \\
&=\pi i\int_{2-i\oo}^{2+i\oo}\frac{d\De}{2\pi i}\bigg(\cC_{\De-1,0}(z_1,z_2,\ptl_{z})\<\cO_{20'}(p)|\mathbb{O}^{+}_{\De,J=-1}(\oo,z)|J(p,z_3)\> \nn \\
&\qquad\qquad\qquad\quad+\cC_{\De-1,1}(z_1,z_2,\ptl_z,\ptl_w)\<\cO_{20'}(p)|(\cD_1\mathbb{O}^{-}_{\De,J=0})(\oo,z,w)|J(p,z_3)\>\bigg) \nn \\
&=-\pi i\int_{2-i\oo}^{2+i\oo}\frac{d\De}{2\pi i}\bigg(C^+_a(\De,-1)\cC_{\De-1,0}(z_1,z_2,\ptl_{z})\<\cO_{20'}(p)|\wL[\cO_{\De,-1}](\oo,z)|J(p,z_3)\>^{(a)}_{+} \nn \\
&\qquad\qquad\qquad\quad+C^-_a(\De,0)\cC_{\De-1,1}(z_1,z_2,\ptl_z,\ptl_w)\<\cO_{20'}(p)|\cD_1\wL[\cO_{\De,0}](\oo,z,w)|J(p,z_3)\>^{(a)}_{-}\bigg),
\ee
where $C^{\pm}_a(\De,J)$ is a coefficient function that encodes the OPE data. It has poles of the form
\be
C^{\pm}_a(\De,J)\sim -\frac{p_{i,J}^{(a)}}{\De-\De_{i,J}}
\ee
where $p_{i,J}^{(a)}$ and $\De_{i,J}$ are the product of OPE coefficients and the scaling dimension of an exchanged operator. Note that there's just one structure label in $C^{\pm}_a(\De,J)$ because the three-point function from the two $\cO_{20'}$'s only has one tensor structure. In the second equality in~\eqref{eq:N=4SYMeventshapeexpression1}, we use the relation between light-ray operators $\mathbb{O}^{\pm}_{\De,J}$ and $C^{\pm}_a(\De,J)$~\cite{Kravchuk:2018htv}
\be
\<\cO_{20'}\mathbb{O}^{\pm}_{\De,J}J\>_{\O}=-C^{\pm}_a(\De,J)\<0|\cO_{20'}\wL[\cO_{\De,J}]J|0\>_{\pm}^{(a)},
\ee
where $\<0|\cO_{20'}\wL[\cO_{\De,J}]J|0\>_{\pm}^{(a)}$ is the analytic continuation of the continuous-spin Wightman function $\<0|\cO_{20'}\wL[\cO_{\De,J}]J|0\>^{(a)}$ from either even or odd spin. It can be defined straightforwardly following section~\ref{sec:Jcontinuation}. We will give their explicit expressions later in section~\ref{sec:celestialblocks}. 

Furthermore, as discussed in section~\ref{sec:directcomputation}, we focus on the $\textbf{175}$ R-symmetry channel event shape defined in~\eqref{eq:175Rsymmetrychannel}. Since the $\textbf{175}$ representation is antisymmetric under the exchange of $\cO_{20'}$'s, the OPE should only contain operators with odd spin, and we have $C^+(\De,J)=0$ for all $J$. The $\textbf{175}$ R-symmetry channel event shape is then given by
\be\label{eq:N=4SYMeventshape}
&\<\cO_{20'}(p)|\wL[\cO_{20'}](\oo,z_1)\wL[\cO_{20'}](\oo,z_2)|J(p,z_3)\>_{\textbf{175}}\nn \\
&=-\pi i\int_{2-i\oo}^{2+i\oo}\frac{d\De}{2\pi i}C'^-_a(\De,0)\cC_{\De-1,1}(z_1,z_2,\ptl_{z},\ptl_{w})\<\cO_{20'}(p)|\cD_{1}\wL[\cO_{\De,0}](\oo,z,w)|J(p,z_3)\>^{(a)}_{-},
\ee
where $C'^-_a(\De,J)$ is simply given by the coefficient function $C_a^-(\De,J)$ without the R-symmetry factor $(y_{24}^2 y_{14}^2 Y_{321})$.

\subsubsection{Celestial blocks}
\label{sec:celestialblocks}
In order to compute \eqref{eq:N=4SYMeventshape}, we first note that by Lorentz invariance and homogeneity,
\be\label{eq:lighttransform3ptfunctionproportionalityrelation}
\<\cO_{20'}(p)|\cD_{1}\wL[\cO_{\De,0}](\oo,z,w)|J(p,z_3)\>^{(a)}_{-} \propto (2w\.z_3z\.p-2w\.pz\.z_3)(-2p\.z)^{-\de-1}.
\ee
Therefore, we need to solve for the $j=1$ celestial block defined by
\be
\cC_{\de,1}(z_1,z_2,\ptl_{z},\ptl_{w})\p{(2w\.z_3z\.p-2w\.pz\.z_3)(-2p\.z)^{-\de-1}}.
\ee
This is the higher transverse spin version of the celestial block computed in \cite{Kologlu:2019mfz}. Lorentz invariance and homogeneity imply that
\be\label{eq:j1celestialblock}
&\cC_{\de,1}(z_1,z_2,\ptl_{z},\ptl_{w})\p{[w,z]\.[z_3,p](-2p\.z)^{-\de-1}}\nn \\
&=\frac{(-p^2)^{\frac{\de_1+\de_2-\de+1}{2}}}{(-2p\.z_1)^{\de_1+1}(-2p\.z_2)^{\de_2+1}}\nn \\
&\qquad\x\p{[z_3,p]\.[z_1,z_2]g(\z)+\{z_3,p\}\.\{z_1,z_2\}h(\z)-\frac{4p\.z_1p\.z_2p\.z_3}{p^2}h(\z)},
\ee
where once again $[a,b]^{\mu\nu}$ is defined in the same way as~\eqref{eq:commutatordefinition}, and similarly
\be
\{a,b\}^{\mu\nu}=a^{\mu}b^{\nu}+b^{\mu}a^{\nu},
\ee
and the cross ratio $\z$ is given by~(\ref{eq:event shape cross ratio}). 
One way to obtain the functions $g(\z)$ and $h(\z)$ is using the fact that \eqref{eq:j1celestialblock} is an eigenvector of the quadratic Casimir of the Lorentz group acting simultaneously on $z_1,z_2$ with eigenvalue $\de(\de-d+2)+d-3$. Proceeding this way, one gets two coupled second-order inhomogeneous differential equations of $g(\z)$ and $h(\z)$, and their boundary conditions are given by the OPE limit of \eqref{eq:j1celestialblock}. However, finding the solutions to these differential equations is a nontrivial task. Furthermore, the system of differential equations gets more and more complicated when one has even higher transverse spin. We would like a method that allows us to compute celestial blocks with general transverse spin. Fortunately, this can be achieved by using weight-shifting operators \cite{Karateev:2017jgd} and the $j=0$ celestial block calculated in \cite{Kologlu:2019mfz}.

First, note that $[w,z]\.[z_3,p](-2p\.z)^{-\de-1}$ can be written in terms of a ``bubble diagram"
\be\label{eq:weightshiftingbubble}
[w,z]\.[z_3,p](-2p\.z)^{-\de-1}=\frac{\cD_{z,w}^{0+}\.\cD_{z,w}^{0-}}{(d-4)(\de-1)(\de-d+3)}\p{[w,z]\.[z_3,p](-2p\.z)^{-\de-1}},
\ee
where $\cD_{z,w}^{0+\mu}$ and $\cD_{z,w}^{0-\mu}$ are weight-shifting operators defined in \cite{Karateev:2017jgd}, with the embedding space coordinates $(X,Z)$ replaced with $(z,w)$. The explicit expression of $\cD_{z,w}^{0+\mu}$ is also given in~\eqref{eq:Dndef}. The operator $\cD_{z,w}^{0+\mu}$ increases the transverse spin and $\cD_{z,w}^{0-\mu}$ decreases the transverse spin, so acting $\cD_{z,w}^{0+}\.\cD_{z,w}^{0-}$ will basically give us the same expression with some overall factor. One can perform crossing on the weight-shifting operator $\cD_{z,w}^{0-\mu}$ such that
\be\label{eq:weightshiftingcrossing1}
\cD_{z,w}^{0-\mu}\p{[w,z]\.[z_3,p](-2p\.z)^{-\de-1}}=\cD_p^{\mu}(-2p\.z)^{-\de},
\ee
where $\cD_p^{\mu}$ is a differential operator acting on $p$. By explicitly evaluating the left-hand side of~\eqref{eq:weightshiftingcrossing1}, we find that $\cD_p^{\mu}$ is given by
\be
\cD_p^{\mu}=-\frac{(d-4)}{\de}\p{(\de-d+3)p^\mu z_3\. \frac{\ptl}{\ptl p}+p^2z_3\. \frac{\ptl}{\ptl p}\frac{\ptl}{\ptl p_\mu}+(\de+1)z_3\.p\frac{\ptl}{\ptl p_\mu}+\de(\de-d+3)z_3^\mu} \, .
\ee
Alternatively, $(-2p\.z)^{-\de}$ and $[w,z]\.[z_3,p](-2p\.z)^{-\de-1}$ can be viewed as spin $0$ and spin $1$ bulk-to-boundary propagators in AdS${}_{d-1}$/CFT${}_{d-2}$, and $\cD_p^{\mu}$ is simply an AdS weight-shifting operator \cite{Costa:2018mcg}. In particular, it is the AdS weight-shifting operator that increases spin by 1 multiplied by a bulk-to-boundary $6j$ symbol.

On the other hand, one can also perform crossing between $\cD_{z,w}^{0+\mu}$ and the OPE differential operator $\cC_{\de,1}$:
\be\label{eq:weightshiftingcrossing2}
\cC_{\de,1}^{(\de_1,0,\de_2,0)}\cD_{z,w}^{0+\mu}&=c_1\cD_{z_1,w_1}^{0-\mu}\cC_{\de,0}^{(\de_1,1,\de_2,0)}+c_2\cD_{z_1,w_1}^{+0\mu}\cC_{\de,0}^{(\de_1-1,0,\de_2,0)}+c_3\cD_{z_1,w_1}^{-0\mu}\cC_{\de,0}^{(\de_1+1,0,\de_2,0)},
\ee
where $\cC_{\de,j}^{(\de_1,j_1,\de_2,j_2)}$ is the OPE differential operator of $\cP_{\de,j}\in \cP_{\de_1,j_1}\x\cP_{\de_2,j_2}$. To obtain the coefficients $c_1,c_2$ and $c_3$, we apply~\eqref{eq:weightshiftingcrossing2} to a scalar two-point function $\<\cP_{\de}\cP_{\de}\>$ and compare the two sides of the equation. The result is given by
\be
c_1&=\frac{\de(\de+\de_1-\de_2-1)}{(d-4)(\de_1-1)(-\de_1+d-3)} \nn \\
c_2&=\frac{\de}{(\de_1-1)(\de_1-2)(-\de_1+d-3)(2\de_1-d+2)} \nn \\
c_3&=\frac{\de(-\de_1-\de_2+\de+d-3)(5-2d+\de+3\de_1-\de_2)}{2(-\de_1+d-3)(2\de_1-d+2)} \, .
\ee
Combining \eqref{eq:weightshiftingbubble}, \eqref{eq:weightshiftingcrossing1} and \eqref{eq:weightshiftingcrossing2}, we have
\be
&\cC_{\de,1}(z_1,z_2,\ptl_{z},\ptl_{w})\p{[w,z]\.[z_3,p](-2p\.z)^{-\de-1}}\nn \\
&=\frac{1}{(d-4)(\de-1)(\de-d+3)}\left(c_1\cD_p\.\cD_{z_1,w_1}^{0-}\cC_{\de,0}^{(\de_1,1,\de_2,0)}\right.\nn \\
&\quad\left.+c_2\cD_p\.\cD_{z_1,w_1}^{+0}\cC_{\de,0}^{(\de_1-1,0,\de_2,0)}+c_3\cD_p\.\cD_{z_1,w_1}^{-0}\cC_{\de,0}^{(\de_1+1,0,\de_2,0)}\right)(-2p\.z)^{-\de} \, .
\ee
Now the calculation is straightforward since $\cC_{\de,0}(-2p\.z)^{-\de}$ is simply the $j=0$ celestial block calculated in \cite{Kologlu:2019mfz}.\footnote{The celestial block $\cC_{\de,0}^{(\de_1,1,\de_2,0)}(-2p\.z)^{-\de}$ was not calculated in \cite{Kologlu:2019mfz}, but it can be easily obtained using the Casimir equation method.}
Finally, we obtain that the functions $g(\z)$ and $h(\z)$ in the $j=1$ celestial block \eqref{eq:j1celestialblock} are given by
\be\label{eq:j1celestialblockexpression_general}
g(\z)&=\z^{\frac{\de-\de_1-\de_2-1}{2}}\left(\tfrac{1}{2}(1+\tfrac{\de_2-\de_1}{\de-d+3}){}_2F_1(\tfrac{\de+\de_1-\de_2-1}{2},\tfrac{\de+\de_2-\de_1+1}{2},\de+2-\tfrac{d}{2},\z)\right. \, \nn \\
&\qquad\qquad\quad\,+\left.\tfrac{1}{2}(1+\tfrac{\de_1-\de_2}{\de-d+3}){}_2F_1(\tfrac{\de+\de_1-\de_2+1}{2},\tfrac{\de+\de_2-\de_1-1}{2},\de+2-\tfrac{d}{2},\z)\right)\, \nn \\
h(\z)&=\frac{\z^{\frac{\de-\de_1-\de_2+1}{2}}}{(1-\z)}\left(\tfrac{1}{2}(1+\tfrac{\de_2-\de_1}{\de-d+3}){}_2F_1(\tfrac{\de+\de_1-\de_2-1}{2},\tfrac{\de+\de_2-\de_1+1}{2},\de+2-\tfrac{d}{2},\z)\right. \, \nn \\
&\qquad\qquad\quad\ -\left.\tfrac{1}{2}(1+\tfrac{\de_1-\de_2}{\de-d+3}){}_2F_1(\tfrac{\de+\de_1-\de_2+1}{2},\tfrac{\de+\de_2-\de_1-1}{2},\de+2-\tfrac{d}{2},\z)\right) .
\ee
One can check that~\eqref{eq:j1celestialblockexpression_general} is indeed the solution to the Casimir differential equations of $g(\z)$ and $h(\z)$ obtained by applying the quadratic Casimir to~\eqref{eq:j1celestialblock}. For $\de_1=\de_2=\de_{\phi}$, we have $h(\z)=0$ and
\be\label{eq:j1celestialblockexpression_identical}
g(\z)&=\z^{\frac{\de-2\de_{\phi}-1}{2}}{}_2F_1\p{\frac{\de-1}{2},\frac{\de+1}{2},\de+2-\frac{d}{2},\z}\, \nn \\
&=\z^{\frac{\De-2\De_{\phi}}{2}}{}_2F_1\p{\frac{\De-2}{2},\frac{\De}{2},\De+1-\frac{d}{2},\z} .
\ee

Having solved for the celestial blocks, we can now calculate $\<\cO_{20'}(p)|\cD_{1}\wL[\cO_{\De,0}]|J(p,z_3)\>^{(a)}$ to obtain the proportionality constant in~\eqref{eq:lighttransform3ptfunctionproportionalityrelation}. For an operator $\cO$ with weights $(\De,J)$, the three-point function $\<0|\cO_{20'}\cO J|0\>^{(a)}$ has two tensor structures. With our choice of conventions in appendix~\ref{app:structureconventions}, their expressions in the embedding space for integer $J$ are given by
\be
\label{eq:O20primeJO_strcuture1}
\<0|\cO_{20'}(X_4)\cO(X_2,Z_2)J(X_3,Z_3)|0\>^{(1)}&=\frac{(-2V_{2,34})^J(-2V_{3,42})}{X_{24}^{\frac{\De+J-2}{2}}X_{23}^{\frac{\De+J+2}{2}}X_{34}^{\frac{6-\De-J}{2}}} \\
\label{eq:O20primeJO_strcuture2}
\<0|\cO_{20'}(X_4)\cO(X_2,Z_2)J(X_3,Z_3)|0\>^{(2)}&=\frac{(-2V_{2,34})^{J-1}(-2H_{23})}{X_{24}^{\frac{\De+J-2}{2}}X_{23}^{\frac{\De+J+2}{2}}X_{34}^{\frac{6-\De-J}{2}}},
\ee
where the structures $V_{i,jk}$ and $H_{ij}$ are defined in~\eqref{eq:tensorstructure_V} and~\eqref{eq:tensorstructure_H}. Note that after setting $J=0$, the first structure \eqref{eq:O20primeJO_strcuture1} is still a valid three-point function of local operators, and hence it should be annihilated by the shortening condition $\cD'_1$ (or equivalently $\cD_1\wL$). So we can just consider the second structure \eqref{eq:O20primeJO_strcuture2}. As explained in section~\ref{sec:Jcontinuation}, its analytic continuation for complex $J$ should be given by
\be
\<0|\cO_{20'}(X_4)\cO(X_2,Z_2)J(X_3,Z_3)|0\>^{(2)}_{\pm}=\mp\frac{(-2V_{2,43})^{J-1}(-2H_{23})}{X_{24}^{\frac{\De+J-2}{2}}X_{23}^{\frac{\De+J+2}{2}}X_{34}^{\frac{6-\De-J}{2}}}.
\ee
where $\mp$ is due to the $(-2V_{2,34})^{J-1}$ factor. Using the algorithm for computing light transform and Fourier transform of three-point functions described in \cite{Kologlu:2019bco} and applying the differential operator $\cD_1$, we obtain
\be\label{eq:onepointeventshape}
&\<\cO_{20'}(p)|\cD_1\wL[\cO_{\De,0}](\oo,z,w)|J(p,z_3)\>^{(2)}_{-}\nn \\
&=16\pi^4\frac{\G(\De-2)}{\G(\frac{\De}{2})^2\G(\frac{\De-2}{2})\G(\frac{4-\De}{2})} \nn \\
&\qquad\x(-2p\. z)^{-\De}(2w\.z_3z\.p-2w\.p z\.z_3)(-p^2)^{\frac{\De-2}{2}}\th(p),
\ee
where $\th(p)\equiv\th(-p^2)\th(p^0)$. Finally, combining \eqref{eq:N=4SYMeventshape}, \eqref{eq:j1celestialblock} and \eqref{eq:onepointeventshape}, we have
\be
&\<\cO_{20'}(p)|\wL[\cO_{20'}](\oo,z_1)\wL[\cO_{20'}](\oo,z_2)|J(p,z_3)\>_{\textbf{175}}=16i\pi^5\frac{[z_3,p]\.[z_1,z_2](-p^2)\th(p)}{(-2p\.z_1)^2(-2p\.z_2)^2}\cF_{JO}(\z),
\ee
where the function $\cF_{JO}(\z)$ is
\be\label{eq:N=4SYMeventshape_crossratiofunction}
\cF_{JO}(\z)=-\int_{2-i\oo}^{2+i\oo}\frac{d\De}{2\pi i}C'^{-}_2(\De,0)\frac{\G(\De-2)}{\G(\frac{\De}{2})^2\G(\frac{\De-2}{2})\G(\frac{4-\De}{2})}g_{\De}^{2,2}(\z),
\ee
and $g^{2,2}_{\De}(\z)$ is the $j=1$ celestial block \eqref{eq:j1celestialblockexpression_identical}
\be
g_{\De}^{2,2}(\z)=\z^{\frac{\De-4}{2}}{}_2F_1\p{\frac{\De-2}{2},\frac{\De}{2},\De-1,\z}.
\ee
Now the remaining task is to find the OPE data $C'^-_2(\De,J)$ at $J=0$ and calculate the event shape. In the next section, we will show that there's a superconformal Ward identity that relates $C'^-_2(\De,J)$ to the OPE data of the $\<\cO_{20'}\cO_{20'}\cO_{20'}\cO_{20'}\>$ 4-point function. Using the identity, we can derive a simple relation between $\cF_{JO}(\z)$ and the energy-energy correlator calculated in \cite{Kologlu:2019mfz}.

\subsubsection{Relation to energy-energy correlator}
Deforming the contour of the $\De$-integral in \eqref{eq:N=4SYMeventshape_crossratiofunction}, we get
\be\label{eq:N=4SYMeventshape_FJO}
\cF_{JO}(\z)=-\sum_{\De}p_{\De,J=0}\frac{\G(\De-2)}{\G(\frac{\De}{2})^2\G(\frac{\De-2}{2})\G(\frac{4-\De}{2})}g_{\De}^{2,2}(\z),
\ee
where $p_{\De,J}$ is the three-point coupling of $\<\cO_{20'}\cO_{20'}\cO_{20'}J\>$ corresponding to the structure \eqref{eq:O20primeJO_strcuture2}, analytically continued to $J=0$, and the sum is over Regge trajectories. 

We can obtain $p_{\De,J}$ from the four-point function $\<\cO_{20'}\cO_{20'}\cO_{20'}J\>$, whose expression is given in \eqref{eq:JOOO4pt}. Note that the function $\Phi(u,v)$ can be written in terms of superconformal blocks as \cite{Beem:2016wfs}
\be
\Phi(u,v)=(2\pi)^4\frac{v}{u^3}\sum_{\De,J}a_{\De,J}g_{\De+4,J}(u,v),
\ee
where $g_{\De,J}$ is the usual 4d conformal block, and $a_{\De,J}$ is the product of the three-point couplings to a given superconformal primary. Plugging this into \eqref{eq:JOOO4pt} and specializing to the configuration $x_1=0,x_3=1, x_4=\oo$, we find\footnote{To obtain this equation, we studied the small $z,\bar{z}$ limit on both sides and matched each term in the series expansion.}
\be
&\<\cO_{20'}(\oo)\cO_{20'}(0)\cO_{20'}(u,v)J(1)\>\nn \\
&=(2\pi)^4\sum_{\De,J} a_{\De,J}\left(-\frac{\De+1}{8}G_{\De+3,J+1}(u,v)+\frac{(J-1)(\De+1)}{8(J+1)}G_{\De+3,J-1}(u,v)\right.\nn \\
&\qquad\qquad\qquad\quad\ \ \, +\frac{(\De+3)(\De+4)(\De+J+4)^2}{128(\De+2)(\De+J+3)(\De+J+5)}G_{\De+5,J+1}(u,v)\nn \\
&\qquad\qquad\qquad\quad\ \ \,\left.-\frac{(J-1)(\De+3)(\De+4)(\De-J+2)^2}{128(J+1)(\De-J+3)(\De-J+1)(\De+2)}G_{\De+5,J-1}(u,v)\right),
\ee
where $G_{\De,J}$ is the conformal block of one conserved current and three scalars with dimension 2, which can be calculated using e.g.\ \cite{Costa:2011dw}. The above expression should agree with the usual conformal block decomposition
\be
\<\cO_{20'}(\oo)\cO_{20'}(0)\cO_{20'}(u,v)J(1)\>=\sum_{\De,J}p_{\De,J}G_{\De,J}(u,v).
\ee
Therefore, there is a superconformal Ward identity that relates the three-point coupling coefficients $p_{\De,J}$ and $a_{\De,J}$:
\be
p_{\De,J}&=(2\pi)^4\left(-\frac{\De-2}{8}a_{\De-3,J-1}+\frac{J(\De-2)}{8(J+2)}a_{\De-3,J+1}\right. \nn \\
&\qquad\qquad\ +\frac{(\De-2)(\De-1)(\De+J-2)^2}{128(\De-3)(\De+J-3)(\De+J-1)}a_{\De-5,J-1}\nn \\
&\qquad\qquad\ \left.-\frac{J(\De-2)(\De-1)(\De-J-4)^2}{128(J+2)(\De-J-3)(\De-J-5)(\De-3)}a_{\De-5,J+1}\right).
\ee
Setting $J=0$ in the above identity and plugging it into \eqref{eq:N=4SYMeventshape_FJO}, we have
\be
\cF_{J\cO}(\z)&=-(2\pi)^4\sum_{\De}\p{-\frac{\De-2}{8}a_{\De-3,-1}+\frac{(\De-2)^3}{128(\De-3)^2}a_{\De-5,-1}}\frac{\G(\De-2)}{\G(\frac{\De}{2})^2\G(\frac{\De-2}{2})\G(\frac{4-\De}{2})}g^{2,2}_{\De}(\z)\nn \\
&=-(2\pi)^4\sum_{\De}a_{\De-4,-1}\frac{\G(\De-2)}{8\G(\frac{\De-1}{2})^3\G(\frac{3-\De}{2})}\left(g^{2,2}_{\De-1}(\z)+\frac{(\De-1)}{4(\De-2)}g^{2,2}_{\De+1}(\z)\right).
\ee
 The sum of $j=1$ celestial blocks in the parentheses satisfies
 \be
 g^{2,2}_{\De-1}(\z)+\frac{(\De-1)}{4(\De-2)}g^{2,2}_{\De+1}(\z)=\z f_{\De}^{4,4}(\z),
 \ee
 where $f_{\De}^{4,4}(\z)$ is the $j=0$ celestial block:
 \be
 f_{\De}^{4,4}(\z)=\z^{\frac{\De-7}{2}}{}_2F_1\p{\frac{\De-1}{2},\frac{\De-1}{2},\De-1,\z}.
 \ee
This gives
\be
\cF_{J\cO}(\z)=-\frac{\z}{2}\p{\sum_{\De}a_{\De-4,-1}\frac{4\pi^4\G(\De-2)}{\G(\frac{\De-1}{2})^3\G(\frac{3-\De}{2})}f_{\De}^{4,4}(\z)}.
\ee
Note that the term in the parentheses is simply the function $\cF_{\cE}(\z)$ related to the energy-energy correlator calculated in \cite{Kologlu:2019mfz}, see (7.11) and (7.16) there. Therefore, we have
\be\label{eq:relationtoEECfromlightray}
\cF_{J\cO}(\z) =-\frac{\z}{2}\cF_{\cE}(\z),
\ee
which agrees with~\eqref{eq:relationtoEEC} from direct computation.

\section{Discussion and future directions}
\label{sec:discussion}

We have seen that a product of light-transformed local operators $\wL[\cO_1]\wL[\cO_2]$ is encoded in a nontrivial way inside the space of light-ray operators. Low transverse spin terms in the product are special linear combinations of light-ray operators with spin $J_1+J_2-1$. Higher transverse spin terms are primary descendants, obtained by acting with the special conformally-invariant differential operators $\cD_n$ on higher-$J$ light-ray operators.

The differential operators $\cD_n$ appear in the general classification of reducible generalized Verma modules described in \cite{Penedones:2015aga}. Most of the operators in this classification act on multiplets with quantum numbers below the unitarity bound. Thus, when they were first identified, it was not obvious a priori what roles they could play in physical unitary CFTs. However, light-ray operators naturally have quantum numbers that violate the unitarity bound, and indeed we have identified a role for $\cD_n$ acting on this space. It is interesting to ask whether there are similar roles in Lorentzian observables for other conformally-invariant differential operators. In addition, it would be interesting to further explore interrelationships between conformally-invariant differential operators, conformally-invariant pairings, conformally-invariant integral transforms, and weight-shifting operators.

One way to motivate the light-ray OPE is by thinking about null-integrated operators as primaries in a fictitious $d{-}2$-dimensional CFT. Now that we have a complete description of the terms in this OPE, can we push the analogy with CFT$_{d-2}$ further? For example, what are the implications of the light-ray OPE for multi-point event-shapes, such as the three-point energy correlators studied in \cite{Chen:2019bpb}? Associativity of the light-ray OPE should give rise to a nontrivial crossing equation satisfied by three-point event shapes, and it would be interesting to study this ``celestial" crossing equation using bootstrap techniques. One of the first lessons of the analytic bootstrap is that the crossing equations imply the existence of ``double-twist" operators with arbitrarily large spin, via the lightcone bootstrap \cite{Komargodski:2012ek,Fitzpatrick:2012yx}. Similar arguments for the light-ray OPE could imply the existence of terms with arbitrarily-large transverse spin $j$. It would be interesting to understand the relationship between these operators and the usual large-spin operators in the lightcone bootstrap.

To fully develop a celestial bootstrap program, one would need to understand OPEs of more general light-ray operators, such as a product of a null-integrated local operator $\wL[\cO]$ with a general light-ray operator $\mathbb{O}_{\De,J}$, or even a product of two general light-ray operators. This is an important problem for the future. A key conceptual question is: do new types of operators appear beyond the ones constructed in \cite{Kravchuk:2018htv}? For the OPE explored in this work, the answer turned out to be ``no" for rather nontrivial reasons. For more general light-ray OPEs, the answer is less clear. It is natural to conjecture, however, that any light-ray operators allowed by symmetries will appear. For example, we expect that the leading light-ray operator in the three-fold OPE of average null energy operators is $\wL[X_4]$, where $X_4$ is the lowest-twist spin-4 local operator. This claim and its implications should be testable using the results of \cite{Chen:2019bpb}.

Currently, the best available data about multi-point event shapes comes from Hofman and Maldacena's calculation of energy correlators in $\cN=4$ Super Yang Mills theory at large 't Hooft coupling \cite{Hofman:2008ar}. They computed the first few terms in the large-$\l$ expansion of a multi-point energy correlator up to order $1/\l^{3/2}$. It would be interesting to understand the structure of their result from the point of view of the light-ray OPE, and also the ``$t$-channel'' expansion of \cite{Kologlu:2019bco}.

Correlators of average null energy operators do not capture complete information about the energy distribution of a state. In particular, they are blind to the value of retarded time when excitations reach future null infinity. To probe this more refined information, it is natural to weight integrals along null infinity by non-constant functions of retarded time. In \cite{Caron-Huot:2020adz}, it was shown that meromorphic weighting functions with carefully-chosen poles can give rise to useful ``dispersive sum rules" that constrain the data of a CFT. Alternative weighting functions may have other useful applications and are worth exploring.

Finally, it would be interesting to explore the structure of the short-angle expansion of energy-energy correlators and other event shapes in non-conformal theories, and in particular how the higher-transverse spin terms arise there~\cite{Dixon:2019uzg}. When the conformal symmetry is present, the operators $\cD_n$ acting on light-ray operators inserted at spatial infinity (in which case $\cD_n$ become expressed in terms of special conformal generators) define some new discrete set of translationally-invariant detectors. As the conformal symmetry is broken, this relation between these detectors and the continuous Regge trajectories gets broken as well. We thus expect that in theories such as QCD some new discrete set of anomalous dimensions should appear in observables for which higher transverse spin is important (such as higher-point event shapes or oriented two-point event shapes).

\section*{Acknowledgements}

We thank Lance Dixon, Adam Levine, Juan Maldacena, Pier Monni, Ian Moult, and Emery Sokatchev for discussions. CHC and DSD are supported by Simons Foundation grant 488657 (Simons Collaboration on the Nonperturbative Bootstrap), a Sloan Research Fellowship, and a DOE Early Career Award under grant no.~DE-SC0019085. PK is supported by DOE grant no.~DE-SC0009988 and the Adler Family Fund at the Institute for Advanced Study. This project has received funding from the European Research Council (ERC) under the European Union's Horizon 2020 research and innovation programme (grant agreement No 787185).

\newpage

\appendix

\section{Comments on existence of light-ray operators}
\label{app:light-ray-apology}

In section~\ref{sec:light-ray-ops} we described a picture in which light-ray operators exist in any CFT and provide analytic continuation of local operators in the sense of equation~\eqref{eq:LR-local-relation}. This picture is supported, for example, by perturbative examples, where explicit expressions can be given for $\hat{\mathbb{O}}^\pm_{i,J,\l}(x,z,\bw)$ in terms of fundamental fields and~\eqref{eq:LR-local-relation} can be verified. However, this picture is not rigorously known to be valid in non-perturbative CFTs as we now review.

In~\cite{Kravchuk:2018htv} a construction was given for light-ray operators in general CFTs, which we reviewed in~\ref{sec:light-ray-ops}. The following statements need to be established before it can be claimed that the story of section~\ref{sec:light-ray-ops} is correct. 

First of all, it needs to be shown that the functions
\be
	\<\Psi|\mathbb{O}^\pm_{\De,J}(x,z)|\Phi\>
\ee
are meromorphic in $\De$ for general $J$. Furthermore, we need to prove that the positions of the poles are independent of the choice of the states $\Psi,\Phi$, as well as of the operators $\f_1,\f_2$ used to define $\mathbb{O}^\pm_{\De,J}$. Finally, we have to argue that the residues of these poles depend on $\f_1,\f_2$ only through an OPE coefficient $f_{12\cO^\dagger}$. 

All these statements are known to be true for non-negative integer $J$ (with $(-1)^J=\pm$)~\cite{Kravchuk:2018htv} but, to the best of our knowledge, no proof is known for other values. In the main text we assume that these are true, since this simplifies the statement of our results. However, even if none of the above statements hold, the results of this paper, and in particular~\eqref{eq:lightrayope_generalops}, continue to hold in the following sense. One needs to take matrix element of~\eqref{eq:lightrayope_generalops} between the states of interest, which for concreteness we take to be created by single insertions of primaries $\cO_3,\cO_4$. The right-hand side is then given by $\de$ integrals of
\be
	\<\cO_3\mathbb{O}^{\pm}_{\de+1,J,\l,(a)}\cO_4\>,
\ee
which can be expressed in terms of the function $C_{ab}(\De,J,\l)$ which is computed by the Lorentzian inversion integral~\cite{Kravchuk:2018htv} for the four-point function
\be
\<\cO_1\cO_2\cO_3\cO_4\>.
\ee
The Lorentzian inversion integral and thus $C_{ab}(\De,J,\l)$ is well-defined on the $\de$ integration contour of~\eqref{eq:lightrayope_generalops}, and so we get a rigorous interpretation of~\eqref{eq:lightrayope_generalops} without assuming any of the above facts about the light-ray operators. Moreover, the entire derivation of~\eqref{eq:lightrayope_generalops} can be carried out in this language, which is what is essentially done in~\cite{Kologlu:2019mfz} and appendix~\ref{app:fixingtheoldderivation}. 

In this language, where the matrix elements 
\be
\<\cO_3\wL[\cO_1]\wL[\cO_2]\cO_4\>
\ee
are expressed as an integral of $C_{ab}(\De,J,\l)$ along principal series in $\De$, the analysis of the small angle limit between the detectors can be carried out in the usual way, by analytically continuing $C_{ab}(\De,J,\l)$ away from the principal series and deforming the integration contour to the right, picking up the singularities that one encounters on the way.

\section{Conventions for two- and three-point structures}
\label{app:structureconventions}

We follow the same conventions for two- and three-point structures of traceless-symmetric operators as in~\cite{Kravchuk:2018htv}. In particular, we define
\be
	\<\cO(x_1,z_1)\cO(x_2,z_2)\>=\frac{(-2z_1\.I(x_{12})\.z_2)^J}{x_{12}^{2\De}}
\ee
where
\be
	I^\mu{}_\nu(x)=\de^\mu{}_\nu-2\frac{x^\mu x_\nu}{x^2}.
\ee
For three-point structures we define
\be\label{eq:ffOconventionPoincare}
	\<\f_1(x_1)\f_2(x_2)\cO(x_3,z)\>=\frac{(2z\.x_{23}\, x_{13}^2-2z\.x_{13}\,x_{23}^2)^J}
	{x_{12}^{\De_1+\De_2-\De+J}x_{13}^{\De_1+\De-\De_2+J}x_{23}^{\De_2+\De-\De_1+J}}.
\ee
In terms of embedding-space formalism these become
\be
	\<\cO(X_1,Z_1)\cO(X_2,Z_2)\>&=\frac{(-2H_{12})^J}{X_{12}^{\De+J}},\label{eq:OOconvetionEF}\\
	\label{eq:ffOconvetionEF}
	\<\f_1(X_1)\f_2(X_2)\cO(X_3,Z_3)\>&=\frac{(-2V_{3,12})^J}
	{X_{12}^{\frac{\De_1+\De_2-\De-J}{2}}
	X_{13}^{\frac{\De_1+\De-\De_2+J}{2}}
	X_{23}^{\frac{\De_2+\De-\De_1+J}{2}}},
\ee
where as usual
\be
X_{ij}&\equiv -2X_i\.X_j\\
V_{i,jk}&\equiv \frac{Z_i\.X_j X_i\.X_k-Z_i\.X_k X_i\.X_j}{X_j\.X_k}, \\
H_{ij}&\equiv -2(Z_i\.Z_j X_i\.X_j-Z_i\.X_jZ_j\.X_i).
\ee
Recall that we project from embedding space using
\be
	(X^+,X^-,X^\mu)=(1,x^2,x^\mu),\quad (Z^+,Z^-,Z^\mu)=(0,2x\.z,z^\mu),
\ee
and the embedding space metric is
\be
	X^2=-X^+X^-+X^\mu X_\mu.
\ee

We use these conventions both for $d$-dimensional structures as well as for $(d-2)$-dimensional celestial structures. For example, replacing $X\to z, Z\to w, \De\to \de$ and $J\to j$ we find
\be
	\<\f_1(z_1)\f_2(z_2)\cO(z,w)\>&=
	\frac{(-2\frac{w\.z_1 z\.z_2-w\.z_2 z\.z_1}{z_1\.z_2})^j}
	{(-2z_1\.z_2)^{\frac{\de_1+\de_2-\de-j}{2}}
	(-2z_1\.z)^{\frac{\de_1+\de-\de_2+j}{2}}
	(-2z_2\.z)^{\frac{\de_2+\de-\de_1+j}{2}}}\nn\\
&=	\frac{(4w\.z_1 z\.z_2-4w\.z_2 z\.z_1)^j}
{(-2z_1\.z_2)^{\frac{\de_1+\de_2-\de+j}{2}}
	(-2z_1\.z)^{\frac{\de_1+\de-\de_2+j}{2}}
	(-2z_2\.z)^{\frac{\de_2+\de-\de_1+j}{2}}}\nn\\
&=\frac{(-4)^j(w\.z_2 z\.z_1-w\.z_1 z\.z_2)^j}
{(-2z_1\.z_2)^{\frac{\de_1+\de_2-\de+j}{2}}
	(-2z_1\.z)^{\frac{\de_1+\de-\de_2+j}{2}}
	(-2z_2\.z)^{\frac{\de_2+\de-\de_1+j}{2}}},
\ee
in agreement with~\eqref{eq:standard3pt_celestial}.

Sometimes we need the standard tensor structures for continuous spin. We define them for the Wightman functions as
\be
\<0|\f_1(x_1)\cO(x_3,z)\f_2(x_2)|0\>=\frac{(2z\.x_{23}\, x_{13}^2-2z\.x_{13}\,x_{23}^2)^J}
{x_{12}^{\De_1+\De_2-\De+J}x_{13}^{\De_1+\De-\De_2+J}x_{23}^{\De_2+\De-\De_1+J}}.
\ee
For non-integer $J$, this is defined to be positive for $x_{ij}^2>0$ and $z\.x_{23}\, x_{13}^2-z\.x_{13}\,x_{23}^2>0$. The values in other configurations are obtained by analytic continuation assuming standard analyticity properties of Wightman functions.

\section{A Lorentzian formula for the light-ray kernel}
\label{app:lorentziankernel}

In this appendix, we prove the Lorentzian formula for the light-ray kernel (\ref{eq:lorentzianformulaforlightraykernel}). We follow the notation of \cite{Kravchuk:2018htv}. Our starting point is the generalized Lorentzian inversion formula
\be
C_{ab}^t(\De,J,\l) &= \frac{-1}{2\pi i}  \int_{\substack{3>4\\1>2}} \frac{d^d x_1\cdots d^d x_4}{\vol(\tl \SO(d,2))} \tsym_2 \tsym_4 \<\O|[\cO_4, \cO_1] [\cO_2,\cO_3]|\O\>  G_\cO,
\label{eq:notsoobvious}
\\
 G_\cO &=  \frac{\p{\tsym_2\<0|\cO_2 \wL[\cO^\dagger]\cO_1|0\>^{(a)}}^{-1}\p{\tsym_4\<0|\cO_4 \wL[\cO]  \cO_3|0\>^{(b)}}^{-1}}{\<\wL[\cO]\wL[\cO^\dagger]\>^{-1}}
.
\label{eq:schematicblock}
\ee
We have written the formula in slightly different conventions relative to \cite{Kravchuk:2018htv}. Firstly, we made the change of variables $x_2\to \tsym_2 x_2$ and $x_4\to \tsym_4 x_4$, so that the causal relationships become $3>4$ and $1>2$ with all other pairs spacelike separated. This choice makes it simpler to apply the Lorentzian two and three-point pairings defined in \cite{Kravchuk:2018htv}. In addition, we only wrote the $t$-channel term in the inversion formula. The treatment of the $u$-channel term is analogous. In our notation, $\cO$ is a representation with quantum numbers $(\De,J,\l)$.

The object $G_\cO$ is a conformal block with internal quantum numbers $(J+d-1,\De-d+1,\l)$. In (\ref{eq:schematicblock}), we have written it in schematic notation, where the three-point structures in the numerator should be glued using the two-point structure in the denominator. In more precise notation, $G_\cO$ is defined by
\be
\label{eq:morepreciseG}
G_\cO = \int_{1>x>2} d^d x D^{d-2} z A_{(a)}(x_1,x_2,x,z) \p{\tsym_4\<0|\cO_4 \wL[\cO]  \cO_3|0\>^{(b)}}^{-1}
\ee
where the kernel $A_{(a)}$ satisfies
\be
\label{eq:morepreciseGagain}
\int_{1>x>2} d^d x D^{d-2} z A_{(a)}(x_1,x_2,x,z) \<\wL[\cO](x,z)\wL[\cO^\dagger](x',z')\>^{-1} &= \p{\tsym_2\<0|\cO_2 \wL[\cO^\dagger](x',z')\cO_1|0\>^{(a)}}^{-1}
\ee
Recall that dual structures $\<\cdots\>^{-1}$ are defined using the Lorentzian two and three-point pairings defined in \cite{Kravchuk:2018htv}.
In \cite{Kravchuk:2018htv}, the block $G_\cO$ was defined by specifying its behavior in the OPE limit $x_1\to x_2$. The above definition in terms of the integral kernel $A_{(a)}$ is equivalent and more convenient for our purposes.

The light-ray kernel should satisfy
\be
\label{eq:kernelshouldsatisfy}
\int_{\substack{1>x>2^{-}}}d^dx_1d^dx_2K^t_{\De,J.\l(a)}(x_1,x_2;x,z)\<\Omega|\cO_4\cO_1\cO_2\cO_3|\Omega\>
&=-C_{ab}^t(\De,J,\l) \<0|\cO_4 \wL[\cO] \cO_3|0\>^{(b)}.
\ee
In the expression (\ref{eq:kernelshouldsatisfy}), we can replace $\<\Omega|\cO_4\cO_1\cO_2\cO_3|\Omega\>$ with a double commutator. The reason is that the kernel $K^t_{\De,J,\l(a)}$ factors through null integrals of $x_1$ and $x_2$, which annihilate the past and future vacuum, see \cite{Kravchuk:2018htv} for details.  Furthermore, let us make the change of variables $x_2\to \tsym_2 x_2$ and $x_4\to \tsym_4 x_4$, so that we have
\be
\int_{\substack{1>x>2}}d^dx_1d^dx_2 (\tsym_2 K^t_{\De,J.\l(a)}) \tsym_2 \tsym_4 \<\Omega|[\cO_4,\cO_1][\cO_2,\cO_3]|\Omega\> =-C_{ab}^t(\De,J,\l) \tsym_4\<0|\cO_4 \wL[\cO] \cO_3|0\>^{(b)}.
\ee
For brevity, here and in the following, we assume the arguments of $\cT_2 K^t_{\De,J.\l(a)}$ are $(x_1,x_2;x,z)$. The arguments of $\wL[\cO]$ will always be $(x,z)$, and we use the notation $\wL[\cO'^\dag]$ to indicate $\wL[\cO^\dag](x',z')$.

Pairing both sides with the dual of the right-hand structure, we obtain
\be
-C_{ab}^t(\De,J,\l) &=\int_{\substack{1>x>2 \\ 3>4}} \frac{d^d x_1 d^d x_2 d^d x_3 d^d x_4 d^d x D^{d-2} z}{\vol \tl \SO(d,2)}
(\tsym_2 K^t_{\De,J.\l(a)})  \p{\tsym_4\<0|\cO_4 \wL[\cO] \cO_3|0\>^{(b)}}^{-1} \nn\\
&\qquad \qquad \x \tsym_2 \tsym_4 \<\Omega|[\cO_4,\cO_1][\cO_2,\cO_3]|\Omega\>.
\ee
Comparing with (\ref{eq:notsoobvious}),  (\ref{eq:morepreciseG}), and (\ref{eq:morepreciseGagain}), we conclude
\be
\int_{1>x>2} d^d x D^{d-2} z (\tsym_2 K^t_{\De,J.\l(a)}) \<\wL[\cO]\wL[\cO'^\dagger]\>^{-1}&= \frac 1 {2\pi i}\p{\tsym_2\<0|\cO_2 \wL[\cO'^\dagger]\cO_1|0\>^{(a)}}^{-1}.
\ee
This is essentially the desired result, written in terms of dual structures. To put it in a more conventional form, we must apply the two- and three-point pairings that appear in the definition of the dual structures.
Pairing both sides with $\tsym_2\<0|\cO_2 \wL[\cO^\dagger](x',z')\cO_1|0\>^{(b)}$, we find
\be
&\int_{\substack{1>x>2 \\ x'\approx 1,2}}\frac{ d^d x_1 d^d x_2 d^d x D^{d-2} z d^d x' D^{d-2} z'}{\vol \tl \SO(d,2)} (\tsym_2 K^t_{\De,J.\l(a)}) \p{\tsym_2\<0|\cO_2 \wL[\cO'^\dagger]\cO_1|0\>^{(b)}}  \<\wL[\cO]\wL[\cO'^\dagger]\>^{-1} \nn\\
&= \frac 1 {2\pi i}\de_{(a)}^{(b)}
\ee
Comparing with the definition of the Lorentzian two-point pairing in \cite{Kravchuk:2018htv}, finally gives
\be
\label{eq:morepreciseGagainagainagain}
\int_{\substack{1>x>2}} \frac{d^d x_1 d^d x_2}{(\vol \SO(1,1))^2} (\tsym_2 K^t_{\De,J.\l(a)}) \tsym_2\<0|\cO_2 \wL[\cO'^\dagger]\cO_1|0\>^{(b)} = \frac 1 {2\pi i}\de_{(a)}^{(b)} \<\wL[\cO]\wL[\cO'^\dagger]\>,
\ee
where the equality is valid if $x'\approx 1,2$. Finally, after changing variables $x_2\to \tsym_2^{-1} x_2$, we obtain (\ref{eq:lorentzianformulaforlightraykernel}).

\section{An alternative derivation for the light-ray OPE formula}
\label{app:fixingtheoldderivation}

In this appendix, we give another derivation for the light-ray OPE formula~\eqref{eq:lightrayope_generalops}. We will first review the derivation for the low transverse spin terms given in \cite{Kologlu:2019mfz}. Then we will derive the ``higher transverse spin" terms by generalizing the derivation in \cite{Kologlu:2019mfz}.

\subsection{Review: derivation of~\cite{Kologlu:2019mfz}}
\label{sec:lowj_deriv}
We first briefly review the proof given in \cite{Kologlu:2019mfz}. We are interested in an expansion for
\be
\wL[\cO_1](x,z_1)\wL[\cO_2](x,z_2).
\ee
For simplicity, we assume $\cO_1,\cO_2$ are traceless symmetric tensors. Generalization to arbitrary representations will become straightforward after finishing the proof. We will study the matrix element
\be
W(z_1,z_2)=\<\O|\cO_4\wL[\cO_1](x,z_1)\wL[\cO_2](x,z_2)\cO_3|\O\>,
\ee
where $\cO_3,\cO_4$ are some local primary operators. Then we can apply harmonic analysis on the celestial sphere to expand $W(z_1,z_2)$ into partial waves. The result is given by~\eqref{eq:W CPW final} and~\eqref{eq:scalarinverse}. The object $W_{\de,j}(z_1,z_2)$ can be further written as
\be\label{eq:Wexpr}
	W_{\de,j}(x,z)&=\a_{\de,j}\int D^{d-2}z_1D^{d-2}z_2\<\tl{\cP}^\dag_{\de_1}(z_1)\tl{\cP}^\dag_{\de_2}(z_2)\cP_{\de,j}(z)\>\<\O|\cO_4\wL[\cO_1](x,z_1)\wL[\cO_2](x,z_2)\cO_3|\O\>\, \nn \\
	&=\int d^dx_1d^dx_2D^{d-2}z_1D^{d-2}z_2~\cL_{\de,j}(x_1,z_1,x_2,z_2;x,z)\<\O|\cO_4\cO_1(x_1,z_1)\cO_2(x_2,z_2)\cO_3|\O\>,
\ee
and the kernel $\cL_{\de,j}$ is
\be\label{eq:Lkerneldef}
&\cL_{\de,j}(x_1,z_1,x_2,z_2;x,z)\, \nn \\
&=\a_{\de,j}\< \tl{\cP}_{\de_1}^\dag(z_1)\tl{\cP}_{\de_2}^\dag(z_2)\cP_{\de,j}(z)\>\, \nn \\
&\quad\times \int_{-\oo}^{\oo}d\a_1d\a_2(-\a_1)^{-\de_1-J_1-1}(-\a_2)^{-\de_2-J_2-1}\de^{(d)}\left(x-z_1/\a_1-x_1\right)\de^{(d)}\left(x-z_2/\a_2-x_2\right).
\ee
Note that we have suppressed the Lorentz indices carried by $W_{\de,j}$ and $\cL_{\de,j}$. They are contracted with the indices carried by $\cC_{\de,j}$ in~\eqref{eq:W CPW final}.

Using \eqref{eq:Wexpr} and the fact that $\wL[\cO_i]$ annihilates the vacuum, we have
\be\label{eq:Wmatrix}
W_{\de,j}(x,z)&=\int d^dx_1d^dx_2D^{d-2}z_1D^{d-2}z_2\cL_{\de,j}(x_1,z_1,x_2,z_2;x,z)\th(4>1)\th(2>3)\, \nn \\
&\qquad\qquad\qquad\qquad\qquad\qquad\x\<\O|[\cO_4,\cO_1(x_1,z_1)][\cO_2(x_2,z_2),\cO_3]|\O\>.
\ee
By conformal invariance, we also have
\be\label{eq:WArelation}
W_{\de,j}(x,z)=A_b(\de,j)\<0|\cO_4\wL[\cO](x,z)\cO_3|0\>^{(b)}_+.
\ee
where $\cO$ has quantum numbers $(\De,J,\l)=(\de+1,J_1+J_2-1,j)$, $(b)$ is the tensor structure index, and $\<0|\cO_4\wL[\cO](x,z)\cO_3|0\>^{(b)}_+$ is the continuous-spin structure analytically continued from even spin.\footnote{The analytic continuation of $\<0|\cO_4\wL[\cO](x,z)\cO_3|0\>^{(b)}$ to complex spin has to be done separately for even and odd spin due to our convention of the three-point function, so we have to make a choice in the right-hand side of~\eqref{eq:WArelation}. However, this choice doesn't affect our final result. It will only change how we relate $C^{\pm}(\De,J)$ and $\mathbb{O}^{\pm}(\De,J)$ in the final step of our derivation. With our choice in~\eqref{eq:WArelation}, we need to identify $C^+(\De,J) \to \mathbb{O}_{\De,J}^+$ and $C^-(\De,J) \to -\mathbb{O}_{\De,J}^-$.}
In order to proceed, we need to introduce conformally-invariant pairings for two-point and three-point continuous spin structures in the Lorentzian signature. The pairings are described in detail in appendix E of \cite{Kravchuk:2018htv}. For a two-point structure of $\cO$ in representation $(\De,J,\l)$, it can be paired with a two-point structure of $\cO^S$ in representation $(d-\De,2-d-J,\l)$. The two-point pairing is defined by
\be
&\frac{\p{\<\cO\cO^{\dag}\>,\<\cO^S\cO^{S\dag}\>}_L}{\vol(\SO(1,1))^2} \nn \\
&\equiv \int_{x_1\approx x_2}\frac{d^dx_1d^dx_2D^{d-2}z_1D^{d-2}z_2}{\vol(\tl{\SO}(d,2))}\<\cO^a(x_1,z_1)\cO^{b\dag}(x_2,z_2)\>\<\cO_b^S(x_2,z_2)\cO_a^{S\dag}(x_1,z_1)\> \nn \\
&=\frac{\<\cO^a(0,z_1)\cO^{b\dag}(\oo,z_2)\>\<\cO_b^S(\oo,z_2)\cO_a^{S\dag}(0,z_1)\>}{2^{2d-2}\vol(\SO(d-2))}\frac{1}{(-2z_1\.z_2)^{2-d}},
\ee
where in the last line we use $\tl{\SO}(d,2)$ transformations to gauge-fix to $x_1=0,x_2=\oo$, and $a,b$ are the indices carried by the representation $\l$. The three-point pairing is defined by
\be
&\p{\<\cO_1\cO_2\cO\>,\<\tl{\cO}_1^{\dag}\tl{\cO}_2^{\dag}\cO^{S\dag}\>}_L \nn \\
&\equiv \int_{\substack{2<1 \\ x\approx 1,2}}\frac{d^dx_1d^dx_2d^dxD^{d-2}z}{\vol(\tl{\SO}(d,2))}\<\cO_1(x_1)\cO_2(x_2)\cO(x,z)\>\<\tl{\cO}_1^{\dag}(x_1)\tl{\cO}_2^{\dag}(x_2)\cO^{S\dag}(x,z)\> \nn \\
&=\frac{1}{2^{2d-2}\vol(\SO(d-2))}\frac{\<\cO_1(e^0)\cO_2(0)\cO(\oo,z)\>\<\tl{\cO}_1^{\dag}(e^0)\tl{\cO}_2^{\dag}(0)\cO^{S\dag}(\oo,z)\>}{(-2z\.e^0)^{2-d}}.
\ee
Similarly, in the last line, we gauge-fixed $x_1=e^0,x_2=0,x=\oo$.

We can then obtain $A_b(\de,j)$ by taking a Lorentzian three-point pairing of both sides with a dual structure
\be\label{eq:Aexpr}
A_b(\de,j) 
&= \p{\p{\cT_4\<0|\cO_4 \wL[\cO](x,z)\cO_3|0\>^{(b)}_+}^{-1},\cT_4W_{\de,j}(x,z)}_L \nn\\
&= \int_{\substack{4>1\\ 2>3}} \frac{d^d x_1 d^d x_2 d^d x_3 d^d x_4 D^{d-2} z_1 D^{d-2} z_2}{\vol\tl\SO(d,2)}  \<\Omega|\big[\cO_4,\cO_1(x_1,z_1)\big]\big[\cO_2(x_2,z_2),\cO_3\big] |\Omega\>\nn\\
&\qquad \x \cT_2^{-1} \cT_4^{-1} \Bigg[\int  d^d x D^{d-2}z \p{\cT_4\<0|\cO_4 \wL[\cO](x,z)\cO_3|0\>^{(b)}_+}^{-1}\nn\\
&\qquad\qquad\qquad\qquad\qquad \x (\cT_2 \cL_{\de,j})(x_1,z_1,x_2,z_2;x,z)\th(4^+>1)\th(2^+>3)\Bigg],
\ee
where $\cT_2,\cT_4$ translate the points $x_2,x_4$ to the next Poincar\'{e} patch on the Lorentzian cylinder. They are introduced so that the causality configuration in the three-point Lorentzian pairing is satisfied. The structure $\p{\cT_4\<0|\cO_4 \wL[\cO](x,z)\cO_3|0\>^{(b)}_+}^{-1}$ is defined by
\be
\p{\p{\cT_4\<0|\cO_4 \wL[\cO](x,z)\cO_3|0\>^{(b)}_+}^{-1},\cT_4\<0|\cO_4 \wL[\cO](x,z)\cO_3|0\>^{(d)}_+}_L=\de_{(b)}^{(d)}.
\ee
The derivation up to this point is true for all transverse spin $j$. In the rest of this section we first finish the derivation assuming $j\leq j_{\mathrm{max}}$, and explain what goes wrong when $j>j_{\mathrm{max}}$, where $j_{\mathrm{max}}$ is the maximal allowed transverse spin in the $\cO_1\x\cO_2$ OPE. We then give the derivation for $j>j_{\mathrm{max}}$ in section~\ref{sec:highj_deriv}. Note that since we assume $\cO_1,\cO_2$ are traceless symmetric tensors, in what follows we will simply use $j_{\mathrm{max}}=J_1+J_2$.

In \eqref{eq:Aexpr}, the term in the bracket is a conformally-invariant four-point structure, and is an eigenfunction of the quadratic Casimir acting on 1,2 (or 3,4). Therefore it is a linear combination of conformal blocks, and we can study it by taking the OPE limit. In the OPE limit $x_3,x_4\to x'$, the 34 three-point structure should be given by a linear operator $B_{34\cO}$ acting on a two-point function:\footnote{For continuous spin, the operator $B_{34\cO}$ should be an integral operator. See appendix H of \cite{Kravchuk:2018htv}.}
\be\label{eq:34OPE_lowj}
&\p{\cT_4\<0|\cO_4 \wL[\cO](x,z)\cO_3|0\>^{(b)}_+}^{-1}\nn \\
&=B_{34\cO}(x_3,x_4,\ptl_{x'},\ptl_{z'})\<\cO^F(x',z')\cO^{F\dag}(x,z)\>,
\ee
where $\cO^F$ has quantum numbers $(\De_F,J_F,j_F)=(J+d-1,\De-d+1,j)$, where $J=J_1+J_2-1$ and $\De=\de+1$. Plugging this into \eqref{eq:Aexpr}, one can then show that the bracketed term is the conformal block that appears in the Lorentzian inversion formula, and therefore we can relate $A_b(\de,j)$ to $C^{\pm}(\de+1,J_1+J_2-1,j)$. Using \eqref{eq:WArelation} and the relation between light-ray operators $\mathbb{O}^{\pm}_{\De,J}$ and $C^{\pm}(\De,J)$, we obtain that
\be\label{eq:W_lightray_lowj}
&\wL[\cO_1](x,z_1)\wL[\cO_2](x,z_2) \nn \\
&=\sum_{j\leq J_1+J_2}\int_{\frac{d-2}{2}-i\oo}^{\frac{d-2}{2}+i\oo}\frac{d\de}{2\pi i}\cC_{\de,j}^{(a)}(z_1,z_2,\ptl_z)\p{\mathbb{O}^+_{\de+1,J_1+J_2-1,j(a)}(x,z)-\mathbb{O}^-_{\de+1,J_1+J_2-1,j(a)}(x,z)} \nn \\
&\qquad+\textrm{higher transverse spin},
\ee
where the differential operator $\cC_{\de,j}^{(a)}(z_1,z_2,\ptl_z)$ can be obtained from a celestial map formula
\be\label{eq:celestialmap_lowj}
&\cC_{\de,j}^{(a)}(z_1,z_2,\ptl_z)\<\wL[\cO](\oo,z_2)\wL[\cO^\dag](0,z_1)\>=\frac{\<0|\wL^{+}[\cO_2](\oo,z_2)\wL[\cO^\dag](0,z)\wL^-[\cO_1](\oo,z_1)|0\>^{(a)}_+}{\vol\SO(1,1)}.
\ee
This result agrees with the first sum in~\eqref{eq:lightrayope_generalops}.

When $j>J_1+J_2$, the invalid step in this derivation is \eqref{eq:34OPE_lowj}. The reason is that the linear operator $B_{34\cO}$ becomes divergent when $j>J_1+J_2$. To see this, consider a conformal block
\be
\cC_{34\cO}(x_3,x_4,\ptl_{x'})\<\cO_1(x_1)\cO_2(x_2)\cO(x')\>.
\ee
It has been shown that this conformal block has simple poles in $\De$, the scaling dimension of $\cO$, due to null descendant states~\cite{Penedones:2015aga,Erramilli:2019aa}.\footnote{In even $d$, some of the poles can become double poles when more than one simple poles are at the same position. However, one can explicitly check that for the case we are discussing (type $\mathrm{II}$ poles with $k=2$), the poles don't overlap with other poles and hence remain simple poles in even $d \geq 4$.} Furthermore, the poles come from the differential operator $\cC_{34\cO}$ and have the form
\be
\cC_{34\cO}\sim\frac{\cN}{\De-\De_*}\cC_{34\cO'}\cD,
\ee
where $\cN$ is some coefficient, $\cO'$ is a primary descendant of $\cO$, $\cD$ is a differential operator such that $\cO'=\cD\cO$ (at $\De=\De_*$), and $\cC_{34\cO'}$ is the OPE operator for $\cO'\in\cO_3\x\cO_4$.
The possible pole positions $\De_*$ are classified in \cite{Penedones:2015aga} using representation theory of the conformal group. For us, the relevant cases are what are called type I and type II poles in \cite{Penedones:2015aga}. If the exchanged operator has representation $(l_1,l_2,\dots,l_N)$, where $l_k$ is the number of boxes of the k-th row of the Young diagram, then the positions of the type I and type II poles are
\be
\De_{\mathrm{I}_k,n}^*&=k-l_k-n \qquad &&(n=1,2,\dots,l_{k-1}-l_k)\, \nn \\
\De_{\mathrm{II}_k,n}^*&=d+l_k-k-n && (n=1,2,\dots,l_k-l_{k+1}),
\ee
and we call $\cD$ for the type I and type II case $\cD_{\mathrm{I}_k,n}$ and $\cD_{\mathrm{II}_k,n}$ respectively.\footnote{$\cD_{\mathrm{I}_2,n}$ is the differential operator $\cD_n$ we use extensively in the main text.}
In \eqref{eq:34OPE_lowj}, the operator $\cO^F$ has scaling dimension $\De_F=J+d-1=d+J_1+J_2-2$. Therefore, if $j=J_1+J_2+n$ for some $n\geq1$, then $\De_{\mathrm{II}_{k=2},n}^*=d+j-2-n=\De_F$. This implies that the linear operator $B_{34\cO}$ is divergent for all $j>J_1+J_2$. To fix \eqref{eq:34OPE_lowj}, we can separate the pole part and the finite part of $B_{34\cO}$ near $\De_F=\De_{\mathrm{II}_2,n}^*$:
\be
B_{34\cO}=\frac{1}{\De_F-\De_{\mathrm{II}_2,n}^*}\cC_{34\cO'}\cD_{\mathrm{II}_{2},n}+B^{\mathrm{finite}}_{34\cO},
\ee
where $n=j-J_1-J_2$, and $\cC_{34\cO'}$ is a new linear operator proportional to the OPE operator of $\cD_{\mathrm{II}_{2},n}\cO^F\in \cO_3\x\cO_4$. Plugging the above expression for $B_{34\cO}$ into \eqref{eq:34OPE_lowj}, we obtain
\be\label{eq:34OPE_highj_0}
&\p{\cT_4\<0|\cO_4 \wL[\cO](x,z)\cO_3|0\>^{(b)}_+}^{-1}\nn \\
&=B_{34\cO}^{\mathrm{finite}}(x_3,x_4,\ptl_{x'},\ptl_{z'})\<\cO^F(x',z')\cO^{F\dag}(x,z)\>\nn \\
&\quad+\cC_{34\cO'}(x_3,x_4,\ptl_{x'},\ptl_{z'})\<\cD_{\mathrm{II}_{2},n}\cO^F(x',z')\cO^{F\dag}(x,z)\>_{\mathrm{lim}},
\ee
where
\be
\<\cD_{\mathrm{II}_{2},n}\cO^F(x',z')\cO^{F\dag}(x,z)\>_{\mathrm{lim}}=\lim_{\De_F\to\De_{\mathrm{II}_2,n}^*}\frac{\<\cD_{\mathrm{II}_{2},n}\cO^F(x',z')\cO^{F\dag}(x,z)\>}{\De_F-\De_{\mathrm{II}_2,n}^*}.
\ee
Note that $\<\cD_{\mathrm{II}_{2},n}\cO^F(x',z')\cO^{F\dag}(x,z)\>\sim O(\De_F-\De_{\mathrm{II}_2,n}^*)$ because when $\De_F=\De_{\mathrm{II}_2,n}^*$ it is a two-point function between two primaries with different scaling dimensions and therefore is zero. So the above limit should be finite. This structure is not conformally-invariant, since it doesn't vanish under the special conformal transformation $K_{\mu}$. However, the result we get by integrating it against $\cT_2\cL_{\de,j}$ is still a conformally-invariant three-point structure. To see this, note that $[K_\mu,\cD_{\mathrm{II}_{2},n}]=O(\De_2-\De_{\mathrm{II}_2,n}^*)$ since $\cD_{\mathrm{II}_{2},n}$ is conformally-invariant at $\De_{\mathrm{II}_2,n}^*$. So, we can define
\be
\cD_{\mu}\equiv \lim_{\De_F\to\De_{\mathrm{II}_2,n}^*}\frac{[K_{\mu},\cD_{\mathrm{II}_{2},n}]}{\De_F-\De_{\mathrm{II}_2,n}^*}.
\ee
Then, we have
\be\label{eq:limstruct_specialconformaltransformation}
&K_{\mu}\int d^dxD^{d-2}z \<\cD_{\mathrm{II}_{2},n}\cO^F(x',z')\cO^{F\dag}(x,z)\>_{\mathrm{lim}}\cT_2\cL_{\de,j}(x_1,z_1,x_2,z_2,x,z) \nn \\
&=\int d^dxD^{d-2}z\<\cD_{\mu}\cO^F(x',z')\cO^{F\dag}(x,z)\>\cT_2\cL_{\de,j}(x_1,z_1,x_2,z_2,x,z).
\ee
This is simply a derivative of the Lorentzian shadow transform of $\cT_2\cL_{\de,j}$. In~\cite{Kologlu:2019mfz} it has been shown that the Lorentzian shadow transform of $\cT_2\cL_{\de,j}$ vanishes for $j>J_1+J_2$, and therefore~\eqref{eq:limstruct_specialconformaltransformation} should vanish as well.

The operator $\cD_{\mathrm{II}_{2},n}\cO^F$ has quantum numbers $(J+d-1+n,\De-d+1,j-n)$. Thus, the appearance of $\cC_{34\cO'}$ in~\eqref{eq:34OPE_highj_0} suggests that for $j=J_1+J_2+n$, $A_b(\de,j)$ should come from exchanged operators with quantum numbers $(\de+1,J_1+J_2-1+n,J_1+J_2)$. As we will see briefly, this is indeed the case.

\subsection{Relation between $\cD_{\mathrm{I}_{2},n}$ and $\cD_{\mathrm{II}_{2},n}$}
We now describe an interesting relation between the differential operators $\cD_{\mathrm{I}_2,n}$ and $\cD_{\mathrm{II}_2,n}$. This relation will be used later in the derivation for the $j>J_1+J_2$ case. First, note that $\cD_{\mathrm{I}_2,n}$ and $\cD_{\mathrm{II}_2,n}$ change the quantum numbers in the following way:
\be
\cD_{\mathrm{I}_2,n}&: (2-j,l_1,j-n) \to (2-j+n,l_1,j) \nn \\
\cD_{\mathrm{II}_2,n}&: (d+j-n-2,2-d-l_1,j) \to (d+j-2,2-d-l_1,j-n),
\ee
and their explicit expressions can be chosen to be
\footnote{After replacing $j-n\to j$ and $l_1\to 1-\De$, the definition of $\cD_{\mathrm{I}_2,n}$ agrees with~\eqref{eq:Dndef}.}
\be
\cD_{\mathrm{I}_2,n}&=\frac{(-1)^n\G(j-n-1-l_1)}{\G(j-1-l_1)\G(n+1)}\p{\ptl_x\.\cD_{z,w}^{0+}}^n\nn \\
\cD_{\mathrm{II}_2,n}&=\frac{(-1)^n\G(j-n-1-l_1)}{\G(j-1-l_1)\G(n+1)}\p{\ptl_x\.\cD_{z,w}^{0-}}^n,
\ee
where $\cD_{z,w}^{0+}$ and $\cD_{z,w}^{0-}$ are weight-shifting operators that increase and decrease the transverse spin, respectively~\cite{Karateev:2017jgd}. As shown in \cite{Karateev:2018oml}, in a pairing between operators one can always integrate weight-shifting operators by parts. In particular, we have
\be
(\cD^{0-}_{z,w}|_{J,j})^*=-2j(h-2+j)\cD^{0+}_{z,w}|_{2-d-J,j-1},
\ee
where $h=\frac{d-2}{2}$, and $\cD|_{J,j}$ indicates that $\cD$ acts on a multiplet with usual spin $J$ and transverse spin $j$. Using this relation, one can show that
\be\label{eq:12adjoint}
(\cD_{\mathrm{II}_{2},n}|_{d-\De-n,2-d-l_1,j})^*=N_{n,j}\cD_{\mathrm{I}_{2},n}|_{\De,l_1,j-n},
\ee
where
\be
N_{n,j}=2^n\frac{\G(j+1)\G(h-1+j)}{\G(j-n+1)\G(h-1+j-n)}.
\ee
Note that this relation holds for general $\De$, but $\cD_{\mathrm{I}_{2},n}$ and $\cD_{\mathrm{II}_{2},n}$ are conformally-invariant only when $\De=2-j$. More explicitly, for an operator $\cO_1$ with quantum numbers $(\De,l_1,j-n)$ and $\cO_2$ with quantum numbers $(d-\De-n,2-d-l_1,j)$, we have
\be
\int d^dxD^{d-2}z~\cO_1(x,z)(\cD_{\mathrm{II}_2,n}\cO_2)(x,z)=N_{n,j}\int d^dxD^{d-2}z~(\cD_{\mathrm{I}_2,n}\cO_1)(x,z)\cO_2(x,z).
\ee
In other words, $\cD_{\mathrm{II}_2,n}$ and $N_{n,j}\cD_{\mathrm{I}_2,n}$ are adjoint to each other.

We end this section by deriving two relations that will be used later in the derivation for the higher-transverse spin terms. First, consider operators $\cO_1$ with quantum numbers $(\De_1,l_1,j-n)$ and $\cO_2$ with quantum numbers $(\De_2,2-d-l_1,j)$, where $\De_1+\De_2=d$. Then by~\eqref{eq:12adjoint}, we have for general $\De_1,\De_2$
\be
&\int_{x\approx x'} d^dxd^dx'D^{d-2}zD^{d-2}z'\<\cD_{\mathrm{II}_2,n}\cO_2\cD_{\mathrm{II}_2,n}\cO'^{\dag}_2\>\<\cO_1\cO'^{\dag}_1\> \nn \\
&=N_{n,j}^2\int_{x\approx x'} d^dxd^dx'D^{d-2}zD^{d-2}z'\<\cO_2\cO'^{\dag}_2\>\<\cD_{\mathrm{I}_2,n}\cO_1\cD_{\mathrm{I}_2,n}\cO'^{\dag}_1\>,
\ee
where we use the short-hand notation that $\cO_i$ is at point $(x,z)$ and $\cO'_i$ is at point $(x',z')$. Setting $\De_1\to \De_{\mathrm{I}_2,n}^*$ and $\De_2 \to \De_{\mathrm{II}_2,n}^*$, this equation can be rewritten as
\be
&(\De_2-\De_{\mathrm{II}_2,n}^*)\int_{x\approx x'} d^dxd^dx'D^{d-2}zD^{d-2}z'\lim_{\De_2 \to \De_{\mathrm{II}_2,n}^*}\frac{\<\cD_{\mathrm{II}_2,n}\cO_2\cD_{\mathrm{II}_2,n}\cO'^{\dag}_2\>}{\De_2-\De_{\mathrm{II}_2,n}^*}\<\cO_1\cO'^{\dag}_1\> \nn \\
&=N_{n,j}^2(\De_1-\De_{\mathrm{I}_2,n}^*)\int_{x\approx x'} d^dxd^dx'D^{d-2}zD^{d-2}z'\<\cO_2\cO'^{\dag}_2\>\lim_{\De_1\to \De_{\mathrm{I}_2,n}^*}\frac{\<\cD_{\mathrm{I}_2,n}\cO_1\cD_{\mathrm{I}_2,n}\cO'^{\dag}_1\>}{\De_1-\De_{\mathrm{I}_2,n}^*}.
\ee
Finally, using $(\De_2-\De_{\mathrm{II}_2,n}^*)=-(\De_1-\De_{\mathrm{I}_2,n}^*)$, we can conclude that
\be\label{eq:12adjoint_twopointrelation}
&\p{\lim_{\De_2 \to \De_{\mathrm{II}_2,n}^*}\frac{\<\cD_{\mathrm{II}_2,n}\cO_2\cD_{\mathrm{II}_2,n}\cO^{\dag}_2\>}{\De_2-\De_{\mathrm{II}_2,n}^*},\<\cO_1\cO^{\dag}_1\>}_L =-N_{n,j}^2\p{\<\cO_2\cO^{\dag}_2\>,\lim_{\De_1\to \De_{\mathrm{I}_2,n}^*}\frac{\<\cD_{\mathrm{I}_2,n}\cO_1\cD_{\mathrm{I}_2,n}\cO^{\dag}_1\>}{\De_1-\De_{\mathrm{I}_2,n}^*}}_L.
\ee
The second relation is about the integral
\be
\int d^dx_1d^dx_2d^dx'D^{d-2}z'\<\cD_{\mathrm{II}_{2},n}\cO^F(x',z')\cO^{F\dag}(x,z)\>F(x_1,x_2,x,z)\cT_2\<0|\cO_2\wL[\cO^\dag](x',z')\cO_1|0\>_{+}\th((1>2)\approx x'),
\ee
where $\cO^F$ has quantum numbers $(\De_F,J_F,j)$, $\cO$ has $(\De,J,j-n)$, and $F(x_1,x_2,x,z)$ is a conformally-invariant kernel that transforms as $\<\tl{\cO}^\dag_1\tl{\cO}^\dag_2\cD_{\mathrm{I}_{2},n}\wL[\cO]\>$. Note that $\De_F+n+(1-J)=d$ in order for the integral to be conformally-invariant. By applying~\eqref{eq:12adjoint}, we have for general $\De_F$
\be
&\int d^dx_1d^dx_2d^dx'D^{d-2}z'\<\cD_{\mathrm{II}_{2},n}\cO^F(x',z')\cO^{F\dag}(x,z)\>F(x_1,x_2,x,z)\cT_2\<0|\cO_2\wL[\cO^\dag](x',z')\cO_1|0\>_{+}\th((1>2)\approx x') \nn \\
&=N_{n,j}\int d^dx_1d^dx_2d^dx'D^{d-2}z'\<\cO^F(x',z')\cO^{F\dag}(x,z)\>F(x_1,x_2,x,z) \nn \\
&\qquad\qquad\qquad\qquad\qquad\qquad\x\cD_{\mathrm{I}_{2},n}\p{\cT_2\<0|\cO_2\wL[\cO^\dag](x',z')\cO_1|0\>_{+}\th((1>2)\approx x')}.
\ee
Now we set $\De_F\to \De_{\mathrm{II}_2,n}^*$ and $J\to J_I^*=\De_{\mathrm{II}_2,n}^*-d+1+n$, then $\cD_{\mathrm{I}_{2},n}$ and $\cD_{\mathrm{II}_{2},n}$ are conformally-invariant, and the above equation becomes
\be
&(\De_F- \De_{\mathrm{II}_2,n}^*)\int d^dx_1d^dx_2d^dx'D^{d-2}z'\lim_{\De_F\to \De_{\mathrm{II}_2,n}^*}\frac{\<\cD_{\mathrm{II}_{2},n}\cO^F(x',z')\cO^{F\dag}(x,z)\>}{\De_F- \De_{\mathrm{II}_2,n}^*}F(x_1,x_2,x,z) \nn \\
&\qquad\qquad\qquad\qquad\qquad\qquad\x\cT_2\<0|\cO_2\wL[\cO^\dag](x',z')\cO_1|0\>_{+}\th((1>2)\approx x') \nn \\
&=N_{n,j}(J-J_I^*)\int d^dx_1d^dx_2d^dx'D^{d-2}z'\<\cO^F(x',z')\cO^{F\dag}(x,z)\>F(x_1,x_2,x,z) \nn \\
&\qquad\qquad\qquad\qquad\qquad\qquad\x\lim_{J\to J_I^*}\frac{1}{J-J_I^*}\cD_{\mathrm{I}_{2},n}\p{\cT_2\<0|\cO_2\wL[\cO^\dag](x',z')\cO_1|0\>_{+}\th((1>2)\approx x')}.
\ee
Moreover, in the right-hand side we can apply \eqref{eq:volSO11nonsense} and finally obtain
\be\label{eq:12adjoint_threepointrelation}
&\int d^dx_1d^dx_2d^dx'D^{d-2}z'\lim_{\De_F\to \De_{\mathrm{II}_2,n}^*}\frac{\<\cD_{\mathrm{II}_{2},n}\cO^F(x',z')\cO^{F\dag}(x,z)\>}{\De_F- \De_{\mathrm{II}_2,n}^*}F(x_1,x_2,x,z) \nn \\
&\qquad\qquad\qquad\qquad\qquad\qquad\x\cT_2\<0|\cO_2\wL[\cO^\dag](x',z')\cO_1|0\>_{+}\th((1>2)\approx x') \nn \\
&=\vol(\SO(1,1))N_{n,j}\int d^dx_1d^dx_2d^dx'D^{d-2}z'\<\cO^F(x',z')\cO^{F\dag}(x,z)\>F(x_1,x_2,x,z) \nn \\
&\qquad\qquad\qquad\qquad\qquad\qquad\x\cD_{\mathrm{I}_{2},n}\p{\cT_2\<0|\cO_2\wL[\cO^\dag](x',z')\cO_1|0\>_{+}\th((1>2)\approx x')},
\ee
where we also use the fact that $(\De_F- \De_{\mathrm{II}_2,n}^*)=(J-J_I^*)$.

\subsection{Derivation for the higher transverse spin case}
\label{sec:highj_deriv}
Now we give the derivation for the higher-transverse spin terms, where $j>J_1+J_2$. We can follow the same steps for the $j\leq J_1+J_2$ until~\eqref{eq:Wmatrix}, but we should change~\eqref{eq:WArelation} to \footnote{Note that this is just a rewriting of \eqref{eq:WArelation}, but in \eqref{eq:WArelation} the operator $\cO$ has quantum numbers $(\de+1,J_1+J_2-1,j)$. Here we use the relation $\wL[\cO_{\de+1,J_1+J_2-1,j}]\propto \cD_{\mathrm{I}_{2},n}\wL[\cO_{\de+1,J_1+J_2-1+n,J_1+J_2}]$ and call the operator $\cO_{\de+1,J_1+J_2-1+n,J_1+J_2}$ as $\cO$. We hope that this does not cause confusion.}
\be
W_{\de,j}(x,z)=A_{b}(\de,j)\<0|\cO_4\cD_{\mathrm{I}_{2},n}\wL[\cO](x,z)\cO_3|0\>_{+}^{(b)},
\ee 
where $\cO$ has quantum numbers $(\De,J,j)=(\de+1,J_1+J_2-1+n,J_1+J_2)$ and $n=j-J_1-J_2$ in order for $\cD_{\mathrm{I}_{2},n}$ to be conformally-invariant. We expect that $A_b(\de,j)$ is related to the OPE data of $\cO_1\x\cO_2$ OPE. In particular, we want to show that $A_b(\de,j)$ is proportional to $C^+(\de+1,J_1+J_2-1+n,J_1+J_2)+C^-(\de+1,J_1+J_2-1+n,J_1+J_2)$. We follow the old derivation and take the Lorentz pairing with a dual structure:
\be\label{eq:Aexpr_highj}
A_b(\de,j)
&= \p{\p{\cT_4\<0|\cO_4 \cD_{\mathrm{I}_{2},n}\wL[\cO](x,z)\cO_3|0\>^{(b)}_+}^{-1},\cT_4W_{\de,j}(x,z)}_L \nn\\
&= \int_{\substack{4>1\\ 2>3}} \frac{d^d x_1 d^d x_2 d^d x_3 d^d x_4 D^{d-2} z_1 D^{d-2} z_2}{\vol\tl\SO(d,2)}  \<\Omega|\big[\cO_4,\cO_1(x_1,z_1)\big]\big[\cO_2(x_2,z_2),\cO_3\big] |\Omega\>\nn\\
&\qquad \x \cT_2^{-1} \cT_4^{-1} \Bigg[\int  d^d x D^{d-2}z \p{\cT_4\<0|\cO_4 \cD_{\mathrm{I}_{2},n}\wL[\cO](x,z)\cO_3|0\>^{(b)}_+}^{-1}\nn\\
&\qquad\qquad\qquad\qquad\qquad \x (\cT_2 \cL_{\de,j})(x_1,z_1,x_2,z_2;x,z)\th(4^+>1)\th(2^+>3)\Bigg]. 
\ee
In the OPE limit where $x_3,x_4\to x'$, the step functions $\th(4^+>1)\th(2^+>3)$ should become $\th((1>2)\approx x')$. Also, from the discussion in section \ref{sec:lowj_deriv}, we know that the three-point function $\p{\cT_4\<0|\cO_4 \cD_{\mathrm{I}_{2},n}\wL[\cO](x,z)\cO_3|0\>^{(b)}}^{-1}$ in the OPE limit is given by
\be\label{eq:34OPE_highj}
&\p{\cT_4\<0|\cO_4 \cD_{\mathrm{I}_{2},n}\wL[\cO](x,z)\cO_3|0\>^{(b)}_+}^{-1}\nn \\
&=B_{34\cO}^{\mathrm{finite}}(x_3,x_4,\ptl_{x'},\ptl_{z'})\<\cO^F(x',z')\cO^{F\dag}(x,z)\>\nn \\
&\quad+\cC_{34\cO'}(x_3,x_4,\ptl_{x'},\ptl_{z'})\<\cD_{\mathrm{II}_{2},n}\cO^F(x',z')\cO^{F\dag}(x,z)\>_{\mathrm{lim}}.
\ee 
As shown in \cite{Kologlu:2019mfz}, one gets zero after integrating the finite part $B_{34\cO}^{\mathrm{finite}}\<\cO^F\cO^{F\dag}\>$ against $\cT_2\cL_{\de,j}$. Therefore, we have
\be\label{eq:shadow_highj}
&\int d^dxD^{d-2}z\p{\cT_4\<0|\cO_4\cD_{\mathrm{I}_{2},n}\wL[\cO]\cO_3|0\>^{(b)}_+}^{-1}(\cT_2\cL_{\de,j}(x_1,z_1,x_2,z_2;x,z))\th(4^+>1)\th(2^+>3)\, \nn \\
&=\cC_{34\cO'}(x_3,z_3,x_4,z_4,\ptl_{x'},\ptl_{z'})\lim_{\De_F\to\De_{\mathrm{II}_2,n}^*}\frac{1}{\De_F-\De_{\mathrm{II}_2,n}^*}\, \nn \\
&\quad\x\int_{x\approx x'} d^dxD^{d-2}z\<\cD_{\mathrm{II}_{2},n}\cO^F(x',z')\cO^{F\dag}(x,z)\>\cT_2\cL_{\de,j}(x_1,z_1,x_2,z_2;x,z)\th((1>2)\approx x')\, \nn \\
&=\cC_{34\cO'}(x_3,z_3,x_4,z_4,\ptl_{x'},\ptl_{z'})(\cD_{\mathrm{II}_{2},n}\bold{S}[\cT_2\cL_{\de,j}])_{\mathrm{lim}}(x_1,z_1,x_2,z_2;x',z') \th((1>2)\approx x'),
\ee
where $\bold{S}$ represents the Lorentzian shadow transform, and we have defined
\be
&(\cD_{\mathrm{II}_{2},n}\bold{S}[\cT_2\cL_{\de,j}])_{\mathrm{lim}}(x_1,z_1,x_2,z_2;x',z') \nn \\
&=\lim_{\De_F\to\De_{\mathrm{II}_2,n}^*}\frac{1}{\De_F-\De_{\mathrm{II}_2,n}^*}\int_{x\approx x'} d^dxD^{d-2}z\<\cD_{\mathrm{II}_{2},n}\cO^F(x',z')\cO^{F\dag}(x,z)\>\cT_2\cL_{\de,j}(x_1,z_1,x_2,z_2;x,z).
\ee
This is a conformally-invariant three-point function that transforms like $\<\tl{\cO}^\dag_1\tl{\cO}^\dag_2\cD_{\mathrm{II}_{2},n}\cO^F\>$.
The expression in~\eqref{eq:shadow_highj} should then give a conformal block. To compute it, let us first act $\cD_{\mathrm{II}_2,n}$ on both sides of \eqref{eq:34OPE_highj}:
\be\label{eq:34OPEdiff_highj}
&\cD_{\mathrm{II}_2,n}\p{\cT_4\<0|\cO_4 \cD_{\mathrm{I}_{2},n}\wL[\cO](x,z)\cO_3|0\>^{(b)}_+}^{-1}\nn \\
&=\cC_{34\cO'}(x_3,x_4,\ptl_{x'},\ptl_{z'})\lim_{\De_F\to\De_{\mathrm{II}_2,n}^*}\frac{\<\cD_{\mathrm{II}_{2},n}\cO^F(x',z')\cD_{\mathrm{II}_{2},n}\cO^{F\dag}(x,z)\>}{\De_F-\De_{\mathrm{II}_2,n}^*}.
\ee
Using \eqref{eq:12adjoint} and the definition of the three-point dual structure $\p{\cT_4\<0|\cO_4\wL[\cO](x,z)\cO_3|0\>^{(b)}}^{-1}$, one can find
\be\label{eq:dualstruct_highj}
\cD_{\mathrm{II}_2,n}\p{\cT_4\<0|\cO_4 \cD_{\mathrm{I}_{2},n}\wL[\cO](x,z)\cO_3|0\>^{(b)}_+}^{-1}=N_{n,j}\p{\cT_4\<0|\cO_4\wL[\cO](x,z)\cO_3|0\>^{(b)}_+}^{-1}.
\ee
Combining \eqref{eq:shadow_highj}, \eqref{eq:34OPEdiff_highj}, and \eqref{eq:dualstruct_highj}, we obtain
\be
&\int d^dxD^{d-2}z\p{\cT_4\<0|\cO_4\cD_{\mathrm{I}_{2},n}\wL[\cO]\cO_3|0\>^{(b)}_+}^{-1}(\cT_2\cL_{\de,j}(x_1,z_1,x_2,z_2;x,z))\, \nn \\
&=N_{n,j}\frac{((\cD_{\mathrm{II}_{2},n}\bold{S}[\cT_2\cL_{\de,j}])_{\mathrm{lim}} \th((1>2)\approx x'))\p{\cT_4\<0|\cO_4\wL[\cO]\cO_3|0\>^{(b)}_+}^{-1}}{\lim_{\De_F\to\De_{\mathrm{II}_2,n}^*}\frac{\<\cD_{\mathrm{II}_{2},n}\cO^F\cD_{\mathrm{II}_{2},n}\cO^{F\dag}\>}{\De_F-\De_{\mathrm{II}_2,n}^*}}.
\ee
This is a conformal block whose exchanged operator has quantum numbers $(J_1+J_2+d-2+n,\de-d+2,J_1+J_2)$. Plugging this into \eqref{eq:Aexpr_highj} and comparing with the Lorentzian inversion formula, we find
\be
A_b(\de,j)&=-2\pi i\x\half(C^{+}_{ab}(\de+1,J_1+J_2-1+n,J_1+J_2)+C_{ab}^{-}(\de+1,J_1+J_2-1+n,J_1+J_2))\nn \\
&\quad\x N_{n,j}\frac{\<\wL[\cO]\wL[\cO^\dag]\>^{-1}}{\lim_{\De_F\to\De_{\mathrm{II}_2,n}^*}\frac{\<\cD_{\mathrm{II}_{2},n}\cO^F\cD_{\mathrm{II}_{2},n}\cO^{F\dag}\>}{\De_F-\De_{\mathrm{II}_2,n}^*}}\nn \\
&\quad\x\p{(\cD_{\mathrm{II}_{2},n}\bold{S}[\cT_2\cL_{\de,j}])_{\mathrm{lim}} \th((1>2)\approx x'),\cT_2\<0|\cO_2\wL[\cO^\dag]\cO_1|0\>^{(a)}_+}_L.
\ee
This result can be simplified by integrating $\cD_{\mathrm{II}_2,n}$ by parts. For the second line, using~\eqref{eq:12adjoint_twopointrelation} we have
\be
N_{n,j}\frac{\<\wL[\cO]\wL[\cO^\dag]\>^{-1}}{\lim_{\De_F\to\De_{\mathrm{II}_2,n}^*}\frac{\<\cD_{\mathrm{II}_{2},n}\cO^F\cD_{\mathrm{II}_{2},n}\cO^{F\dag}\>}{\De_F-\De_{\mathrm{II}_2,n}^*}}=N_{n,j}^{-1}\frac{\p{\lim_{J\to J_{\mathrm{I}}^*}\frac{\<\cD_{\mathrm{I}_2,n}\wL[\cO]\cD_{\mathrm{I}_2,n}\wL[\cO^\dag]\>}{J-J_{\mathrm{I}}^*}}^{-1}}{\<\cO^F\cO^{F\dag}\>},
\ee
where $J_{\mathrm{I}}^*=J_1+J_2-1+n$. For the third line, using~\eqref{eq:12adjoint_threepointrelation} we have
\be
&\p{(\cD_{\mathrm{II}_{2},n}\bold{S}[\cT_2\cL_{\de,j}])_{\mathrm{lim}} \th((1>2)\approx x'),\cT_2\<0|\cO_2\wL[\cO^\dag]\cO_1|0\>^{(a)}_+}_L \nn \\
&=\vol(\SO(1,1))N_{n,j}\int_{x\approx x'}\frac{d^d x_1 d^d x_2 d^d x' d^d x D^{d-2} z_1  D^{d-2} z_2 D^{d-2} z' D^{d-2} z}{\vol \tl \SO(d,2)}\<\cO^F(x',z')\cO^{F\dag}(x,z)\>\nn \\
&\quad\x\cT_2\cL_{\de,j}(x_1,z_1,x_2,z_2;x,z)\cD_{\mathrm{I}_2,n}\p{\cT_2\<0|\cO_2(x_2,z_2)\wL[\cO^\dag](x',z')\cO_1(x_1,z_1)|0\>^{(a)}_+\th((1>2)\approx x')}.
\ee
Plugging in the definition of $\cL_{\de,j}$ in \eqref{eq:Lkerneldef}, we find that $A_b(\de,j)$ is given by
\be
A_b(\de,j)&=-\pi i(C^{+}_{ab}(\de+1,J_1+J_2-1+n,J_1+J_2)+C_{ab}^{-}(\de+1,J_1+J_2-1+n,J_1+J_2))\nn \\
&\quad\x\frac{\p{\p{\lim_{J\to J_{\mathrm{I}}^*}\frac{\<\cD_{\mathrm{I}_2,n}\wL[\cO]\cD_{\mathrm{I}_2,n}\wL[\cO^{\dag}]\>}{J-J_{\mathrm{I}}^*}}^{-1},Q_{\de,j}^{(a)}}_L}{\vol \SO(1,1)},
\ee
where
\be
Q_{\de,j}^{(a)}(x,z,x',z')&=\a_{\de,j}\int D^{d-2}z_1 D^{d-2}z_2~\<\tl\cP_{\de_1}^\dag(z_1) \tl \cP_{\de_2}^\dag(z_2)\cP_{\de,j}(z)\>\nn \\
&\qquad\quad\x\<0|\wL^+[\cO_2](x,z_2)\cD_{\mathrm{I}_2,n}\wL[\cO^\dag](x',z')\wL^-[\cO_1(x,z_1)]|0\>^{(a)}_+,
\ee
where $\wL^+[\cO_2]$ indicates that the light transform contour is restricted to $2>x'$, and $\wL^{-}[\cO_1]$ is restricted to $1\approx  x'$. Finally, comparing the expression of $A_b(\de,j)$ with the old derivation for the lower transverse spin case, we find that the for higher transverse spin $j>J_1+J_2$, we have
\be
&\wL[\cO_1](x,z_1)\wL[\cO_2](x,z_2) \nn \\
&=\sum_{n=1}^{\oo}(-1)^n\int_{\frac{d-2}{2}-i\oo}^{\frac{d-2}{2}+i\oo}\frac{d\de}{2\pi i}\cC_{\de,j}^{(a)}(z_1,z_2,\ptl_z)\nn \\
&\qquad\qquad\qquad\qquad\x\p{\cD_{\mathrm{I}_2,n}\mathbb{O}^+_{\de+1,J_1+J_2-1+n,J_1+J_2(a)}(x,z)-\cD_{\mathrm{I}_2,n}\mathbb{O}^-_{\de+1,J_1+J_2-1+n,J_1+J_2(a)}(x,z)} \nn \\
&\qquad+\textrm{lower transverse spin},
\ee
where $n=j-J_1-J_2$. The celestial map formula for $\cC_{\de,j}^{(a)}$ is given by
\be
&\cC_{\de,j}^{(a)}(z_1,\bw_1,z_2,\bw_2,\ptl_{z_2},\ptl_{\bw_2})\p{\lim_{J\to J_{\mathrm{I}}^*}\frac{\<\cD_{\mathrm{I}_2,n}\wL[\cO](\oo,z_2,\bw_2)\cD_{\mathrm{I}_2,n}\wL[\cO^\dag](0,z_0,\bw_0)\>}{J-J_{\mathrm{I}^*}}}\nn \\
&=(-1)^n\<0|\wL^+[\cO_2](\oo,z_2,\bw_2)\cD_{\mathrm{I}_2,n}\wL[\cO^\dag](0,z_0,\bw_0)\wL^-[\cO_1](\oo,z_1,\bw_1)|0\>^{(a)}_+.
\ee
This result agrees with the second sum in~\eqref{eq:lightrayope_generalops}.

\section{Kernel of the celestial map}
\label{app:tensorstructureselectionrule}

In the main text we have given the celestial map formulas~\eqref{eq:ltsDefinitionSimple} and~\eqref{eq:htsDefinitionSimple} which map $d$-dimensional three-point tensor structures to OPE differential operators in $(d-2)$-dimensional space. The latter are in turn in one-to-one correspondence with three-point tensor structures in $(d-2)$-dimensional space. In general the space $T_{d}$ of three-point structures in $d$ and the space $T_{d-2}$ of three-point structures in $d-2$ dimensions have different dimensionality. Therefore, the celestial map $T_d\to T_{d-2}$ in general has non-trivial kernel or cokernel. In this appendix we identify a part $K_0\subseteq K$ of the kernel $K\subseteq T_d$ of this map and conjecture that it is in fact the entire kernel ($K=K_0$) and that the cokernel is trival, i.e.\ that the celestial map is surjective. We give support to this conjecture by matching the dimension of $T_d/K_0$ with the dimension of $T_{d-2}$. Finally, we consider the $\SO(d-2)$ representations $\l$ that can be generated by the celestial map and show that they cover all the representations appearing in the $(d-2)$-dimensional OPE.

We begin by identifying $K_0$. Let us start with the low transverse spin case~\eqref{eq:ltsDdefinition}. We are instructed to evaluate the structure from $T_d$ in configuration~\eqref{eq:celestiallocus} after multiplying by $V_{0,12}$. Since in this configuration $V_{0,12}$ vanishes, the only structures that survive are those which contain $V_{0,12}^{-1}$. Structures with more negative powers of $V_{0,12}$ will be singular and the structures with non-negative powers will vanish. We claim that no structures have more singular power of $V_{0,12}$ and that there is a simple rule for counting those with $V_{0,12}^{-1}$.

To see this, let us label the $\SO(d-1,1)$ irreps of the operators as $\r_1=(J_1,\l_1),\r_2=(J_2,\l_2)$ and $\r=(J,\l)$. For large $J$ the number of three-point tensor structures is $J$-independent. This is because the number of structures is given~\cite{Kravchuk:2018htv} by the dimension of
\be
	(\r_1\otimes\r_2\otimes\r)^{\SO(d-1)}.
\ee
The dependence on $J$ can be exhibited by first computing the $\SO(d-1)$ content of $\r_1\otimes\r_2$ and matching $\SO(d-1)$ irreps there to dual irreps in $\SO(d-1)$ content of $\r$. As $J$ is increased, all that happens is that new irreps appear in $\SO(d-1)$ decomposition of $\r$, but any given irrep appears at most once.\footnote{In what follows we use facts about dimensional reduction of $\SO(N)$ irreps. See, e.g., \cite{Kravchuk:2017dzd} for a review.} Since $\r_1\otimes\r_2$ contains finitely many $\SO(d-1)$ irreps, at some point the number of matching dual pairs stabilizes. 

As $J$ is decreased to sufficiently low values, some structures disappear from this counting. In terms of explicit expressions this happens because the structures at large $J$ depend on $V_{0,12}^{J-n}$ for various $n$, and as $J$ becomes less than $n$ such structures cease to be polynomial in $z$ and have to disappear from the above counting. We are interested in the maximal $n$ among all structures, which is therefore the same as the value of $J$ at which the number of structures stabilizes. As $J$ is increased by 1, the $\SO(d-1)$ content of $\r$ is appended by representations with the first row of length $J$. Since the maximal length of the first row of $\SO(d-1)$ representations in $\r_1\otimes\r_2$ is $J_1+J_2$,\footnote{To see this, note that the length of the first row of the Young diagram gives the maximal eigenvalue under any given boost. Since it is $J_i$ for $\r_i$, it must be $J_1+J_2$ for $\r_1\otimes \r_2$. Choosing a boost in (a complexification of) $\SO(d-1)$ yields the desired result.} it follows that the stable number of representations is achieved starting from at most $J=J_1+J_2$. This implies that $n\leq J_1+J_2$. This means that the power of $V_{0,12}$ is no smaller than
\be
	V_{0,12}^{J-J_1-J_2}.
\ee
This finishes the proof of the claim that structures analytically continued to $J=J_1+J_2-1$ have at most $V^{-1}_{0,12}$ singularity.

We thus find that the map~\eqref{eq:ltsDefinitionSimple} is well-defined and the structures with $V_{0,12}$ to non-negative powers get mapped to $0$. These structures constitute the set $K_0$. We conjecture that~\eqref{eq:ltsDefinitionSimple} is non-degenerate on the remaining structures in $T_{d}$, i.e.\ those which contain $V_{0,12}^{-1}$. To support this conjecture, let us count these structures and match their number to the number of structures in $T_{d-2}$.

To do that, note that the above discussion implies that the number of structures with $V_{0,12}^{-1}$ is precisely the difference between the number of polynomial structures for $J=J_1+J_2$ and the number of polynomial structures for $J=J_1+J_2-1$, i.e.\ the dimension of\footnote{Here we use the formal difference $\ominus$. To make this precise one can interpret all identities involving it as character identities.}
\be
	&(\r_1\otimes\r_2\otimes(J_1+J_2,\l))^{\SO(d-1)}\ominus (\r_1\otimes\r_2\otimes(J_1+J_2-1,\l))^{\SO(d-1)}\nn\\
	&=(\r_1\otimes\r_2\otimes\mathrm{Res}^{\SO(d-1,1)}_{\SO(d-1)}((J_1+J_2,\l)\ominus (J_1+J_2-1,\l)))^{\SO(d-1)}.
\ee
We can simplify this by noting that
\be
\mathrm{Res}^{\SO(d-1,1)}_{\SO(d-1)}((J_1+J_2,\l)\ominus (J_1+J_2-1,\l))=\bigoplus_{\tau\in \mathrm{Res}^{\SO(d-2)}_{\SO(d-3)}\l}(J_1+J_2,\tau).
\ee
In particular, this only involves $\SO(d-1)$ irreps with the first row of length $J_1+J_2$. As mentioned above, this is the maximal length of the first row in $\SO(d-1)$ irreps in $\r_1\otimes\r_2$, and this part of the tensor product $\r_1\otimes\r_2$ simplifies\footnote{One can see this by treating $\SO(d-1,1)$ irreps as shortened parabolic Verma modules of $\SO(d-1,1)$, in which case it is analogous to the statement that all primaries of dimension $\De_1+\De_2$ one can build out of primaries $\cO_1^a,\cO_2^b$ of dimensions $\De_1,\De_2$ are those given by decomposing $\cO_1^a\cO_2^b$ into irreducible Lorentz irreps.}
\be\label{eq:r1r2reduction}
\mathrm{Res}^{\SO(d-1,1)}_{\SO(d-1)}\r_1\otimes\r_2 = \bigoplus_{\tau\in \mathrm{Res}^{\SO(d-2)}_{\SO(d-3)}\l_1\otimes \l_2}(J_1+J_2,\tau)\oplus\cdots,
\ee
where the dots represent irreps with shorter first row.
By comparing the last two equations we see that 
\be\label{eq:ltsirrepresult}
&(\r_1\otimes\r_2\otimes(J_1+J_2,\l))^{\SO(d-1)}\ominus (\r_1\otimes\r_2\otimes(J_1+J_2-1,\l))^{\SO(d-1)}\nn\\
&=(\l_1\otimes\l_2\otimes \l)^{\SO(d-3)}
\ee
which is the same rule as for counting the structures in $T_{d-2}$.

We now turn to the higher transverse spin case. In this case the celestial map is applied to three-point structures with representations $(J_i,\l_i)$ and $(J,\l)$ where $J=J_1+J_2-1+n$ and $\l=(J_1+J_2,\g)$.
As discussed in section~\ref{sec:deltafunction}, the structures that contain $V_{0,12}$ to powers higher than the minimal possible $J-J_1-J_2$ are mapped to zero by the celestial map~\eqref{eq:htsDefinitionSimple}. Similarly to the above, we can determine the number of structures which contain $V_{0,12}^{J-J_1-J_2}$ by taking the difference between the number of polynomial structures at $J=J_1+J_2$ and $J=J_1+J_2-1$. However, since $\l=(J_1+J_2,\g)$, there are no polynomial structures for $J=J_1+J_2-1$. Therefore, all tensor structures with such $\l$ contain $V_{0,12}^{J-J_1-J_2}$. In this case $K_0$ is trivial and we simply would like to match the dimensions of $T_{d}$ and $T_{d-2}$.

Since $\l=(J_1+J_2,\g)$, the minimal length of the first row in $\SO(d-1)$ irreps contained in $(J,\l)$ with sufficiently large $J$\footnote{We need the number of analytically-continued tensor structures, which is the same as for very large $J$.} is $J_1+J_2$, in particular
\be
\mathrm{Res}^{\SO(d-1,1)}_{\SO(d-1)}(J,\l)=\bigoplus_{\tau\in \mathrm{Res}^{\SO(d-2)}_{\SO(d-3)}\l}(J_1+J_2,\tau)+\cdots,
\ee
where the dots represent irreps with longer first row. Taking into account~\eqref{eq:r1r2reduction}, we find
\be
&(\r_1\otimes\r_2\otimes(J,\l))^{\SO(d-1)}=(\l_1\otimes\l_2\otimes \l)^{\SO(d-3)}.
\ee
Since $\l_i$ has first row that is no larger than $J_i$, it follows that $\l_1\otimes \l_2$ only contains $\SO(d-3)$ irreps with first row at most of length $J_1+J_2$. Since the first row of $\l$ is already $J_1+J_2$, it makes no difference to increase it by $n$,
\be\label{eq:htsirreps}
&(\r_1\otimes\r_2\otimes(J,\l))^{\SO(d-1)}=(\l_1\otimes\l_2\otimes \l(+n))^{\SO(d-3)}.
\ee
This establishes the equality of dimensions of $T_d$ and $T_{d-2}$.

Finally, let us discuss which representations $\l$ can be generated through celestial map. The low transverse spin terms contain $\l$'s which are parts of the $\r=(J,\l)$ representations which appear in $\cO_1\times\cO_2$ OPE. We claim that they cover all $\l$'s that can appear in $(d-2)$-dimensional OPE and have first row length at most $J_1+J_2$. Indeed, using any such $\l$ in~\eqref{eq:ltsirrepresult} we find that
\be
	\dim (\r_1\otimes\r_2\otimes(J,\l))^{\SO(d-1)}\geq \dim (\l_1\otimes\l_2\otimes \l)^{\SO(d-3)}
\ee
for $J=J_1+J_2$ and thus for any larger $J$. Since the right-hand side is non-zero whenever $\l$ appears in $(d-2)$-dimensional OPE, and the left-hand side being non-zero implies that $(J,\l)$ appears in $d$-dimensional OPE, we obtain the desired result.

It remains to establish that the higher-transverse spin terms cover all $\l$'s in $(d-2)$-dimensional OPE with first row of length more than $J_1+J_2$. Any such $\l'$ can be represented as $\l(+n)$ where $\l$ has first row $J_1+J_2$. This can be then used in~\eqref{eq:htsirreps} to conclude that $(J,\l)$ appears in $\cO_1\x\cO_2$ OPE for generic $J$.

\section{Spinor Conventions}
\label{sec:spinor conventions}

We use conventions from \cite{Belitsky:2014zha} but switch to mostly plus signature, see appendix A in that paper. A four-dimensional vector $z^\mu = (z^0 , \vec z)$ is represented by $2 \times 2$ matrix
\be
z_{\alpha \dot \alpha} &= z_\mu (\bar \sigma^{\mu})_{\alpha \dot \alpha} =z^0 \sigma^0 + \vec z \cdot \vec \sigma , \nn \\
z^{\dot \alpha \alpha} &= z_\mu (\sigma^{\mu})^{\dot \alpha \alpha} =- z^0 \sigma^0 + \vec z \cdot \vec \sigma
\ee
where $\sigma^{\mu} =(1, \vec \sigma)$, $\bar \sigma^{\mu} = (-1, \vec \sigma)$ and $\vec \sigma = (\sigma^1, \sigma^2, \sigma^3)$.

Convention for lowering and raising of indices is
\be
x^{\dot \beta \beta} = \e^{\beta \alpha} x_{\alpha \dot \alpha} \e^{\dot \alpha \dot \beta} , \nn \\
a^{\alpha \beta} = \e^{\alpha \gamma} a_{\gamma \delta} \e^{\delta \beta} \ .
\ee
The Levi-Civita tensors are normalized as follows
\be
\e^{12} = \e_{12} = \e_{\dot 1 \dot 2} = \e^{\dot 1 \dot 2} = 1 \ .
\ee

The $X$ matrices are defined as follows
\be
(X_{i j k})_{\alpha \dot \alpha} = {(x_{ij})_{\alpha \dot \beta} \over x_{ij}^2} (x_{j k})^{\dot \beta \gamma}  {(x_{ki})_{\gamma \dot \alpha} \over x_{ik}^2} \ .
\ee
Using sigma matrix identities we can simplify
\be
X_{ijk} =- \frac{x_{ij}\cdot x_{jk} x_{ki\mu} + x_{jk}\cdot x_{ki} x_{ij\mu} - x_{ij}\cdot x_{ki} x_{jk\mu}}{x_{ij}^2 x_{ki}^2} \bar \sigma^\mu \, .
\ee 

\section{Distributional formulas}

In this appendix we formally derive some of the expressions involving distributions that were used in the main text.

\subsection{Analytic continuation of distributions}
\label{app:deltapole}

In this section we prove~\eqref{eq:deltasandgammas}, i.e.\ we study the analytic continuation of the distribution
\be\label{eq:stdistribution}
	(s+t)^a s^b t^c \theta(s)\theta(t)
\ee
from the region $a,b,c>0$, where it is represented by a locally-integrable function, to general complex $a,b,c$.

First, let us clarify the idea of analytic continuation of a distribution. Let us restrict to one-variable case with one parameter, i.e.\ we consider a distribution $g_a(x)$ defined for values of parameter $a\in U\subseteq \C$. Assume that this distribution depends on $a$ holomorphically. That is, for any test function $f(x)$ the pairing
\be
	\<g_a,f\> \equiv \int dx g_a(x)f(x)
\ee
is a holomorphic function of $a$. We say that a distribution $h_a(x)$ defined and holomorphic for $a\in V\subseteq \C$ is an analytic continuation of $g_a(x)$ if $U\subseteq V$ and for any test function $f(x)$ we have
\be
	\<h_a,f\> = \<g_a,f\>
\ee
for all $a\in U$. Similarly, we say that $g_a(x)$ is meromorphic for $a\in U$ if $\<g_a,f\>$ is meromorphic for any test function $f$ and the set of poles is independent of $f$ (of course, some poles may disappear for a specially chosen $f$). We say that $h(x)$ is the residue of $g_a(x)$ at $a_*$ if $\<h,f\>=\mathrm{res}_{a=a_*}\<g_a,f\>$ for any $f$, etc. All these notions generalize straightforwardly to the case of several variables and several parameters.

Before studying~\eqref{eq:stdistribution}, let us consider a simpler example,
\be
	g_a(x)=x^a\theta(x).
\ee
For $\mathrm{Re}\,a>-1$ this is an integrable function of $x$, and is holomorphic in $a$ as a distribution. We claim that it admits analytic continuation to $\C\setminus \Z_{<0}$ that is meromorphic in $\C$ with simple poles at negative integer $a$. 
As a simple example, consider a test function $f(x)$ that is equal to $e^{-x}$ for $x\geq 0$ and for $x<0$ is completed in some smooth way so that it decays quickly at $x\to -\oo$.\footnote{Everywhere in this section we can work with tempered distributions, so that $e^{-x}$ is an appropriate test function for $x>0$ (i.e.\ it is Schwartz).} We have then
\be
	\<g_a,f\> = \int_0^\oo e^{-x} x^a dx = \G(a+1),
\ee
which is indeed meromorphic and has simple poles at negative integer $a$. To see that this statement holds for more general test functions, recall that 
\be
	(x\pm i\e)^{a}
\ee
is a distribution that is an entire function of $a$.\footnote{The notation $(x\pm i\e)^{a}$ means the boundary value of $x^a$ on real line, approached either from above or below. Since $x^a$ has at most power-law singularity for any $a$, Vladimirov's theorem~\cite{Vladimirov} ensures that it's boundary values are well-defined tempered distributions in $x$, analytic in the parameter $a$. See, e.g.~\cite{Kravchuk:2020scc} for a review of these facts.} For $\mathrm{Re}\,a>-1$ we can write the equality of distributions
\be
	x^a\theta(x)=\frac{(x-i\e)^{a}e^{i\pi a}-(x+i\e)^{a}e^{-i\pi a}}{2i\sin \pi a},
\ee
where the right-hand side is in fact analytic for all $a\in \C\setminus \Z$. We can compute the residue at $a=-n$ as
\be
	&\mathrm{res}_{a=-n}\frac{(x-i\e)^{a}e^{i\pi a}-(x+i\e)^{a}e^{-i\pi a}}{2i\sin \pi a}=
	\frac{(x-i\e)^{-n}-(x+i\e)^{-n}}{2i\pi}\nn\\
	&=\frac{(-1)^{n-1}}{(n-1)!}\de^{(n-1)}(x).
\ee
In particular, this vanishes for $n\leq 0$, consistently with $g_a(x)$ being analytic for $\mathrm{Re}\,a>-1$. 

We conclude that $g_a(x)=x^a\theta(x)$ can be analytically continued so that it has simple poles at $a=-n$ with residues
\be
	\mathrm{res}_{a=-n} x^{a}\theta(x) = \frac{(-1)^{n-1}}{(n-1)!}\de^{(n-1)}(x).
\ee
Let us perform a simple check with the $f(x)$ defined above. We have
\be
	\mathrm{res}_{a=-n}\<g_a,f\>=\mathrm{res}_{a=-n}\G(a+1) = \frac{(-1)^{n-1}}{(n-1)!}
\ee
and this is indeed equal to
\be
	\<\mathrm{res}_{a=-n} g_a,f\>=\frac{(-1)^{n-1}}{(n-1)!}\<\de^{(n-1)},f\> = \frac{(-1)^{n-1}}{(n-1)!}.
\ee
Note that the non-trivial part of this analytic continuation is taking care of the singularity at $x=0$, since for $x>x_0>0$ the function $x^a\theta(x)$ is locally-integrable for any $a\in \C$.

This simple result for $x^a\theta(x)$ can be extended a more general setup: consider a finite set of $k$ smooth functions $q_i(x),\,x\in \R^n$ and consider the function
\be
	g_a(x)=\prod_{i=1}^k \theta(q_i(x))q_i(x)^{a_i}.
\ee
For $\mathrm{Re}\,\a_i>0$ this defines a locally-integrable function. Provided that the functions $q_i$ are in general position (clarified below), we claim that the distribution $g_a(x)$ can be analytically continued to a distribution meromorphic for $a\in \C^{k}$. To see this, suppose we want to define the analytic continuation in a neighborhood of some point $x_0$ where $r$ functions $q_{i_1}(x),\cdots,q_{i_r}(x)$ vanish. Provided that the matrix of derivatives
\be
	M_{jl}=\ptl_l q_{i_j}(x_0)
\ee
has rank $r$ (in particular, $r\leq n$), we can use $y_j\equiv q_{i_j}(x)$ as the first $r$ coordinates in a neighborhood $U$ of $x_0$. This condition is what we mean by ``general position'' above. With this choice of coordinates, we simply have
\be
	g_a(y)=\tl g_a(y)\prod_{j=1}^r y_j^{a_{i_j}}\theta(y_j),
\ee
where $\tl g_a(y)$ is a smooth function in $U$. Each of $y_j^{a_{i_j}}\theta(y_j)$ can be analytically continued as above, and we can take their product because they are distributions in different variables $y_j$. We can then finally multiply the resulting distribution by the smooth function $\tl g_a(y)$.

This more general result still doesn't apply to~\eqref{eq:stdistribution} because in~\eqref{eq:stdistribution} we have functions
\be
	q_1(s,t)=s,\quad q_2(s,t)=t,\quad q_3(s,t)=s+t
\ee
which are not in general position near $s,t=0$. The conclusion about analytic continuation, however, still holds. To establish it, one needs
to use a general result about resolution of singularities. We do not reproduce this argument, and instead refer to~\cite{AtiyahDistr,BGDistr}. Here we simply work out the required resolution in the concrete example of~\eqref{eq:stdistribution}. Our goal is to define the integral
\be\label{eq:todefine}
	\int ds dt (s+t)^a s^b t^c\theta(s)\theta(t) f(s,t)
\ee
for test functions $f$. To do so, we define new coordinates $u,v$ by
\be
	s=uv,\quad t=u(1-v).
\ee
The region $s,t>0$ is mapped one-to-one onto the region $u>0, 0<v<1$, and the integral~\eqref{eq:todefine}
can be written as
\be\label{eq:uvintegral}
	\int du dv u^{a+b+c+1}v^b(1-v)^c \theta(u)\theta(v)\theta(1-v) \tl f(u,v)
\ee
where 
\be\label{eq:uvtestfunction}
	\tl f(u,v)\equiv f(uv,u(1-v)).
\ee
Importantly, $\tl f(u,v)$ is a test function in $u,v$. Therefore, if we manage to define 
\be
u^{a+b+c+1}v^b(1-v)^c \theta(u)\theta(v)\theta(1-v)
\ee
as a distribution meromorphic in $a,b,c$, we are done. In this case, setting
\be
	q_1(u,v)=u, \quad q_2(u,v)=v,\quad q_3(u,v)=1-v,
\ee
we find that $q_i$ are in general position at all points $x_0$ where at least one function vanishes. (The coordinates $u,v$ ``resolve the singularity'' that we had at $s,t=0$.)
In particular, we find poles at
\be
	a+b+c+1=-n
\ee
with residues proportional to $\de^{(n)}(u)$, poles at
\be
	b=-n
\ee
with residues proportional to $\de^{(n)}(v)$, and poles at
\be
	c=-n
\ee
with residues proportional to $\de^{(n)}(1-v)$. As explained in the main text, we are interested in the pole near
\be
	a+b+c+1=-1,
\ee
in which case from the analysis above we get
\be
	u^{a+b+c+1}v^b(1-v)^c \theta(u)\theta(v)\theta(1-v)\sim \frac{1}{a+b+c+2}\de(u)v^b(1-v)^c\theta(v)\theta(1-v).
\ee
Now we only need to pull this back to $s,t$ coordinates, i.e.\ evaluate~\eqref{eq:uvintegral} with $\tl f$ given by~\eqref{eq:uvtestfunction}. We find
\be
&\int ds dt (s+t)^a s^b t^c\theta(s)\theta(t) f(s,t)\nn\\
&=\int du dv u^{a+b+c+1}v^b(1-v)^c \theta(u)\theta(v)\theta(1-v) \tl f(u,v)\nn\\
&\sim \frac{1}{a+b+c+2}\int du dv \de(u)v^b(1-v)^c \theta(v)\theta(1-v) \tl f(u,v)\nn\\
&=\frac{1}{a+b+c+2}\int dv v^b(1-v)^c \theta(v)\theta(1-v) \tl f(0,v)\nn\\
&=\frac{f(0,0)}{a+b+c+2}\int dv v^b(1-v)^c \theta(v)\theta(1-v) \nn\\
&=\frac{f(0,0)}{a+b+c+2}\frac{\G(b+1)\G(c+1)}{\G(b+c+2)}.
\ee
This implies that
\be
(s+t)^a s^b t^c\theta(s)\theta(t)\sim \frac{1}{a+b+c+2}\frac{\G(b+1)\G(c+1)}{\G(b+c+2)}\de(s)\de(t),
\ee
as stated in~\eqref{eq:deltasandgammas}.

\subsection{An identity}
\label{sec:distributionidentity}

In this section, we show
\be
\label{eq:thingweevaluate}
	\int_{-\oo}^{+\oo} \frac{dx}{(xy+1+i\e)^a} =-\frac{2\pi i}{a-1}\de(y)
\ee
We define the integral by analytic continuation from the region $\Re a > 1$. Thus, let us evaluate it assuming $\Re a > 1$. Suppose first that $y$ is nonzero. If $y>0$, the integrand is holomorphic in the upper-half plane for $x$. Furthermore, because $\Re a>1$, it decays sufficiently quickly at infinity that the integration contour can be deformed into the upper half-plane, giving zero. If $y<0$, a similar argument shows that the integral can be deformed into the lower half-plane, giving zero. It follows that the distribution (\ref{eq:thingweevaluate}) is supported at $y=0$.

Now consider the integral against a test function $f(y)$. Because (\ref{eq:thingweevaluate}) is supported at $y=0$, we can restrict the $y$ integral to the range $[-1,1]$ (or any finite-size interval containing the origin). We furthermore substitute the Taylor expansion of $f(y)$ and integrate term by term:
\be
\label{eq:substtaylor}
	\int_{-1}^1 dy \int_{-\oo}^{+\oo} \frac{dx}{(xy+1+i\e)^a} \sum_{n=0}^\oo \frac{f^{(n)}(0)}{n!} y^n
\ee
Let us evaluate the term proportional to $f(0)$. Swapping the order of integration, we have
\be
f(0) \int_{-\oo}^{+\oo} dx \int_{-1}^1 dy\frac{1}{(xy+1+i\e)^a} 
&=
f(0) \int_{-\oo}^{+\oo} dx \frac{(1+i\e+x)^{1-a}-(1+i\e-x)^{1-a}}{(1-a)x}
\ee
Because the integrand on the right-hand side is holomorphic at $x=0$, we can deform the contour so that it moves slightly above the origin (staying below the singularity at $x=1+i\e$). We denote this by $\int_{-\oo}^\oo \to \int_\curvearrowright$. After this deformation, we split the integrand into two terms
\be
\frac{f(0)}{1-a} \int_\curvearrowright dx \p{\frac{(1+i\e+x)^{1-a}}{x}-\frac{(1+i\e-x)^{1-a}}{x}}
\ee
The first term is holomorphic in the positive imaginary direction for $x$, so we can deform the contour that direction and obtain zero. The second term is holomorphic in the negative imaginary direction for $x$, except for the pole at $x=0$. Thus, we can deform the contour that direction and pick up only the residue at $x=0$. We obtain
\be
\label{eq:termf0}
\frac{2\pi i}{1-a} f(0).
\ee

Finally, consider the terms in (\ref{eq:substtaylor}) proportional to $f^{(n)}(0)$. For these terms, note that
\be
\frac{y^n}{(xy+1+i\e)^a} &\propto \ptl_x^n \frac{1}{(xy+1+i\e)^{a-n}}
\ee
This is a total derivative in $x$, and hence integrates to zero. Together with (\ref{eq:termf0}), this establishes (\ref{eq:thingweevaluate}).

\bibliographystyle{JHEP}
\bibliography{refs}

\end{document}